\documentclass[twocolumn,floatfix,times]{aastex63}
\accepted{June 20, 2020}
\submitjournal{ApJ}

\shorttitle{Hot Corinos in Orion PGCCs}
\graphicspath{{./}{figure/}}

\begin{document}

 \title{ALMA Survey of Orion Planck Galactic Cold Clumps (ALMASOP)\\ 
 I. Detection of New Hot Corinos with ACA}

\author[0000-0002-1369-1563]{Shih-Ying Hsu}
\email{seansyhsu@gmail.com}
\affiliation{National Taiwan University (NTU), Taiwan (R.O.C.)}
\affiliation{Institute of Astronomy and Astrophysics, Academic Sinica, Taipei, Taiwan (R.O.C.)}
\author[0000-0012-3245-1234]{Sheng-Yuan Liu}
\email{syliu@asiaa.sinica.edu.tw}
\affiliation{Institute of Astronomy and Astrophysics, Academic Sinica, Taipei, Taiwan (R.O.C.)}
\author[0000-0002-5286-2564]{Tie Liu}
\affiliation{Key Laboratory for Research in Galaxies and Cosmology, Shanghai Astronomical Observatory, Chinese Academy of Sciences, 80 Nandan Road, Shanghai 200030, People’s Republic of China}
\author[0000-0002-4393-3463]{Dipen Sahu}
\affiliation{Institute of Astronomy and Astrophysics, Academic Sinica, Taipei, Taiwan (R.O.C.)}
\author[0000-0001-9304-7884]{Naomi Hirano}
\affiliation{Institute of Astronomy and Astrophysics, Academic Sinica, Taipei, Taiwan (R.O.C.)}
\author[0000-0002-3024-5864]{Chin-Fei Lee}
\affiliation{Institute of Astronomy and Astrophysics, Academic Sinica, Taipei, Taiwan (R.O.C.)}
\author[0000-0002-8149-8546]{Ken'ichi Tatematsu}
\affiliation{Nobeyama Radio Observatory, National Astronomical Observatory of Japan, National Institutes of Natural Sciences, 462-2 Nobeyama, Minamimaki, Minamisaku, Nagano 384-1305, Japan}
\affiliation{Department of Astronomical Science, SOKENDAI (The Graduate University for Advanced Studies), 2-21-1 Osawa, Mitaka, Tokyo 181-8588, Japan}
\author[0000-0003-2011-8172]{Gwanjeong Kim}
\affiliation{Nobeyama Radio Observatory, National Astronomical Observatory of Japan, National Institutes of Natural Sciences, 462-2 Nobeyama, Minamimaki, Minamisaku, Nagano 384-1305, Japan}
\author[0000-0002-5809-4834]{Mika Juvela}
\affiliation{Department of Physics, P.O.Box 64, FI-00014, University of Helsinki, Finland}
\author[0000-0002-7125-7685]{Patricio Sanhueza}
\affiliation{National Astronomical Observatory of Japan, National Institutes of Natural Sciences, 2-21-1 Osawa, Mitaka, Tokyo 181-8588, Japan}
\author[0000-0002-3938-4393]{Jinhua He}
\affiliation{Yunnan Observatories, Chinese Academy of Sciences, 396 Yangfangwang, Guandu District, Kunming, 650216, P. R. China}
\affiliation{Chinese Academy of Sciences South America Center for Astronomy, National Astronomical Observatories, CAS, Beijing 100101, China}
\affiliation{Departamento de Astronom\'{i}a, Universidad de Chile, Casilla 36-D, Santiago, Chile}
\author[0000-0002-6773-459X]{Doug Johnstone}
\affiliation{NRC Herzberg Astronomy and Astrophysics, 5071 West Saanich Rd, Victoria, BC, V9E 2E7, Canada}
\affiliation{Department of Physics and Astronomy, University of Victoria, Victoria, BC, V8P 5C2, Canada}
\author[0000-0003-2302-0613]{Sheng-Li Qin}
\affiliation{Department of Astronomy, Yunnan University, and Key Laboratory of Astroparticle Physics of Yunnan Province, Kunming, 650091, People’s Republic of China}
\author[0000-0002-9574-8454]{Leonardo Bronfman}
\affiliation{Departamento de Astronom\'{i}a, Universidad de Chile, Casilla 36-D, Santiago, Chile}
\author[0000-0002-9774-1846]{Huei-Ru Vivien Chen}
\affiliation{Institute of Astronomy and Department of Physics, National Tsing Hua University, Hsinchu 30013, Taiwan }
\author[0000-0002-2338-4583]{Somnath Dutta}
\affiliation{Institute of Astronomy and Astrophysics, Academic Sinica, Taipei, Taiwan (R.O.C.)}
\author[0000-0002-5881-3229]{David J. Eden}
\affiliation{Astrophysics Research Institute, Liverpool John Moores University, iC2, Liverpool Science Park, 146 Brownlow Hill, Liverpool, L3 5RF, UK.}
\author[0000-0003-2069-1403]{Kai-Syun Jhan}
\affiliation{National Taiwan University (NTU), Taiwan (R.O.C.)}
\affiliation{Institute of Astronomy and Astrophysics, Academic Sinica, Taipei, Taiwan (R.O.C.)}
\author[0000-0003-2412-7092]{Kee-Tae Kim}
\affiliation{Korea Astronomy and Space Science Institute (KASI), 776 Daedeokdae-ro, Yuseong-gu, Daejeon 34055, Republic of Korea}
\affiliation{University of Science and Technology, Korea (UST), 217 Gajeong-ro, Yuseong-gu, Daejeon 34113, Republic of Korea}
\author{Yi-Jehng Kuan}
\affiliation{Department of Earth Sciences, National Taiwan Normal University, Taipei, Taiwan (R.O.C.)}
\affiliation{Institute of Astronomy and Astrophysics, Academic Sinica, Taipei, Taiwan (R.O.C.)}
\author[0000-0003-4022-4132]{Woojin Kwon}
\affiliation{Department of Earth Science Education, Seoul National University (SNU), 1 Gwanak-ro, Gwanak-gu, Seoul 08826, Republic of Korea}
\affiliation{Korea Astronomy and Space Science Institute (KASI), 776 Daedeokdae-ro, Yuseong-gu, Daejeon 34055, Republic of Korea}
\author[0000-0002-3179-6334]{Chang Won Lee}
\affiliation{Korea Astronomy and Space Science Institute (KASI), 776 Daedeokdae-ro, Yuseong-gu, Daejeon 34055, Republic of Korea}
\affiliation{University of Science and Technology, Korea (UST), 217 Gajeong-ro, Yuseong-gu, Daejeon 34113, Republic of Korea}
\author[0000-0003-3119-2087]{Jeong-Eun Lee}
\affiliation{School of Space Research, Kyung Hee University, 1732, Deogyeong-Daero, Giheung-gu Yongin-shi, Gyunggi-do 17104, Korea}
\author{Anthony Moraghan}
\affiliation{Institute of Astronomy and Astrophysics, Academic Sinica, Taipei, Taiwan (R.O.C.)}
\author[0000-0002-6529-202X]{M. G. Rawlings}
\affiliation{East Asian Observatory, 660 N. A’ohōkū Place, University Park, Hilo, HI 96720, USA}
\author[0000-0001-8385-9838]{Hsien Shang}
\affiliation{Institute of Astronomy and Astrophysics, Academic Sinica, Taipei, Taiwan (R.O.C.)}
\author[0000-0002-6386-2906]{Archana Soam}
\affiliation{SOFIA Science Center, USRA, NASA Ames Research Center, M.S.-12, N232, Moffett Field, CA 94035, USA}
\author{M.A. Thompson}
\affiliation{Centre for Astrophysics Research, School of Physics Astronomy \& Mathematics, University of Hertfordshire, College Lane, Hatfield, AL10 9AB, UK.}
\author[0000-0003-1665-6402]{Alessio Traficante} 
\affiliation{INAF-IAPS, via Fosso del Cavaliere, 100. 00133, Rome, IT}
\author{Yuefang Wu}
\affiliation{Department of Astronomy, School of Physics, Peking University, Beijing, 1000871, PR China}
\affiliation{Kavli Institute for Astronomy and Astrophysics, Peking University, Beijing, 100871, PR China}
\author[0000-0001-8227-2816]{Yao-Lun Yang}
\affiliation{Department of Astronomy, University of Virginia, Charlottesville, VA 22904, USA}
\author[0000-0003-2384-6589]{Qizhou Zhang}
\affiliation{Center for Astrophysics | Harvard \& Smithsonian, 60 Garden Street, Cambridge, MA 02138, USA}

\begin{abstract}
We report the detection of four new hot corino sources, G211.47-19.27S, G208.68-19.20N1, G210.49-19.79W and G192.12-11.10 from a survey study of Planck Galactic Cold Clumps in the Orion Molecular Cloud Complex with the Atacama Compact Array (ACA).
Three sources had been identified as low mass Class 0 protostars in the Herschel Orion Protostar Survey (HOPS). One source in the $\lambda$ Orionis region is firstly reported as a protostellar core.
We have observed abundant complex organic molecules (COMs), primarily methanol but also other oxygen-bearing COMs (in G211.47-19.27S and G208.68-19.20N1) and the molecule of prebiotic interest NH$_2$CHO (in G211.47-19.27S), 
signifying the presence of hot corinos.
While our spatial resolution is not sufficient for resolving most of the molecular emission structure, the large linewidth and high rotational temperature of COMs suggest that they likely reside in the hotter and innermost region immediately surrounding the protostar.
In G211.47-19.27S, the D/H ratio of methanol ([CH$_2$DOH]/[CH$_3$OH]) and the $^{12}$C/$^{13}$C ratio of methanol ([CH$_3$OH]/[$^{13}$CH$_3$OH]) are comparable to those of other hot corinos.
Hydrocarbons and long carbon-chain molecules such as c-C$_3$H$_2$ and HCCCN are also detected in the four sources, likely tracing the outer and cooler molecular envelopes. 
\end{abstract}


\section{Introduction\label{sec:Intro}}
Many Class 0/I low- (and also intermediate-) mass protostellar cores show considerable chemical diversity.
Those cores characterized by the presence of abundant saturated complex organic molecules (COMs) within a warm ($\sim$100~K) and compact ($<$ 100~au) region around the central protostar are called  ``hot corinos" \citep{2004Ceccarelli}.
Hot corinos seem qualitatively similar to ``hot cores'' also characterized by abundant COMs but associated with larger ($\sim$1000~au) and warmer ($\sim$300~K) regions in high-mass star formation sites; however, the abundances of the COMs in hot corinos can be, in some cases, orders of magnitude higher (fractional abundance respect to the hydrogen molecule $X\sim10^{-7}-10^{-9}$) than those in hot cores \citep{2018Ospina-Zamudio_Cep-E-mm}.

The current hot corino formation scenario can be described by three phases \citep{2009Herbst_COM_review}.
Initially, atoms and molecules in the gas phase of prestellar cores accrete onto dust grains and the 0th generation COMs (e.g. the COM precursor H$_2$CO, methanol CH$_3$OH, and possibly other COMs) are produced by grain-surface chemistry during this ``cold phase."
In the  ``warm-up phase", the photodissociated radicals (e.g. methoxy radical CH$_3$O and formyl radical HCO of CH$_3$OH and H$_2$CO, respectively) form larger complex molecules (e.g. methyl formate HCOOCH$_3$ and formic acid HCOOH) as the first-generation molecules via radical-radical reactions in the ice mantles.
The  ``hot corino phase" takes place when the temperature reaches $\sim$100~K from 10~K in the cold phase.
The icy mantles completely sublimate into the gas and the second-generation molecules are consequently produced via gas-phase reactions.

So far a good number of hot corinos have been identified, including 
IRAS 16293-2422 in the Rho Ophiuchi cloud complex \citep{2003Cazaux_IRAS16293-2422}, 
IRAS4A2, IRAS2A, IRAS4B and SVS13-A in the Perseus molecular cloud \citep{2004Bottinelli_IRAS4A, 2005Jorgensen_IRAS2, 2017Lopez-Sepulcre_IRAS4A2, 2007Bottinelli_IRAS4B, 2018Bianchi_SVS13-A}, 
B335 in Lynd L663 \citep{2016Imai_B335}, 
B1b in Barnard 1 \citep{2018Lefloch_B1b}, 
Cep E-mm in the Cepheus E molecular cloud \citep{2018Ospina-Zamudio_Cep-E-mm},
L483 in Serpens-Aquila Rift \citep{2017Oya_L483, 2019Jacobsen_L483_infalling},
Ser-emb 1, Ser-emb 8 and Ser-emb 17 in the Serpens Cluster B \citep{2019Martin-Domenech_Ser-emb-1, 2019Bergner_Ser-emb-8-_Ser-emb-17}, 
HH-212 in the Orion Molecular Cloud Complex  \citep{2016Codella_HH212, 2017Lee_HH212}
and BHR-71 IRS1 in the BHR-71 Bok globule  \citep{2020Yang_BHR71}.
In addition, a hot-corino-like atmosphere was implied toward IRAS4A1 in the Perseus 103 molecular cloud \citep{2019Sahu_IRAS4A1}.
In particular, the hot corino nature of most of these corinos were found through case studies toward specific sources rather than from surveys.

In addition to the hot corino chemistry, there is also the so-called  ``warm-carbon-chain chemistry (WCCC)" associated with protostellar warm ($\sim$ 30 K) infalling envelopes being abundant in unsaturated carbon-chain molecules \citep{2008Sakai_IRAS04368+2557, 2009Sakai_IRAS15398-3359}.
Although the hot corino chemistry and the warm-carbon chain chemistry exhibit two distinct behaviors, there are sources bearing signatures of both COMs and long-carbon chain molecules. \citet{2018Higuchi} carried out a survey toward 36 Class 0/I protostars in the Perseus molecular cloud complex and showed that the abundance ratios between CH$_3$OH, a hot corino tracer, and ethynyl (C$_2$H) and cyclopropenylidene (c-C$_3$H$_2$), both being proxies of WCCC, range within a continuous spectrum of 1--2 orders of magnitude.
While there is no distinct separation between the two types of chemical signatures, the column density ratios between ethynyl and methanol ([C$_2$H]/[CH$_3$OH]) appear to be correlated with the core locations in the cloud complex \citep{2018Higuchi}.

The Planck Galactic Cold Clumps (PGCC) catalog is an all-sky catalog consisting of cold clump candidates \citep{2016Planck_PGCC}.
At an angular scale of $\sim$ 5$'$, the Planck's resolution, these PGCCs appear dense (with molecular hydrogen column density $N(\mathrm{H}_2) >10^{20}\,\,\mathrm{cm}^{-2}$) and cold (10-20 K) regions potentially harboring star formation at their very early stages. 
Survey observations in 850~$\mu$m were previously conducted with the Submillimetre Common User Bolometer Array-2 (SCUBA-2) at the 15m James Clerk Maxwell Telescope (JCMT) toward 96 dense PGCCs (clump-averaged column density larger than $\mathrm{5\times10^{20} cm^{-2}}$) in Orion A, B and $\lambda$ Orionis \citep{2018Yi}, and as part of the JCMT large program  ``SCOPE: SCUBA-2 Continuum Observations of Pre-protostellar Evolution” \citep{2018Liu_TOP-SCOPE, 2019Eden_SCOPE}.
This 850~$\mu$m (dust) continuum survey identified 119 protostellar and starless cores within 40 Orion PGCCs \citep{2018Yi}.
These cores were further observed with the Nobeyama Radio Observatory (NRO) 45m telescope for studying their evolutionary stages gauged by, for example, their N$_2$D$^+$ abundances \citep[][Kim et al. 2020, submitted]{2017Tatematsu}.
On the basis of these surveys, we selected 72 dense and compact 850 $\mu$m continuum cores at early stages (starless cores with intense N$_2$D$^+$ emission and Class 0 protostellar cores with and without intense N$_2$D$^+$ emission) and carried out observations with the Atacama Compact Array (ACA) within an Atacama Large Millimeter/submillimeter Array (ALMA) project. This project, ALMA Survey of Orion Planck Galactic Cold Clumps (ALMASOP), aims to probe the onset of star formation (See Sect. \ref{sec:Obs}).
We conducted chemical studies on the sample and report in this paper the finding of four sources, being rich in molecular lines and harboring saturated COMs. Information about the ALMA observations over the full sample will be reported in a separate paper (Dutta et al. in preparation).

\section{Observations\label{sec:Obs}}

\begin{deluxetable*}{ccccccc}
\setlength{\tabcolsep}{2pt} 
\tablecaption{\label{tab:Obs}The parameters of the observations.}
\tabletypesize{\small}
\tablewidth{6pt}
\tablehead{
\colhead{Source Name} &
\colhead{$\alpha_{J2000}\;\;\delta_{J2000}$} &
\colhead{\begin{tabular}[c]{@{}c@{}}Total On-source\\Integration  Time\end{tabular}} &
\colhead{\begin{tabular}[c]{@{}c@{}}Beam \\ $\theta_\mathrm{max}$, $\theta_\mathrm{min}$, PA\end{tabular}} &
\colhead{\begin{tabular}[c]{@{}c@{}}$\sigma_\mathrm{CONT}$,~~$\sigma_\mathrm{Chn}$ \\ mJy~beam$^{-1}$ (mK)\end{tabular}} &
\colhead{Date} &
\colhead{\begin{tabular}[c]{@{}c@{}}Calibrator\\ Bandpass \& Flux, Phase\end{tabular}}
}
\startdata
G211.47-19.27S &
05:39:56.097 -07:30:28.403 &
210 sec &
8.79$''$, 4.20$''$, -76.5$^\circ$ &
~8.6 ~(5.4),~~37.1 (23.3) &
11/19 &
J0522-3627, J0607-0834
\\
\hline
\begin{tabular}[c]{@{}c@{}}G208.68-19.20N1\\ G210.49-19.79W\end{tabular} &
\begin{tabular}[c]{@{}c@{}}05:35:23.486 -05:01:31.583\\ 05:36:18.860 -06:45:28.035\end{tabular} &
\begin{tabular}[c]{@{}c@{}}500 sec\\ 300 sec\end{tabular} &
\begin{tabular}[c]{@{}c@{}}7.60$''$, 4.00$''$, -83.3$^\circ$\\ 7.65$''$, 4.02$''$, -82.4$^\circ$\end{tabular} &
\begin{tabular}[c]{@{}c@{}} 13.5 (10.3), 27.5 (20.9)  \\ ~1.9 ~(1.5),~~27.3 (20.5) \end{tabular} &
\begin{tabular}[c]{@{}c@{}}11/21\\ 11/27\\ 11/27\\ 11/27\\ 11/28\end{tabular} &
\begin{tabular}[c]{@{}c@{}}J0854+2006, J0607-0834\\ J0423-0120, J0542-0913\\ J0854+2006, J0607-0834\\ J0522-3627, J0542-0913\\ J0423-0120, J0542-0913\end{tabular}
\\
\hline
G192.12-11.10 &
05:32:19.540 +12:49:40.190&
480 sec &
6.86$''$, 4.81$''$, 80.7$^\circ$ &
~1.2 ~(0.9),~~20.1 (14.1) &
\begin{tabular}[c]{@{}c@{}}11/21\\ 11/26\end{tabular} &
\begin{tabular}[c]{@{}c@{}} J0423-0120, J0530+1331 \\ J0423-0120, J0530+1331 \end{tabular}
\\
\enddata
\tablecomments{
All the measurements were made in 2018.
The observations toward G208 and G210 were separated into five executions and G192 observation toward were separated into two executions.
The $\theta_\mathrm{max}$ and $\theta_\mathrm{min}$ are the FWHM of the synthesized beam along the major and minor axes, respectively. 
The PA is the beam position angle.
The $\sigma_\mathrm{CONT}$ and $\sigma_\mathrm{Chn}$ are the root-mean-square (RMS) noises of the measurement spanning over 7.5 GHz (continuum) and 1.129 MHz (spectral resolution)}, respectively.
The brightness temperatures of $\sigma_\mathrm{CONT}$ and $\sigma_\mathrm{Chn}$ are derived with 230 GHz and the synthesized beam area of each source (see Table \ref{tab:CONTResult}).  
\end{deluxetable*}

The observations were carried out with the ACA, which is an interferometer composed of twelve 7-meter antennas. 
They were conducted as a part of the  ALMA Cycle 6 project, (\#2018.1.00302.S; PI: Tie Liu), aiming to study the fragmentation of dense cores in the Orion Molecular Cloud Complex. 

The longest baseline was 49 m, which was about 37.7~k$\lambda$. The HPBW of the synthesized beam was $\sim$ 5.6${''}$ and the field of view was $\sim$ 43.2${''}$. Four spectral windows, centered at 216.6, 218.9,  231.0 and 233.0 GHz with an uniform bandwidth of 1875 MHz ($\sim$2500 km~s$^{-1}$) and a resolution of $\sim$1.129 MHz ($\sim$1.5 km~s$^{-1}$) were set.

The data were calibrated with the Common Astronomy Software Applications package \citep[CASA, ][]{2007McMullin_CASA} version 5.4.0-68 and its pipeline version 42030M (Pipeline-CASA54-P1-B). 
In the pipeline, the images were processed through tclean with auto-masking and the robust parameter of the Briggs weighting set to 0.5.
See ALMA Science Pipeline website\footnote{https://almascience.eso.org/processing/science-pipeline} for other parameters of auto-masking (e.g. sidelobethreshold=1.25, noisethreshold=5.0 and negativethreshold=0.0).

We report the result of four sources, which we identified as hot corinos within the whole sample (See Sect. \ref{sec:DiscHotCorinoID}). 
They are G211.47-19.27S, G208.68-19.20N1, G210.49-19.79W and G192.12-11.10, hereafter G211, G208, G210 and G192, respectively.
See Table \ref{tab:Obs} for the coordinates, date, on-source integration time, calibrators of the observations, the sizes and position angles of beams, and the root-mean-square (RMS) noise of the continuum and line images. 

We note that in this paper we use only the 7m ACA data for the hot corino identification. The ALMA 12m array data, which provide better angular resolutions, will be employed for further analyses, such as the structures of the line emitting regions, in a forthcoming paper.

\section{Results}
\subsection{Dust Continuum \label{sec:ResultCONT}}
The continuum images of the four sources are shown in Figure \ref{fig:Image_CONT_Mol}. 
All four sources are well detected with a single dominant components in the map centers and their peak brightness temperatures are all less than 1~K.
Such brightness temperatures are much less than the expected physical temperature (greater than 10 K in protostars) so their continuum emission is optically thin.
The images for the two brighter targets (G211 and G208) are severely dynamical-range limited (i.e. the residual sidelobe features dominate over the ideal theoretical noise, $5\times$ to $10\times$ in our cases), hence noisier than maps for the other two targets as indicated in Table \ref{tab:Obs}.
In addition to the compact component, there exists also an extended component in G210 and it is possibly the remnant of the envelope.
We applied a 2D Gaussian fit in CASA to the continuum and based on the fitting result, including the source-averaged flux, the peak positions, the integrated flux density and the apparent and de-convolved angular sizes, we calculated the molecular hydrogen column density ($N_\mathrm{H_2}$), the mass ($M$) and the radius ($R$) of the sources with formulae modified from Eq. A.26 and A.30 in \citet{2008Kauffmann_CONTFormula} for optically thin continuum emission.

The source-averaged molecular hydrogen column density $N_\mathrm{H_2}$ can be derived via:
\begin{equation}
N_\mathrm{H_2}=\frac{F^\mathrm{Int}_\mathrm{CONT}}{\Omega_S\,\mu_\mathrm{H_2}\,m_\mathrm{H}\,\kappa_\nu\,B_\nu(T_\mathrm{d})}
\end{equation}
where $F^\mathrm{Int}_\mathrm{CONT}$ is the total flux, $\Omega_S$ is the size of the deconvolved source image, $\mu_\mathrm{H_2}$ is the molecular weight per hydrogen molecule $\sim 2.8$, $m_\mathrm{H}$ is the mass of the hydrogen atom, $B_\nu(T_\mathrm{d})$ is the black-body radiation function evaluated at the dust temperature $T_\mathrm{d}$.
$\kappa_\nu$ is the dust opacity in the form of $\kappa_\nu=0.1(\nu/1\,\mathrm{THz})^\beta\,\mathrm{cm^2\,g^{-1}}$ where $\beta$ is the the dust opacity index \citep{1990Beckwith_DustOpacity}. 
The indices $\beta$ were assumed to be 1.70, which is the typical opacity index of cold clumps in the submillimeter band \citep{2018Juvela_dustIndex} and the resulting $\kappa_\nu$ at 1.3 mm is 0.0083 cm$^2$ g$^{-1}$.
Since the continuum emission of all the four sources are marginally resolved (i.e. the observed source size is larger but comparable to the synthesized beam size), the derived column densities could be lower limits.

The mass is estimated via:
\begin{equation}
M=\frac{F_\nu\,D^2}{\kappa_\nu\,B_\nu(T_\mathrm{d})}
\end{equation}
where $D$ is the distance to the source adopted from the PGCC catalog \citep{2016Planck_PGCC}.
The sources are marginally resolved and the radius of the sources were estimated via
\begin{equation}
R=\frac{1}{2}\,D\,\theta_S
\end{equation}
where $\theta_S$ is the deconvolved FWHM size of the source derived from the 2D fit to the continuum data. All the derived quantities are presented in Table \ref{tab:CONTResult}.
We note that the gas column density and mass are estimated with a range of dust temperatures at 50K, 100K, and 150K.

\begin{deluxetable*}{lrrrrcccccccc}
\setlength{\tabcolsep}{5pt} 
\tablecaption{The parameters and result of the continuum analysis.\label{tab:CONTResult}}
\tablewidth{2pt}
\tablehead{
\colhead{Source Name} & \colhead{$\theta_A$} & \colhead{$\theta_S$} & \colhead{$F_\mathrm{CONT}^\mathrm{Int}$} & \colhead{$D$} & \colhead{} & \colhead{ $N_\mathrm{H_2}$ ($\mathrm{cm^{-2}}$)} & \colhead{} & \colhead{} & \colhead{} & \colhead{$M$ ($\mathrm{M_{\odot}}$)} & \colhead{} & \colhead{$R$}  \vspace{-2mm} \\
\cline{6-8}
\cline{10-12}
\colhead{} & \colhead{($\mathrm{{}^{\prime\prime}}$)} & \colhead{($\mathrm{{}^{\prime\prime}}$)} & \colhead{($\mathrm{mJy}$)} & \colhead{($\mathrm{pc}$)} & \colhead{50 K} & \colhead{100 K} & \colhead{150 K} & \colhead{} & \colhead{50 K} & \colhead{100 K} & \colhead{150 K} & \colhead{($\mathrm{au}$)} 
}
\startdata
G211.47-19.27S & 6.07 & 2.22 & 552 & 415 & 1.63e+24 & 7.73e+23 & 5.06e+23 &  & 0.82 & 0.39 & 0.25 & 461 \\
G208.68-19.20N1 & 5.51 & 2.51 & 1530 & 415 & 3.55e+24 & 1.68e+24 & 1.10e+24 &  & 2.28 & 1.08 & 0.71 & 520 \\
G210.49-19.79W & 5.55 & 3.42 & 185 & 415 & 2.31e+23 & 1.09e+23 & 7.16e+22 &  & 0.28 & 0.13 & 0.09 & 709 \\
G192.12-11.10 & 5.74 & 2.87 & 199 & 400 & 3.52e+23 & 1.66e+23 & 1.09e+23 &  & 0.28 & 0.13 & 0.09 & 575 \\
\enddata
\tablecomments{
$\theta_A$ is the FWHM of the synthesized beam.
$\theta_S$ is the FWHM of the deconvolved source image.
$\theta_A$ and $\theta_S$ are the geometric mean of their maximum and minimum FWHM values, i.e. $\theta=\sqrt{\theta_\mathrm{max}\theta_\mathrm{min}}$.
$F_\mathrm{CONT}^\mathrm{Int}$ is the total flux integrated over the source. 
$N_\mathrm{H_2}$, $M$ and $R$ are the calculated column density of molecular hydrogen, the mass, and the radius of the sources, respectively.
The $N_\mathrm{H_2}$ and $M$ are estimated at three different dust temperatures, 50 K, 100 K and 150 K.
Note that column densities listed here could be lower limits. See Sect. \ref{sec:ResultCONT}
}
\end{deluxetable*}

According to the coordinates of the continuum peaks, we searched for known young stellar objects (YSOs) within $5{''}$ using SIMBAD\footnote{http://simbad.u-strasbg.fr/simbad/} \citep{2000Wenger_SIMBAD}.
We found that three of the four sources are likely to be associated with the Herschel Orion Protostar Survey (HOPS) objects \citep{2013Manoj_HOPS}, HOPS 288 for G211, HOPS 87 for G208 and HOPS 168 for G210, which are all located in the Orion A molecular cloud (see Table \ref{tab:SrcCoord}).
First, the offsets between each source and its corresponding HOPS object are small ($0.7{''}$, $2.17{''}$ and $0.84{''}$ for G211, G208 and G210, respectively).

Second, there are no other compact sources within each of their fields of view.
In contrast, G192 in $\lambda$ Orionis is firstly identified as a YSO. 
See Sect. \ref{sec:DiscSrcOverview} for more discussions.

\begin{deluxetable*}{lccclccccc}
\setlength{\tabcolsep}{6pt} 
\tablecaption{\label{tab:SrcCoord}The coordinates of the continuum peaks and the information of their corresponding HOPS objects and the protostellar classifications.  }
\tablewidth{2pt}
\tablehead{
\colhead{} & \colhead{} & \colhead{} & \colhead{} & \multicolumn{6}{c}{HOPS \citep{2016Furlan_HOPS_Class}} \\ [-2ex]
\colhead{Source Name} & \colhead{Cloud} & \colhead{$\mathrm{\alpha_{J2000}^{peak}}$} & \colhead{$\mathrm{\delta_{J2000}^{peak}}$} \\ [-2ex]
\cline{5-10}
\colhead{} & \colhead{} & \colhead{} & \colhead{} & \colhead{Index} & \colhead{$\mathrm{\alpha_{J2000}}$} & \colhead{$\mathrm{\delta_{J2000}}$} & \colhead{Class} & $L_\mathrm{bol}$ (L$_\odot$ ) & $T_\mathrm{bol}$ (K)
}

\startdata
G211.47-19.27S & Orion A & 05:39:55.988 & -07:30:27.593 & HOPS 288 & 05:39:55.95 & -07:30:28.0 & 0 & 135.47  & 48.6 \\
G208.68-19.20N1 & Orion A & 05:35:23.434 & -05:01:30.803 & HOPS 87 & 05:35:23.47 & -05:01:28.7 & 0 & 36.49  & 38.1 \\
G210.49-19.79W & Orion A & 05:36:18.965 & -06:45:23.355 & HOPS 168 & 05:36:18.93 &  -06:45:22.7 & 0 & 48.07  & 54.0 \\
G192.12-11.10 & $\lambda$ Orionis & 05:32:19.345 & +12:49:41.140 & \nodata & \nodata & \nodata & \nodata & \nodata & \nodata \\
\enddata
\end{deluxetable*}

\subsection{Molecular Emission \label{sec:ResultMolEmission.tex}}
We extracted the spectra at the continuum peak in the four spectral datacubes for each of the 72 sources in the project (see Sect. \ref{sec:Intro}).
We then scanned all the samples and identified only those containing rich spectral features including particularly multiple CH$_3$OH lines with excitation temperature higher than 100 K for further analysis.
With the above criteria, we identified four sources (G211, G208, G210, and G192) and present their spectra, respectively, in Fig. \ref{fig:Spectra_G211}, \ref{fig:Spectra_G208}, \ref{fig:Spectra_G210}, and \ref{fig:Spectra_G192}.
In total, around 154 molecular lines in G211, 80 lines in G208, 38 lines in G210 and 30 lines in G192 above 3$\sigma_\mathrm{Chn}$ were recognized.
We further applied Gaussian fitting over those transitions to obtain the observed frequency ($f^\mathrm{obs}$), peak brightness temperature ($T_\mathrm{p}$), velocity width ($\Delta v$), and the integrated intensity ($w$) of the identified lines.
The result of the Gaussian fitting to each line of each source is in Table \ref{tab:MolLineList_G211}, \ref{tab:MolLineList_G208}, \ref{tab:MolLineList_G210} and \ref{tab:MolLineList_G192} for G211, G208, G210, and G192, respectively.

For line identifications, we searched the Splatalogue\footnote{https://www.cv.nrao.edu/php/splat/}, which is an online database including the Jet Propulsion Laboratory Molecular Spectroscopy \citep[JPL\footnote{https://spec.jpl.nasa.gov/},][]{1988JPL}  and the Cologne Database for Molecular Spectroscopy \citep[CDMS\footnote{https://cdms.astro.uni-koeln.de/},][]{2005CDMS} catalog for astronomical spectroscopy, for possible molecule candidates.
We detected 25 molecular species (including isotopologues) in G211, 18 in G208, 18 in G210 and 16 in G192.
Some species were detected in all the four sources including CO, C$^{18}$O, OCS, $^{13}$CS, H$_2$S, HCCCN, DCN, DCO$^+$, D$_2$CO, H$_2$CO, c-C$_3$H$_2$, CCD, $^{34}$SO, HCOOH and CH$_3$OH. 
Two molecules were detected in three of the four sources. They are HNCO (not in G208) and SiO (not in G192).
CH$_3$CHO and HCOOCH$_3$ were detected in G211 and G208, but were not found in the other two sources.
SO$_2$ is detected in G211 and G208.
Among the four sources, G211 is the most line-rich target in which C$_2$H$_5$OH, NH$_2$CHO, and isotopologue species including $^{13}$CH$_3$OH, CH$_2$DOH, $^{13}$CH$_3$CN, and HC$^{13}$CCN were identified.
G210 and G192 seem to have similar molecular composition.
We consolidate and present in Table \ref{tab:TransList} the molecular line parameters, such as their rest frequencies, quantum numbers, excitation energies, etc., for the identified species and transitions.

Spectra of the molecule candidates were then modeled with the eXtended CASA Line Analysis Software Suite \citep[XCLASS,\footnote{https://xclass.astro.uni-koeln.de/}][]{2017Moller_XCLASS}, which is a CASA toolbox for the molecular line synthesis.
In myXCLASS, a function of XCLASS, each (gaseous) component of a molecule is assumed to be in its individual local-thermodynamic equilibrium (LTE).
It means that the source function is in the form of Planck function with an rotational temperature ($T_\mathrm{rot}$) which is expected to be the same as its kinetic temperature if all the levels are thermalized \citep{1999Goldsmith_popdiagram, 2017Moller_XCLASS}.
To execute myXCLASS, users need to define the rotational temperature ($T_\mathrm{rot}$), the size ($\theta_C$), the column density ($N$), the local-standard-of-rest velocity ($v_\mathrm{LSR}$) and the velocity width ($\Delta v$) of each component.
These parameters are assumed to be identical for all transitions of each molecular component.
By solving the radiative transfer equation for an isothermal object in one dimension with the given molecular component parameters, myXCLASS generates the synthetic spectrum and provides the transition list, intensities profiles and optical depth profiles of each component.
We further employed Modeling and Analysis Generic Interface for eXternal numerical codes \citep[MAGIX; ][]{2013Moller_MAGIX}, another CASA package, to optimizes these molecular component parameters within the given ranges through the XCLASS interface.
XCLASS also provides the transition information based on the CDMS and JPL databases.

In our study, we fixed the component sizes ($\theta_C$) to be the same as the deconvolved dust continuum (i.e. $\theta_C=\theta_S$) except for carbon monoxide C$^{18}$O, and optimized the other parameters using MAGIX.
We note that this size may be overestimated for molecules (e.g. COMs) with compact emission and lead to underestimation of their column densities.
For the extended molecules (e.g. carbon-chain molecules), this size may be underestimated but the column density averaged within the beam should be correctly estimated.
Some molecules (e.g. hydrogen sulfide H$_2$S and sulfur monoxide $^{34}$SO), which are commonly detected in star-forming regions, were detected with only one line so that their (rotation) temperature could not be estimated and were fixed to be 100 K for their molecular column density calculation.
We executed MAGIX for each molecule individually to find their correspondingly best-fit. With those numbers as inputs, we then executed a final run including all species for global optimization.
Table \ref{tab:MolList} show the results from MAGIX optimization for G211, G208, G210 and G192, respectively.
The parameter uncertainties estimated by the error estimation function of MAGIX are also presented.
We also mark in Table \ref{tab:MolList} the molecules which are unresolved in their moment 0 map. 
The hydrocarbons and the carbon-chain molecules are resolved while saturated COMs are not, indicating that they are tracing the cold and the warm/hot regions, respectively. 
The S-bearing molecules are in general unresolved as well and they might be tracing the central region of the outflows due to their large linewidths. 
See Sect. \ref{sec:Disc} for more discussions.

The population diagram (or rotational diagram) is a common tool for estimating the rotational temperature, and the column density \citep{1999Goldsmith_popdiagram}.
Figure \ref{fig:pop} shows the population diagram of CH$_3$OH in each source and the derived column densities and rotational temperatures are all consistent with the estimation of MAGIX. 
See Appendix \ref{sec:rotDiagram} for the detail.

\setlength{\tabcolsep}{2pt} 
\begin{deluxetable*}{c|cccc|cccc|cccc|cccc}
\tablecaption{\label{tab:MolList} Molecule component list calculated by MAGIX.}
\tablewidth{6pt}
\tabletypesize{\scriptsize}
\tablehead{
\colhead{} & \multicolumn{4}{c}{G211.47-19.27S} & \multicolumn{4}{c}{G208.68-19.20N1} & \multicolumn{4}{c}{G210.49-19.79W} & \multicolumn{4}{c}{G192.12-11.10}   \\[-2ex]
\colhead{Formula} & \colhead{$T_\mathrm{rot}$} & \colhead{$N$} & \colhead{ $v_\mathrm{LSR}$} & \colhead{$\Delta v$} & \colhead{$T_\mathrm{rot}$} & \colhead{$N$} & \colhead{ $v_\mathrm{LSR}$} & \colhead{$\Delta v$} & \colhead{$T_\mathrm{rot}$} & \colhead{$N$} & \colhead{ $v_\mathrm{LSR}$} & \colhead{$\Delta v$} & \colhead{$T_\mathrm{rot}$} & \colhead{$N$} & \colhead{ $v_\mathrm{LSR}$} & \colhead{$\Delta v$} \\ [-2ex]
\colhead{} & \colhead{$\mathrm{K}$} & \colhead{$\mathrm{cm^{-2}}$} & \colhead{\scriptsize$\mathrm{km\,s^{-1}}$} & \colhead{\scriptsize$\mathrm{km\,s^{-1}}$} & \colhead{$\mathrm{K}$} & \colhead{$\mathrm{cm^{-2}}$} & \colhead{\scriptsize$\mathrm{km\,s^{-1}}$} & \colhead{\scriptsize$\mathrm{km\,s^{-1}}$} & \colhead{$\mathrm{K}$} & \colhead{$\mathrm{cm^{-2}}$} & \colhead{\scriptsize$\mathrm{km\,s^{-1}}$} & \colhead{\scriptsize$\mathrm{km\,s^{-1}}$} & \colhead{$\mathrm{K}$} & \colhead{$\mathrm{cm^{-2}}$} & \colhead{\scriptsize$\mathrm{km\,s^{-1}}$} & \colhead{\scriptsize$\mathrm{km\,s^{-1}}$} 
}
\startdata
CH$_3$OH & 185$\pm$4 & 8.5E+16 (0.1)$^\dagger$ & 2.2 & 6.2 & 298$\pm$3 & 4.9E+15 (0.1)$^\dagger$ & 10.4 & 4.1 & 155$\pm$3 & 7.1E+15 (0.1)$^\dagger$ & 8.1 & 9.8 & 173$\pm$5 & 7.2E+15 (0.2)$^\dagger$ & 9.6 & 9.2 \\
$^{13}$CH$_3$OH & 156$\pm$6 & 9.2E+15 (0.1)$^\dagger$ & 2.0 & 7.7 & \nodata & \nodata & \nodata & \nodata & \nodata & \nodata & \nodata & \nodata & \nodata & \nodata & \nodata & \nodata \\
CH$_2$DOH & 67$\pm$5 & 2.3E+16 (0.1)$^\dagger$ & 1.8 & 6.0 & \nodata & \nodata & \nodata & \nodata & \nodata & \nodata & \nodata & \nodata & \nodata & \nodata & \nodata & \nodata \\
CH$_3$CHO & 127$\pm$4 & 1.4E+15 (0.1)$^\dagger$ & 2.4 & 4.9 & 223$\pm$5 & 5.6E+14 (0.1)$^\dagger$ & 11.0 & 2.8 & \nodata & \nodata & \nodata & \nodata & \nodata & \nodata & \nodata & \nodata \\
CH$_3$OCHO & 197$\pm$6 & 8.5E+15 (0.1)$^\dagger$ & 3.3 & 7.8 & 143$\pm$3 & 1.2E+15 (0.1)$^\dagger$ & 11.4 & 1.7 & \nodata & \nodata & \nodata & \nodata & \nodata & \nodata & \nodata & \nodata \\
C$_2$H$_5$OH & 304$\pm$6 & 1.7E+16 (0.1)$^\dagger$ & 1.5 & 7.2 & \nodata & \nodata & \nodata & \nodata & \nodata & \nodata & \nodata & \nodata & \nodata & \nodata & \nodata & \nodata \\
$^{13}$CH$_3$CN & 295$\pm$6 & 8.2E+13 (0.1)$^\dagger$ & 2.4 & 7.0 & \nodata & \nodata & \nodata & \nodata & \nodata & \nodata & \nodata & \nodata & \nodata & \nodata & \nodata & \nodata \\
CCD & 62$\pm$6 & 2.3E+14 (0.1) & 4.8 & 2.2 & 24$\pm$3 & 8.1E+13 (0.1) & 11.1 & 1.4 & 24$\pm$3 & 4.3E+13 (0.1) & 8.7 & 2.0 & 19$\pm$4 & 5.5E+13 (0.2) & 10.3 & 2.3 \\
c-C$_3$H$_2$ & 32$\pm$4 & 1.2E+14 (0.1) & 5.1 & 3.1 & 22$\pm$3 & 5.7E+13 (0.1) & 11.0 & 1.0 & 20$\pm$3 & 2.2E+13 (0.1) & 8.6 & 1.7 & 13$\pm$5 & 3.7E+13 (0.2) & 10.4 & 1.8 \\
H$_2$CO & 189$\pm$4 & 8.2E+15 (0.1) & 4.6 & 4.9 & 39$\pm$3 & 5.9E+14 (0.1) & 11.1 & 1.6 & 51$\pm$3 & 2.7E+14 (0.1) & 8.4 & 2.6 & 45$\pm$5 & 2.9E+14 (0.2) & 10.5 & 2.8 \\
D$_2$CO & 136$\pm$5 & 3.5E+14 (0.1)$^\dagger$ & 3.9 & 4.6 & 38$\pm$3 & 5.7E+13 (0.1) & 11.2 & 1.9 & 40$\pm$3 & 2.4E+13 (0.1) & 8.4 & 2.2 & 22$\pm$5 & 3.3E+13 (0.2) & 10.5 & 2.3 \\
DCO$^+$ & $100$ & 2.6E+13 (0.2) & 4.8 & 2.4 & $100$ & 2.9E+13 (0.1) & 11.3 & 1.8 & $100$ & 9.1E+12 (0.1) & 8.3 & 1.3 & $100$ & 1.8E+13 (0.2) & 10.4 & 2.0 \\
N$_2$D$^+$ & \nodata & \nodata & \nodata & \nodata & $100$ & 1.9E+12 (0.1) & 11.9 & 1.4 & \nodata & \nodata & \nodata & \nodata & \nodata & \nodata & \nodata & \nodata \\
HCOOH & $100$ & 9.2E+14 (0.1)$^\dagger$ & 1.5 & 4.3 & $100$ & 3.5E+13 (0.1)$^\dagger$ & 10.8 & 0.8 & $100$ & 7.7E+13 (0.1)$^{\dagger}$ & 8.1 & 6.5 & $100$ & 1.2E+14 (0.2)$^\dagger$ & 9.9 & 7.2 \\
DCN & $100$ & 4.0E+13 (0.1) & 4.3 & 4.8 & $100$ & 1.2E+13 (0.1) & 11.1 & 1.7 & $100$ & 7.3E+12 (0.1) & 8.5 & 3.1 & $100$ & 6.8E+12 (0.2) & 10.5 & 3.2 \\
HCCCN & $100$ & 5.0E+13 (0.1) & 4.3 & 5.3 & $100$ & 1.3E+13 (0.1)$^\dagger$ & 11.3 & 3.4 & $100$ & 7.1E+12 (0.1)$^\dagger$ & 8.2 & 4.9 & $100$ & 6.0E+12 (0.2)$^\dagger$ & 11.0 & 5.3 \\
HC$^{13}$CCN & $100$ & 1.6E+13 (0.2)$^\dagger$ & 5.0 & 5.9 & \nodata & \nodata & \nodata & \nodata & \nodata & \nodata & \nodata & \nodata & \nodata & \nodata & \nodata & \nodata \\
HNCO & 205$\pm$4 & 1.2E+15 (0.1)$^\dagger$ & 2.5 & 7.8 & \nodata & \nodata & \nodata & \nodata & 161$\pm$4 & 1.2E+14 (0.1)$^\dagger$ & 8.4 & 10.0 & $100$ & 3.8E+13 (0.2)$^\dagger$ & 10.8 & 5.5 \\
NH$_2$CHO & 295$\pm$6 & 3.6E+14 (0.1)$^\dagger$ & 2.5 & 7.7 & \nodata & \nodata & \nodata & \nodata & \nodata & \nodata & \nodata & \nodata & \nodata & \nodata & \nodata & \nodata \\
H$_2$S & $100$ & 2.1E+15 (0.1)$^\dagger$ & 3.2 & 5.8 & $100$ & 1.6E+15 (0.1)$^\dagger$ & 11.2 & 1.9 & $100$ & 1.7E+14 (0.1)$^\dagger$ & 9.2 & 3.8 & $100$ & 3.5E+14 (0.2)$^\dagger$ & 10.6 & 6.7 \\
$^{13}$CS & $100$ & 5.5E+13 (0.1)$^\dagger$ & 4.3 & 5.3 & $100$ & 3.2E+13 (0.1) & 11.1 & 1.8 & $100$ & 3.6E+12 (0.1)$^\dagger$ & 8.4 & 2.1 & $100$ & 6.3E+12 (0.2)$^\dagger$ & 10.4 & 3.4 \\
OCS & 70$\pm$4 & 2.1E+15 (0.1)$^\dagger$ & 3.0 & 5.8 & 122$\pm$3 & 4.3E+14 (0.1)$^\dagger$ & 11.5 & 2.1 & 127$\pm$6 & 2.3E+14 (0.1)$^\dagger$ & 8.5 & 7.3 & 86$\pm$5 & 2.4E+14 (0.2)$^\dagger$ & 10.5 & 7.7 \\
$^{34}$SO & $100$ & 2.1E+14 (0.1)$^\dagger$ & 4.8 & 6.9 & $100$ & 6.8E+13 (0.1) & 11.2 & 2.8 & $100$ & 3.8E+13 (0.1)$^\dagger$ & 11.5 & 13.9 & $100$ & 2.8E+13 (0.2)$^\dagger$ & 10.6 & 3.1 \\
SO$_2$ & $100$ & 3.0E+15 (0.2)$^\dagger$ & 4.1 & 12.7 & \nodata & \nodata & \nodata & \nodata & $100$ & 6.2E+14 (0.1)$^\dagger$ & 10.2 & 9.6 & \nodata & \nodata & \nodata & \nodata \\
SiO & $100$ & 5.9E+13 (0.1)$^\dagger$ & 8.9 & 13.1 & $100$ & 4.3E+13 (0.1)$^\dagger$ & 30.2 & 40.5 & $100$ & 3.3E+12 (0.1)$^\dagger$ & 8.9 & 5.1 & \nodata & \nodata & \nodata & \nodata \\
  &   &   &  & & $100$ & 2.9E+13 (0.1)$^\dagger$ & -16.1 & 16.7  &   &    &   &  &  &  &  &  \\
C$^{18}$O & $100$ & 1.3E+16 (0.1) & 5.0 & 3.2 & $100$ & 2.0E+16 (0.1) & 11.2 & 1.5 & $100$ & 1.4E+16 (0.1) & 8.3 & 1.7 & $100$ & 1.9E+16 (0.2) & 10.5 & 2.4 \\
\enddata
\tablecomments{
The $T_\mathrm{rot}$ is the rotational temperature and the $T_\mathrm{rot}$=100 K without any error interval is fixed in the MAGIX simulation.
The $N$ is the column density and the value in the brackets is their standard deviations in $\log_{10}N$. Their values 0.1-0.2 correspond to  fractional errors ranging around 25\%-70\%. 
The $v_\mathrm{LSR}$ is the local standard of rest velocity and the $\Delta v$ is the velocity width.
Their standard deviation values are around 0.4 km s$^{-1}$.
The component size of C$^{18}$O is assumed to be 8${''}$.
The SiO profile in G208 was assumed to be constituted of two components.
The daggers mark the unresolved molecules (i.e. observed source sizes is smaller than the synthesized beam size on the moment 0 map) and note that the column densities of these spatially unresolved molecules may be underestimated.
}
\end{deluxetable*}

\section{Discussion\label{sec:Disc}}

\subsection{Source Overview\label{sec:DiscSrcOverview}\label{sec:DiscOutflow}}
All four sources, except for G192, have been identified as YSOs in previous studies.
G211, G208, and G210 were first identified as YSOs (MGM2012 518, 2433, and 777, respectively) in a mid-IR (spanning 3-24 $\mu$m) survey toward the Orion A and B clouds conducted with the Spitzer Space Telescope \citep{2012Megeath_survey_MGM2012}.
G211 and G210 were also identified as YSOs (FKV2013 771 and 593, respectively) in a survey from $0.4$ to 24 $\mu$m toward the LYNDS 1641 cloud which is a star forming region in Orion A \citep{2013Fang_FKV2013_survey}.
G208, on the other hand, also coincides with OMC3/MMS6 \citep{1997Chini_HOPS87_OMC3-MMS6}.
These three sources were further classified as Class 0 protostars \citep{2016Furlan_HOPS_Class}. 

The three sources are also subjects of outflow studies in the literature \citep[e.g. ][]{2016Manoj_HOPS288_CO, 2016vanKempen_HOPS288_outflow, 2016Watson_HOPS87_MassOutflow, 2010Fischer_HOPS168_Infall, 2014Velusamy_HOPS168_outflow}.
From our observations, the carbon monoxide CO $J$=2-1 images show a clear bi-conical outflow morphology in G211 and G210. (See Figure \ref{fig:Image_CONT_Mol}).
In addition, silicon monoxide SiO is known as a good probe of shocks.
The spatial distribution of the SiO $J$=5-4 emission in G211 and G210 is consistent with their outflows (See Figure \ref{fig:Image_CONT_Mol}.).
Although no obvious outflow signature appears in our G208 CO $J$=2-1 image, \cite{2012Takahashi_HOPS87_outflow} and \cite{2019Takahashi_HOPS87_young} have detected a very compact ($<$ 5${''}$) bipolar molecular outflow in the CO $J$=3-2 and 2-1 lines, respectively, indicating the object being in an extremely young stage of formation.
Our detection of compact SiO emission is consistent with what was inferred by \citet{2012Takahashi_HOPS87_outflow}.

G192 is the only object in $\lambda$ Orionis among the four sources.
While it has not been associated with any known YSOs in the literature, our observations indicate 
the presence of bi-conical outflow in its CO $J$=2-1 emission.  

\subsection{Hot Corino Identification\label{sec:DiscHotCorinoID}}
A hot corino is normally identified according to their compact ($<$ 100~au), abundant and presumably passively thermal-evaporated COMs \citep{2009Herbst_COM_review}. 
Methanol (CH$_3$OH), as the root of the COMs in the grain origin scheme in particular, is considered as an important indicator of the presence of hot corinos.
Toward all the four sources (G211, G210, G192, and G208), we detected 
CH$_3$OH with very compact distribution (See Figure \ref{fig:Image_CONT_Mol}).
Similar to the continuum analysis, we tried to used CASA 2D Gaussian fitting to fit the CH$_3$OH integrated intensity maps.
The distributions of CH$_3$OH emission, even for the transition with the low excitation energy (E$_u$ $\sim$ 55~K) we have, are very compact as the fitting routine suggested the sources to be unresolved.
Additionally, the high rotational temperatures ($> 165$ K) are also consistent with and indicative of thermal evaporation of icy CH$_3$OH.
Unlike those, often extended, CH$_3$OH emission seen in cold, dark clouds \citep{1988Friberg_CH3OH_ISM}, cold, massive prestellar clumps \citep{2013Sanhueza_CH3OH_clump}, or outflow \citep{2008Araya_CH3OH_outflow}, the evidences we present suggest the four objects be hot corinos. 
Further studies with higher angular and spectral resolutions would help to solidify (or falsify) this classification.

\begin{deluxetable*}{ccccccccc}
\setlength{\tabcolsep}{6pt} 
\tablecaption{\label{tab:X_H2} The fractional column density with respect to $\mathrm{H_2}$ of the COMs in hot corinos. }
\tablewidth{6pt}
\tabletypesize{\normalsize}
\tablehead{
\colhead{$\times10^{-9}$} & \colhead{$\mathrm{CH_3OH}$} & \colhead{$\mathrm{CH_3CHO}$} & \colhead{$\mathrm{HCOOCH_3}$} & \colhead{HNCO} & \colhead{HCOOH} & \colhead{$\mathrm{NH_2CHO}$} & \colhead{$\mathrm{C_2H_5OH}$} & \colhead{Ref.} 
}
\startdata
G211.47-19.27S & 110 & 2.09 & 13.1 & 1.64 & 1.32 & 0.51 & 26.9 &  \\
G208.68-19.20N1 & 2.92 & 0.41 & 0.66 & \nodata & 0.03 & \nodata & \nodata &  \\
G210.49-19.79W & 64.9 & \nodata & \nodata & 1.16 & 0.76 & \nodata & \nodata &  \\
G192.12-11.10 & 43.5 & \nodata & \nodata & 0.26 & 0.83 & \nodata & \nodata &  \\
IRAS 16293-2422 B & 833 & 10.0 & 21.7 & \nodata & 4.67 & \nodata & 19.2 & 1, 2 \\
B335 & 380 & 2.40 & 4.60 & 17.0 & 4.70 & 0.40 & 3.80 & 3 \\
HH-212 & 160 & 3.90 & 8.40 & \nodata & 5.3 & 0.42 & 7.10 & 4 \\
\enddata
\tablecomments{
This fractional column density of CH$_3$OH may be affected by its opacity (See Sect. \ref{sec:DiscHotCorinoID}).}
\tablerefs{
1.~\citet{2018Jorgensen_IRAS16293-2422B_COM}; 
2.~\citet{2016Jorgensen_IRAS16293_NH2}
3.~\citet{2016Imai_B335}; 
4.~\citet{2019Lee_HH212_COM_atm}
}
\end{deluxetable*}

\begingroup
\setlength{\tabcolsep}{3pt} 
\begin{deluxetable}{ccccc}
\tablecaption{\label{tab:X_CH3OH} The fractional column density of COMs relative to CH3OH, [COM]/[CH$_3$OH].}
\tablewidth{6pt}
\tabletypesize{\small}
\tablehead{
\colhead{$\times10^{-3}$} & \colhead{$\mathrm{CH_3CHO}$} & \colhead{$\mathrm{HCOOCH_3}$} & \colhead{$\mathrm{C_2H_5OH}$} & \colhead{Ref.} 
}
\startdata
G211.47-19.27S & 18.9 & 119 & 243 &  \\
G211.47-19.27S\tablenotemark{$\dagger$} & 2.0 & 12.9 & 26.4 \\
G208.68-19.20N1 & 141 & 225 & \nodata &  \\
IRAS 16293-2422 B & 12.0 & 26.0 & 23.0 & 1 \\
B335 & 6.32 & 12.1 & 10.0 & 2 \\
HH-212 & 24.4 & 52.5 & 44.4 & 3\\
\enddata
\tablecomments{$^\dagger$ The column density of CH$_3$OH is derived from the column density of $^{13}$CH$_3$OH assuming $^{12}$C/$^{13}$C ratio is 70.}
\tablerefs{1.~\citet{2018Jorgensen_IRAS16293-2422B_COM}; 2.~\citet{2016Imai_B335}; 3.~\citet{2019Lee_HH212_COM_atm}}
\end{deluxetable}
\endgroup

Both COMs and carbon-chain molecules are detected in our four sources. \citet{2018Lefloch_B1b} proposed a classification scheme based on the ratio of the number of O-bearing molecular species over that of the hydrocarbon species.
While the scheme was not related to any chemical models but rather phenomenological, the definitions of the O-bearing and hydrocarbon molecules are in the type of $\mathrm{C_xH_yO_z}$ and $\mathrm{C_xH_y}$, respectively. 
In our samples, the values of this ratio are all larger than 1.9, which is the lower limit suggested by \citet{2018Lefloch_B1b} for a hot corino.
Note that this classification is the N- and S- bearing molecules are not involved in the definitions.

Table \ref{tab:X_H2} shows the fractional column density, $X$, of COMs relative to hydrogen. 
The fractional column densities of methanol $X$(CH$_3$OH) are larger than $10^{-8}$, which is typical of a hot corino, except in G208.
In addition to methanol, there are more COMs (e.g. acetaldehyde CH$_3$CHO and methyl formate HCOOCH$_3$) detected in G211 and G208.
In G208, $X$(CH$_3$CHO) and $X$(HCOOCH$_3$) are $4.1\times10^{-10}$ and $6.6\times10^{-10}$, respectively, and they are 5 and 20 times larger, respectively, in G211.
Ethanol C$_2$H$_5$OH is also detected in G211 and $X$(C$_2$H$_5$OH) is $2.69\times10^{-8}$, comparable to the value in IRAS 16293-2422 B \citep[][see Table \ref{tab:X_H2}]{2018Jorgensen_IRAS16293-2422B_COM}.
The brightness temperatures of all these COM emissions are well below $<1$ K, implying these emissions are optically thin. However, if some of these lines suffer severely from beam dilution, their opacities could be higher, which will lead to higher fractional column densities.

Table \ref{tab:X_CH3OH} shows the fractional column densities of COMs with respect to CH$_3$OH ([COM]/[CH$_3$OH]).
In IRAS 16293-2422 B, B335, and HH-212, [CH$_3$CHO]/[CH$_3$OH] is roughly half of [HCOOCH$_3$]/[CH$_3$OH] and [C$_2$H$_5$OH]/[CH$_3$OH].
In our sources, CH$_3$CHO and HCOOCH$_3$ are detected in G211 and G208, and C$_2$H$_5$OH is detected in G211.
The ratio [CH$_3$CHO]/[CH$_3$OH] in G211 is comparable to B335 and HH-212; however, [CH$_3$CHO]/[CH$_3$OH], [HCOOCH$_3$]/[CH$_3$OH], and [C$_2$H$_5$OH]/[CH$_3$OH] in G211 and G208 (without C$_2$H$_5$OH) are an order of magnitude larger than the typical values in other hot corinos.

The discrepancy in these ratios could result from a few possibilities. 
First, it may be due to different beam dilution factors of different COM emissions. 
In the current analyses, we have assumed that all COM emissions originate from the same extent as the dust continuum.
As noted earlier, the COM emissions in fact appear to be more compact than the continuum emission.
The COM emission size may have been over-estimated, leading to underestimations of the COM column densities.
We speculate a more compact CH$_3$OH emission size (than the continuum as well as HCOOCH$_3$ and C$_2$H$_5$OH), which may correspond to a higher CH$_3$OH column density and hence reduces [HCOOCH$_3$]/[CH$_3$OH], and [C$_2$H$_5$OH]/[CH$_3$OH].
Our current spatial resolution is not yet sufficient to fully resolve the region from where the COM emission is emitted.
Future investigations using the 12m ALMA data may help to validate this scenario.

Second, CH$_3$OH abundances in our sources may have been underestimated due to opacity.
A hint of this is the column density ratio of methanol and its isotopologues.
In G211, the ratio of methanol isotopologues $^{12}$C/$^{13}$C derived from [CH$_3$OH]/[$^{13}$CH$_3$OH] is $\sim$10, which is lower than the typical $\mathrm{^{12}C/^{13}C}\sim70$ in the local interstellar medium (ISM) \citep{2011Wirstrom_12C13C_70}.
The ratio appears comparable to [CH$_3$OH]/[$^{13}$CH$_3$OH] of other Class 0 hot corinos, e.g. 24 in HH-212 \citep[][see Table \ref{tab:Isotope}]{2019Lee_HH212_COM_atm}.
CH$_3$OH transitions with high optical depths have been observed in \cite{2015Taquet_IRAS2A_IRAS4A_COM_12C13C} and \cite{2019Lee_HH212_COM_atm} for the cases of IRAS 2A and HH-212, respectively.
If we assume that CH$_3$OH column density in G211 is seven times larger in order to make $^{12}$C/$^{13}$C compatible with the local ISM, the high [COM]/[CH$_3$OH] observed in our sources will be reduced.
In this case, however, the deuteration ratios of CH$_3$OH ([CH$_2$DOH]/[CH$_3$OH], Table \ref{tab:Isotope}) and [CH$_3$CHO]/[CH$_3$OH] become small as compared to others.
Instead, we can estimate those [COM]/[CH$_3$OH] ratios in G211 by utilizing the column density of $^{13}$CH$_3$OH as a proxy of CH$_3$OH, assuming $^{13}$CH$_3$OH emission is optically thin and a 
$^{12}$C/$^{13}$C of 70.
As shown in Table \ref{tab:X_CH3OH}, the [COM]/[CH$_3$OH] ratios in G211 appear comparable to the other hot corino sources.

Finally, \citet{2019Bergner_Ser-emb-8-_Ser-emb-17} compared the [COM]/[CH$_3$OH] ratio in hot corinos Ser-emb 1, Ser-emb 8, and Ser-emb 17 with other hot corinos and found that it spanned two orders of magnitude. 
They suggested this variation is the result of the local environment and/or the time-dependent warm-up chemistry.

The only COM detected in G210 and G192 is CH$_3$OH. 
Assuming other COMs such as CH$_3$CHO and HCOOCH$_3$ bear similar [COM]/[CH$_3$OH], their column densities would be below $\sim10^{14}$ cm$^{-2}$ and become hardly detected as is the case we have.

\subsection{Formaldehyde (H\texorpdfstring{$_2$}{Lg}CO) \label{sec:DiscH2CO}}
Once CO is frozen onto grain surfaces, formaldehyde (H$_2$CO) can be formed via hydrogen addition reactions from CO and then further forms CH$_3$OH, and their photodissociated radicals are parents of some COMs synthesized via both ice and gas-phase chemistry \citep{2009Herbst_COM_review}.
H$_2$CO and CH$_3$OH are detected in all our four sources and they seem to show similarly compact distributions (See Fig. \ref{fig:Image_CONT_Mol}).
Although H$_2$CO and CH$_3$OH are detected in similar regions, the excitation temperatures of the two species are different in G208, G210 and G192.
Assuming that their kinetic temperatures are the same, it indicates that their energy levels may not be thermalized.

H$_2$CO and its isotopologue, double-deuterated formaldehyde (D$_2$CO), are both detected in all our hot corino sources.
Their rotational temperatures are intermediate among the species detected with more than one transition and hence with temperatures estimated from XCLASS, and their linewidths are intermediate as well. 

\subsection{Formamide (NH\texorpdfstring{$_2$}{Lg}CHO) \label{sec:DiscNH2CHO}}
Studies show that formamide NH$_2$CHO is a potential key species in pre-biotic evolution.
It has been proposed as one of the main components of both (pre)genetic and (pre)metabolic processes \citep{2012Saladino}.
In G211, the fractional abundance with respect to H$_2$ of formamide ($X$(NH$_2$CHO)) and of isocyanic acid ($X$(HNCO)) are $1.3\times10^{-10}$ and $4.1\times10^{-10}$, respectively.
They are within the ranges (from $10^{-11}$ to $10^{-9}$ and from $10^{-12}$ to $10^{-8}$, respectively) in star-forming regions with H$_2$CO detections \citep{2015Lopez-Sepulcre_NH2CHO}.

The [NH$_2$CHO])/[HNCO] in G211 is about 0.32, which is comparable to 0.14 in the hot corino L483 \citep{2017Oya_L483}.
Furthermore, it was suggested that NH$_2$CHO is chemically related to HNCO and their abundances follow a tight correlation, $X(\mathrm{NH_2CHO})=0.04\,X(\mathrm{HNCO})^{0.93}$ in star-forming regions with H$_2$CO detection \citep{2015Lopez-Sepulcre_NH2CHO}.
$X$(NH$_2$CHO) in G211 is $5.1\times10^{-10}$, which is comparable to the value estimated by the above formula ($2.7\times10^{-10}$).
The formation mechanism of NH$_2$CHO supported by this tight correlation is the hydrogenation of HNCO in icy grain mantles \citep{2004Raunier_NH2CHO_HNCO, 2008Garrod_gas-grain-model, 2011Jones_NH2CHO_HNCO, 2015Lopez-Sepulcre_NH2CHO}.
However, a recent laboratory experiment suggested that this pathway of NH$_2$CHO forming from the hydrogenation of HNCO is insufficient \citep{2015Noble_NH2CHO}.

Another mechanism of NH$_2$CHO formation is via the reactions between NH$^+_4$ and H$_2$CO in the gas-phase: 
(1) the radiative association reaction and the followed dissociative recombination of NH$_2$CHO$^+$ \citep{2007Quan_NH2CHO_H2CO},
(2) the ion–molecule reaction and the followed subsequent electron recombination of NH$_3$CHO$^+$ \citep{2011Halfen_NH2CHO_H2CO} and 
(3) the reaction between amidogen and formaldehyde \citep[NH$_2$+H$_2$CO, ][]{2015Barone_NH2CHO_gas}.
In HH-212, the NH$_2$CHO and the D$_2$CO have similar spatial distributions.
It indicates the possible correlation between NH$_2$CHO and H$_2$CO and further supports the formation of NH$_2$CHO from H$_2$CO in gas-phase \citep{2017Lee_HH212}.
Due to the limitation of the spatial resolution, we are not able to examine the spatial correlation between them.

Finally, a laboratory experiment show that NH$_2$CHO is also possibly formed via the icy mixtures of carbon monoxide and ammonia (i.e. CO:NH$_3$) irradiated by energetic particles such as electrons \citep{2011Jones_NH2CHO_ice}.
Moreover, the UV photoprocessing of CO:NH$_3$ and CO:CH$_4$ ice samples produces NH$_2$CHO and CH$_3$CHO, respectively, and the predicted abundance of NH$_2$CHO is 2 to 16 times larger than CH$_3$CHO \citep{2020Martin-Domenech_NH2CHO_CH3CHO_UV}. 
It is opposite to the current observations via the other hot corinos as well as our studies \citep[ e.g. $\sim$0.1 in HH-212, ][]{2019Lee_HH212_COM_atm} so an additional chemical mechanism may be required. 

In G210 and G192, $X$(HNCO) are $1.2\times10^{-9}$ and $2.6\times10^{-10}$, respectively, which are within the range presented by \citet{2015Lopez-Sepulcre_NH2CHO}.
Their $X$(NH$_2$CHO) are estimated to be $\sim10^{-10}$ and their column densities of NH$_2$CHO are therefore estimated to be in the order of $\sim10^{13}$, which may be insufficient to be detected.

\subsection{Hydrocarbons and Long Carbon-Chain Molecules \label{sec:DiscWCCC}}
Deuterated ethynyl (CCD), cyclopropenylidene (c-C$_3$H$_2$) and cyanoacetylene (HCCCN and HC$^{13}$CCN in G211), which bear carbon-chains or carbon-rings, are also detected.
The estimated rotational temperatures of CCD and c-C$_3$H$_2$ are distinctly cooler: $\sim$25~K and $\sim$15~K, respectively, in G208, G210 and G192, and slightly warmer ($\sim$63~K and $\sim$32~K, respectively) in G211.
It is consistent with the trend that temperatures are warmer in G211 for all the species.
The cooler temperatures of these species are also compatible with their more extended emission presented in Fig. \ref{fig:Image_CONT_Mol}.
Meanwhile, their linewidths are narrow ($\Delta v<$2.5 km~s$^{-1}$ in 208, G210 and G192 and $\Delta v <$3.5 km~s$^{-1}$ in G211) compared with the linewidths of the COMs and the outflow tracers.

HCCCN, a long carbon-chain molecule, is detected in all four sources, but its linewidths are wider than that of CCD and c-C$_3$H$_2$.
Assuming the rotational temperature of HCCCN to be 100K, the column density of HCCCN is in general higher than that of c-C$_3$H$_2$ and CCD within an order of magnitude.

The HC$^{13}$CCN, which is an isotopologue of HCCCN, is detected in G211 only. 
Under the assumption of $T_\mathrm{rot}=100$ K, the ratio [HC$^{13}$CCN]/[HCCCN]$\sim$0.33.
While here we assumed an excitation temperature higher than the calculated value of the other carbon-chain molecule (i.e. CCD), the derived isotopic ratio is meaningful. This is because the column density or abundance ratios will be simply very close to the intensity ratio of the 
$J$=24-23 (with similar molecular parameters) seen in both isotopologues.
It is an order of magnitude higher than what was reported by \cite{2016Araki_12C13CRatio} in the low-mass star-forming region L1527.

\subsection{Deuterations \label{sec:DiscDHRatio}}
The deuterium fractionation (D/H ratio) has been suggested as an indicator of the gas temperature in (low mass) star formation.
This D/H ratio is in general anti-correlated with the temperature during the cold prestellar phase \citep{2000Roberts_D_MolCloud, 2018Persson_H2CO_D2CO_IRAS16293-2422B}.
The deuterium fractionation then decreases through the protostellar evolution \citep{2014Taquet_DHModel, 2017Bianchi_D_stage_SVS13-A}.
Our measurements of D/H from H$_2$CO appear to follow this trend since the [D$_2$CO]/[H$_2$CO] of the hottest source (G211) is the smallest. 
The [D$_2$CO]/[H$_2$CO]=0.039, 0.091, 0.089 and 0.112 for G211, G208, G210 and G192, respectively.
These values are comparable to those of other hot corinos \citep[e.g. 0.046 in IRAS 4B;][]{2006Parise_D_H2CO_CH3OH_Class0} except IRAS 16293-2422 B, which is an order of magnitude lower than others (see Table \ref{tab:Isotope}).
Since the temperature of H$_2$CO in IRAS 16293-2422 B is higher \citep[$\sim$107~K;][]{2018Persson_H2CO_D2CO_IRAS16293-2422B} compared to the sources we presented, it seemingly suggests that the evolution stage of the sources we presented is in general earlier than that of IRAS 16293-2422 B, which is estimated to be $\sim 10^5$ yr \citep{2018Persson_H2CO_D2CO_IRAS16293-2422B}.
We note, however, that deuterated methanol is also detected in G211, and the D/H ratio of methanol ([CH$_2$DOH]/[CH$_3$OH]) is 0.27, slightly lower than that of IRAS 16293-2422 B.
This trend is the opposite to the case for formaldehyde.
The use of D/H for diagnosing or differentiating evolutionary ages of objects thus needs to be applied with caution.
We note that correcting for the optical thickness of CH$_3$OH may further reduce the derived [CH$_2$DOH]/[CH$_3$OH] values.

\begin{deluxetable}{ccccc}
\tablecaption{\label{tab:Isotope}The column densities ratio between the isotopes of the formaldehyde and methanol in other hot corinos.}
\tablewidth{6pt}
\tabletypesize{\normalsize}
\tablehead{
\colhead{} & \colhead{$\mathrm{\frac{[D_2CO]}{[H_2CO]}}$} & \colhead{$\mathrm{\frac{[CH_2DOH]}{[CH_3OH]}}$} & \colhead{$\mathrm{\frac{[CH_3OH]}{[^{13}CH_3OH]}}$} & Ref.
}
\startdata
G211.47-19.27S & 0.043 & 0.27 & 9.2 & \\
IRAS 16293-2422 B & 0.006 & 0.37 & \nodata & 1 \\ 
HH-212 & \nodata & 0.12 & 24 & 2 \\
IRAS 4B & 0.046 & 0.43 & \nodata & 3 \\
IRAS 2A & 0.052 & 0.62 & 26 & 3, 4 \\
\enddata
\tablerefs{1.~\citet{2017Lopez-Sepulcre_IRAS4A2}; 
2.~\citet{2019Lee_HH212_COM_atm}; 
3.~\citet{2006Parise_D_H2CO_CH3OH_Class0}; 
4.~\citet{2015Taquet_IRAS2A_IRAS4A_COM_12C13C}; 
}
\end{deluxetable}

\subsection{{S-bearing Molecules} \label{sec:DiscSulfur}}
Several S-bearing molecules are commonly detected toward the four sources.
They mostly have wide linewidths except for carbon monosulfide $^{13}$CS.
The abundance ratio [$^{13}$CS]/[H$_2$S] is $\sim$0.02 in all the four sources.
In \citet{2018Drozdovskaya_sulfur} and \citet{2019LeGal_sulfur_disk}, the abundance ratio of sulfur-bearing molecules with respect to CS in IRAS 16293-2422 B and other environments was presented but the uncertainties made it difficult to make the comparison.

The formation or synthesis processes of those S-bearing molecules remains unclear.
Models proposed that sulfur monoxide SO and sulfur dioxide SO$_2$ are formed in the gas phase from hydrogen sulfide H$_2$S and carbonyl sulfide OCS evaporated from grain mantles \citep{1997Palumbo_OCS, 1998Hatchell_SBearing, 2014Esplugues_H2S}.

Recently, \citet{2019Luo_S} presented opposite temperature dependencies in the abundances of carbon-sulfur compounds (e.g. $^{13}$CS and OCS) and carbon-free sulfur-bearing species (e.g. H$_2$S, $^{34}$SO and SO$_2$) in the Orion KL region.
Unfortunately our sample of four hot corinos is too small to investigate these effects within the Orion Molecular Cloud and larger observational samples are required \citep{2019LeGal_sulfur_disk}.

\subsection{{Trends and Comparisons} \label{sec:DiscComparison}}
Toward the four targets, tracers like CO and SiO show the broadest linewidths, clearly marking energetic outflow activities.
While COMs emission are not spatially resolved, their linewidths are also wide and it is especially true for CH$_3$OH. 
\citet{2017Lee_HH212} and \citet{2019Lee_HH212_COM_atm} imaged with ALMA a set of COM emission from the atmosphere of a Keplerian rotating disk around the central YSO. 
All four sources we are studying, as alluded earlier, are associated with molecular outflows. 
They therefore could have circumstellar disks that mediate the accretion and the launching of the outflows.
It is tempting to speculate if the COMs we detected are of a similar origin.
If the linewidths are really as a result of rotating motion in the putative circumstellar disks, the broader linewidths of CH$_3$OH suggest that it could originate from the inner part of an incipient disk, as compared to, for example, H$_2$CO.
This is in general consistent with the rotational temperature of CH$_3$OH being higher than that of H$_2$CO.
On the other hand, the very presence of the outflows also implies the alternative possibility that our observed COMs like CH$_3$OH are in fact related to the shock activities at the very base of the molecular outflows in the vicinity of the central YSO.

\begin{figure*}
\centering
\caption{\label{fig:Image_CONT_Mol} The moment 0 images of (from left to right, from top to bottom) 
continuum, 
CO (2-1, $f_\mathrm{rest}$=230538 MHz, $E_u$=16 K), 
SiO (5-4, $f_\mathrm{rest}$=217105 MHz, $E_u$=31 K),
CH$_3$OH (5(1,4)-4(2,3) E, $f_\mathrm{rest}$=216946 MHz, $E_u$=55 K),
H$_2$CO (3(2,2)-2(2,1), $f_\mathrm{rest}$=218476 MHz, $E_u$=68 K)
and c-C$_3$H$_2$ (3(3,0)-2(2,1), $f_\mathrm{rest}$=216279 MHz, $E_u$=68 K). 
The contours represent the continuum at [5,10,20,40,60,80]$\sigma_\mathrm{CONT}$ where $\sigma_\mathrm{CONT}$ is the RMS noise of each continuum (see Table \ref{tab:CONTResult}).
The color scale of the continuum and the other moment 0 images are in the unit of Jy~beam$^{-1}$ and Jy~beam$^{-1}$~km~s$^{-1}$.
The magenta marker is the location of each corresponding HOPS object (see Table \ref{tab:SrcCoord}).
Note that there is no SiO $J$=5-4 transition detection in G192.12-11.10.
}
\setlength{\tabcolsep}{0pt} 
\renewcommand{\arraystretch}{0} 
\begin{tabular}{ccc}

\multicolumn{3}{c}{G211.47-19.27S} \\ [1ex]
\hline
\includegraphics[width=.32\textwidth]{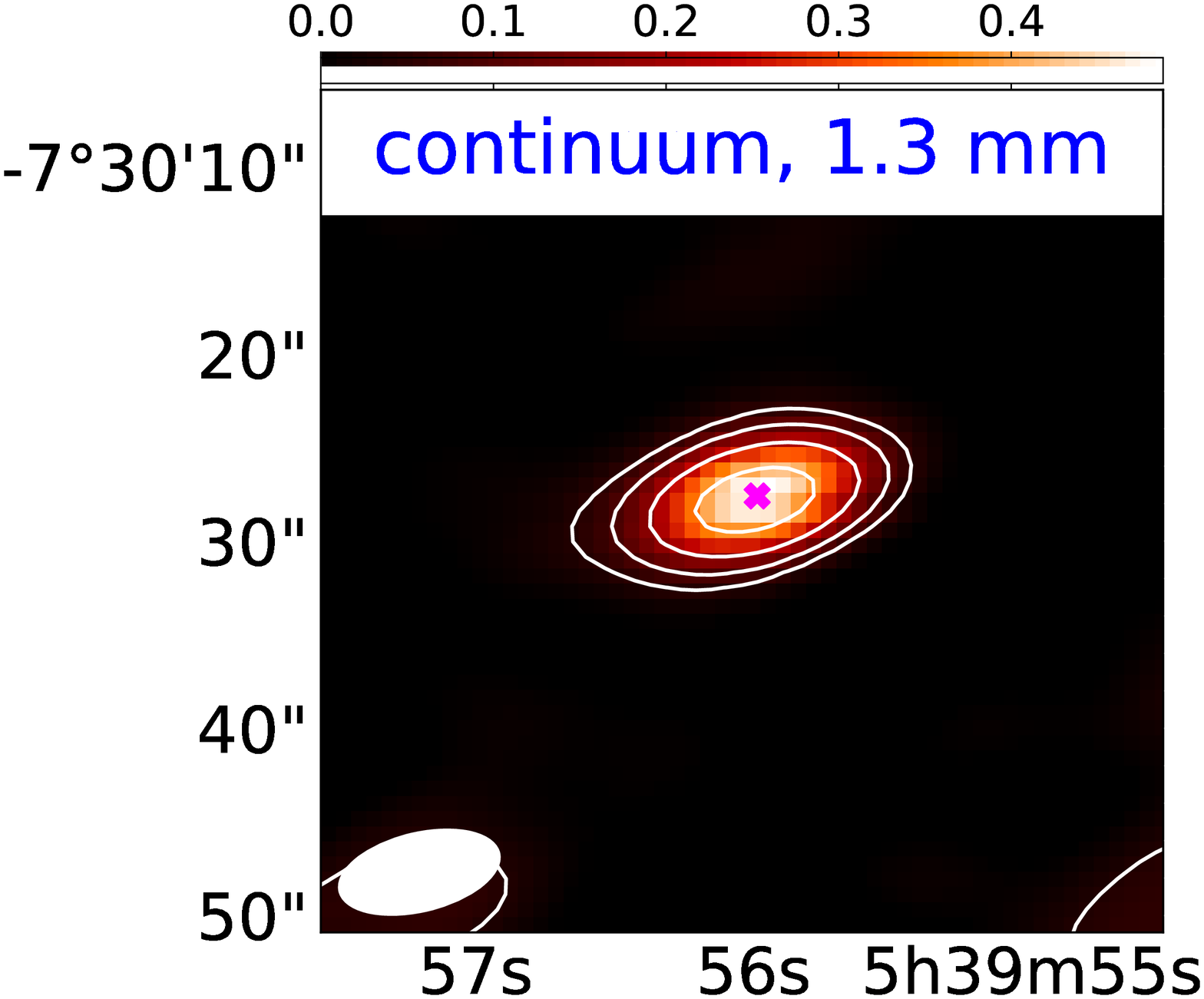} & \includegraphics[width=.32\textwidth]{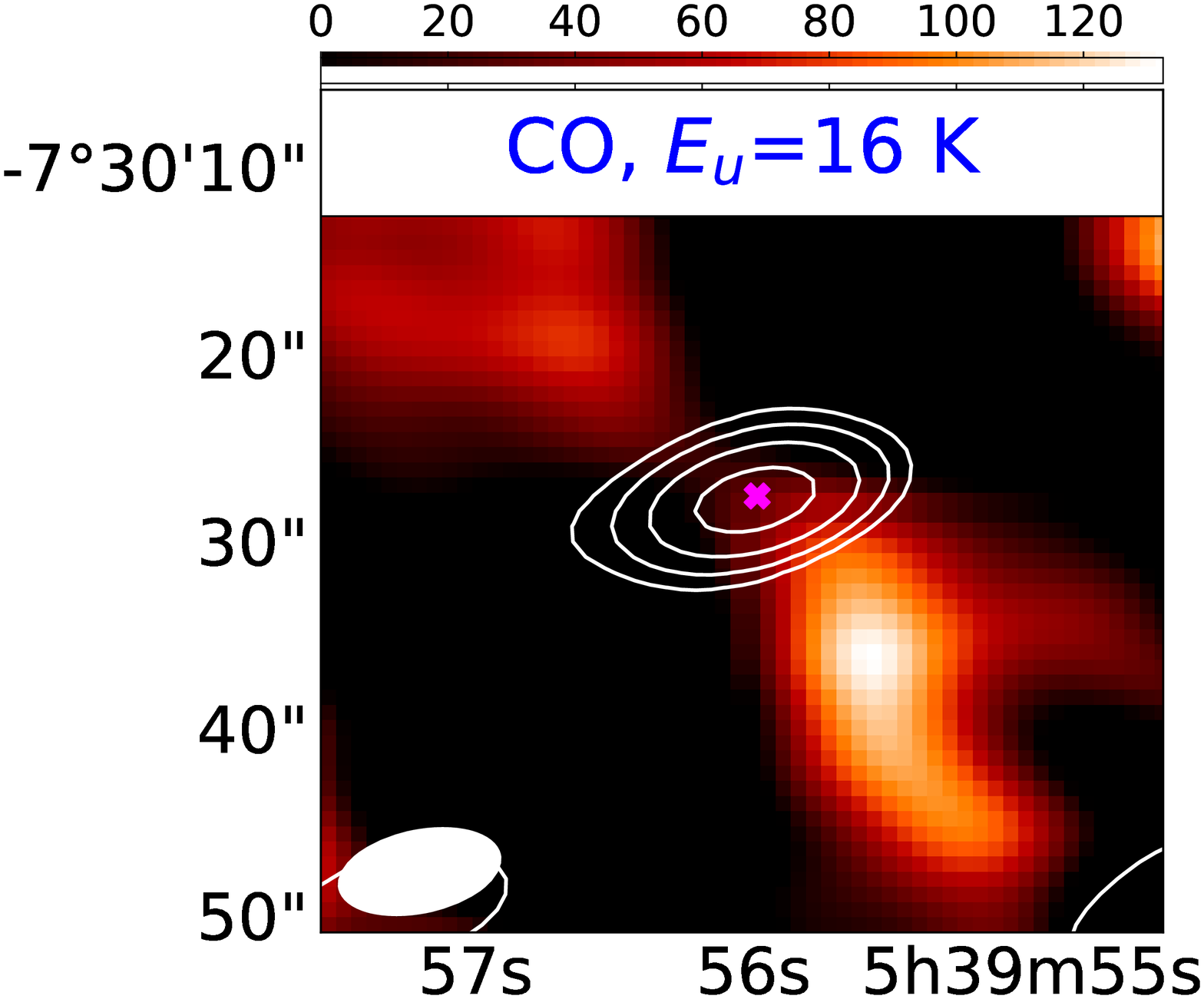} & \includegraphics[width=.32\textwidth]{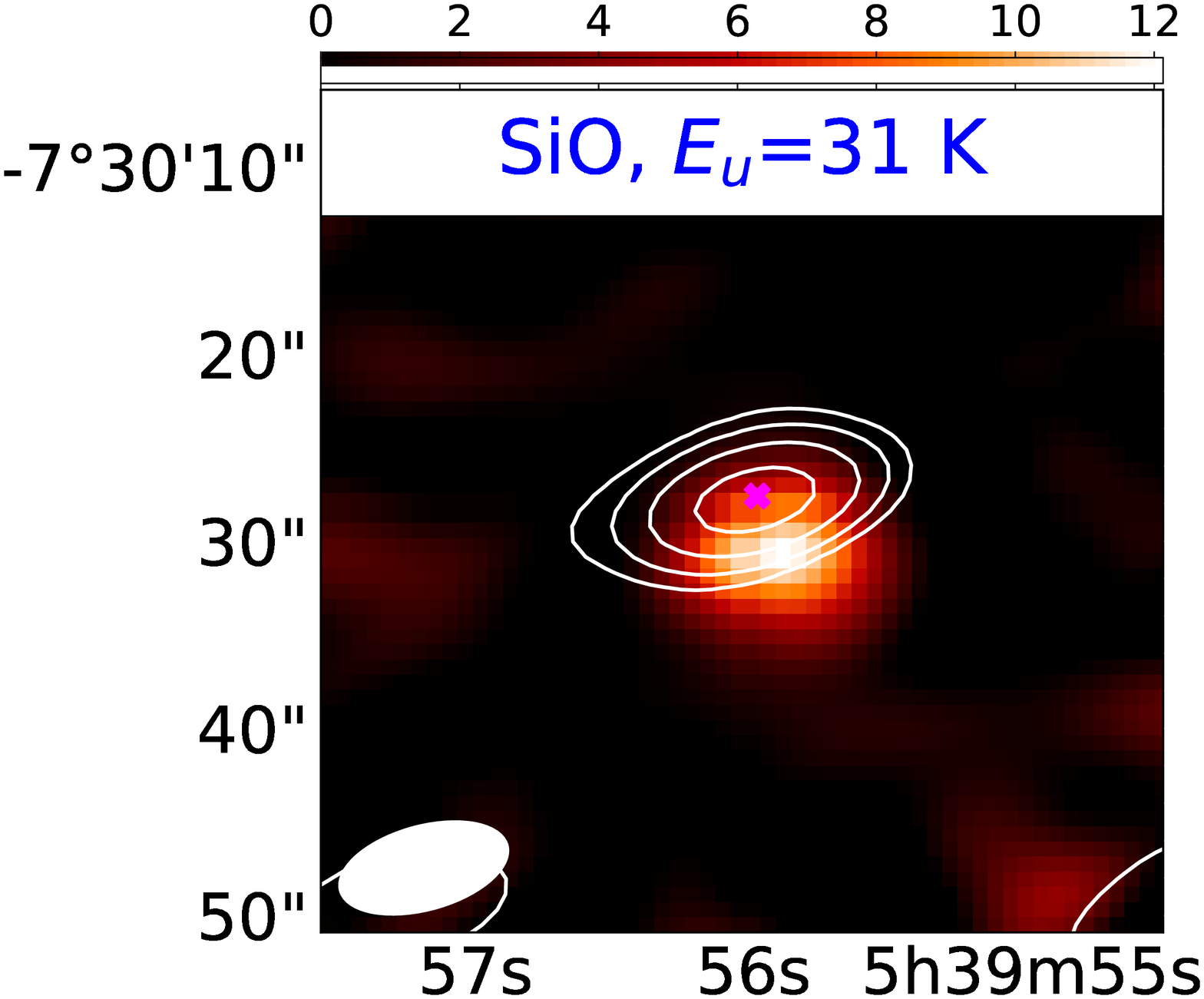} \\ [2ex]
\includegraphics[width=.32\textwidth]{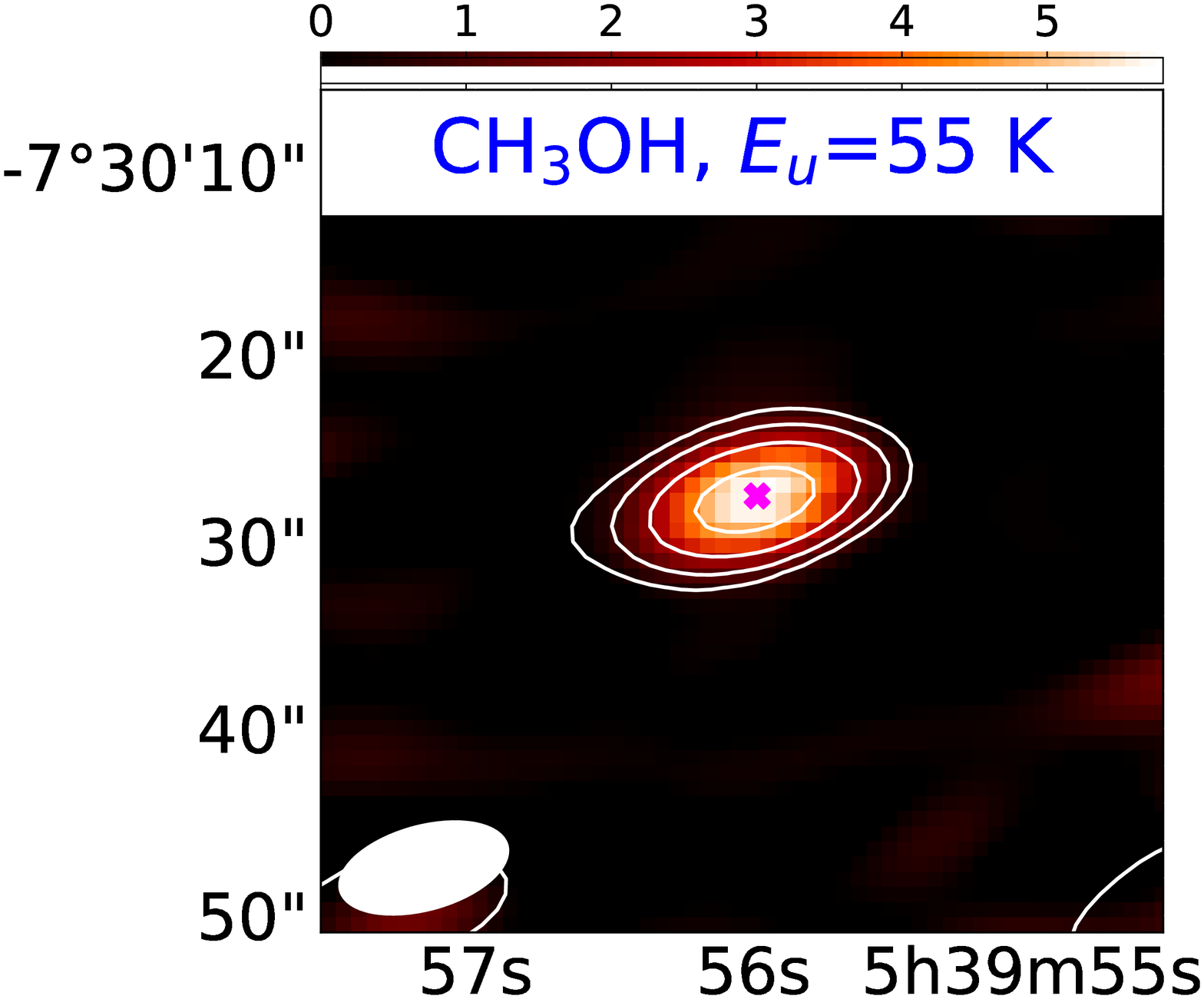} & \includegraphics[width=.32\textwidth]{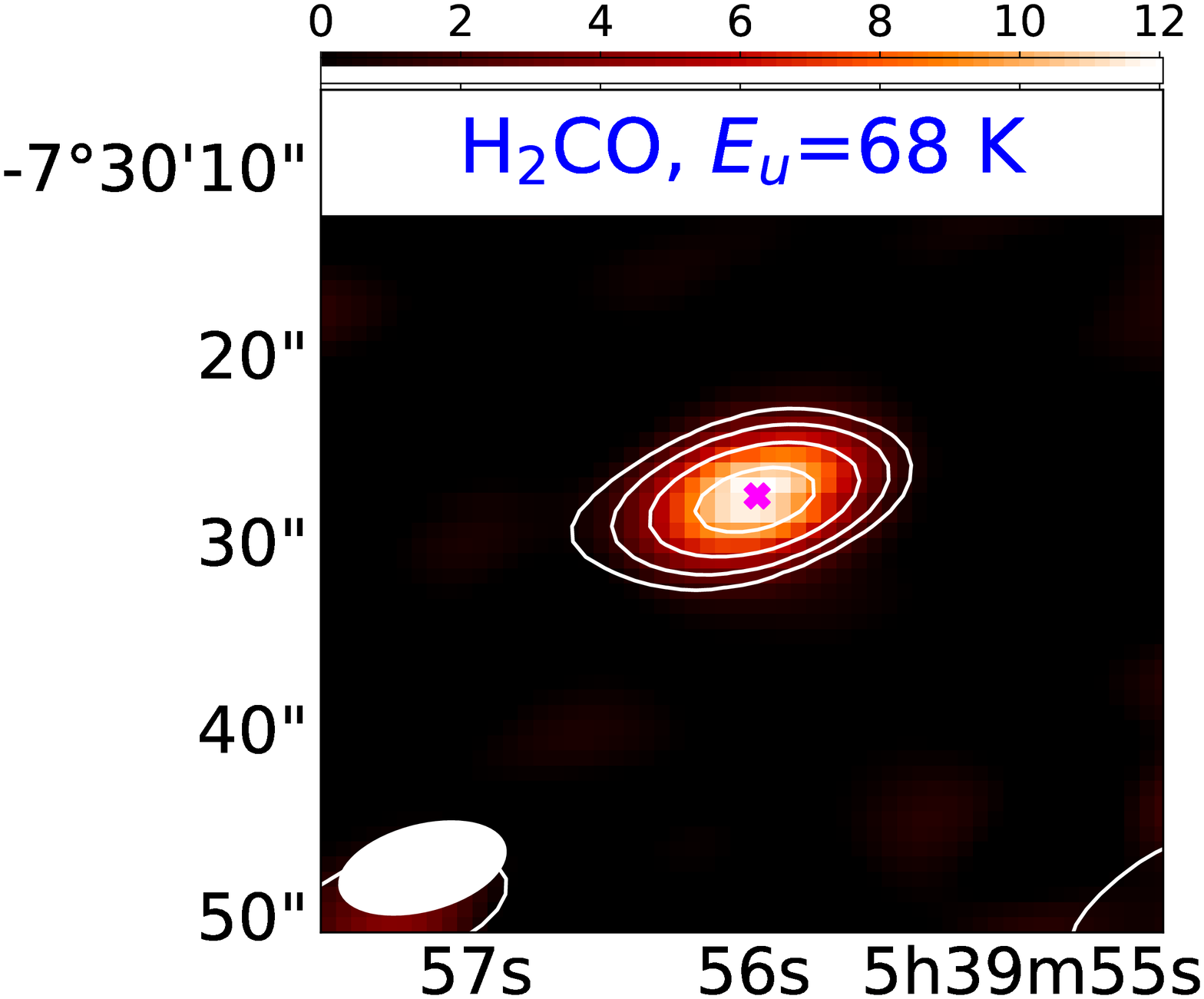} & \includegraphics[width=.32\textwidth]{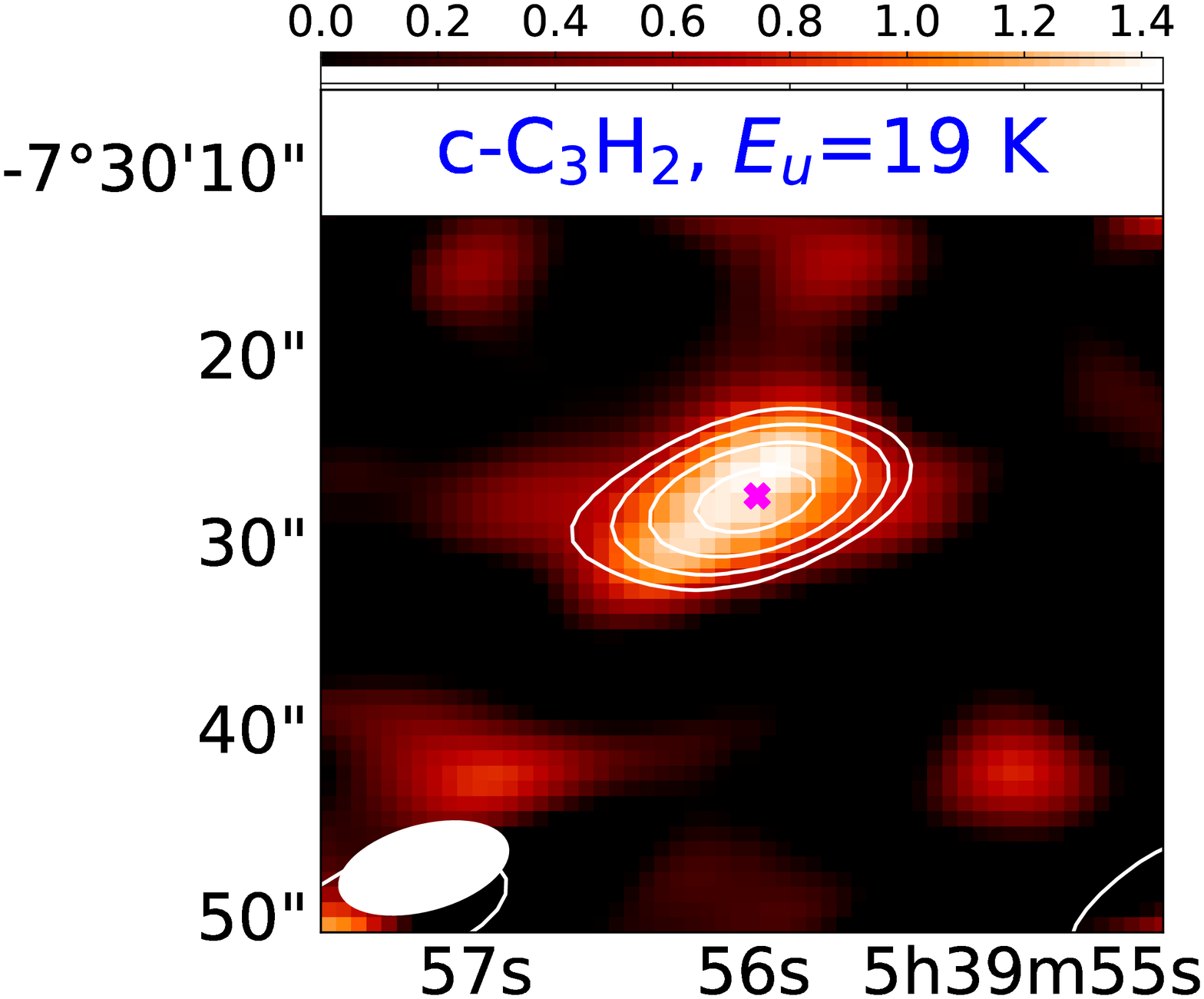}\\ [10ex]

\multicolumn{3}{c}{G208.68-19.20N1} \\ [1ex]
\hline
\includegraphics[width=.32\textwidth]{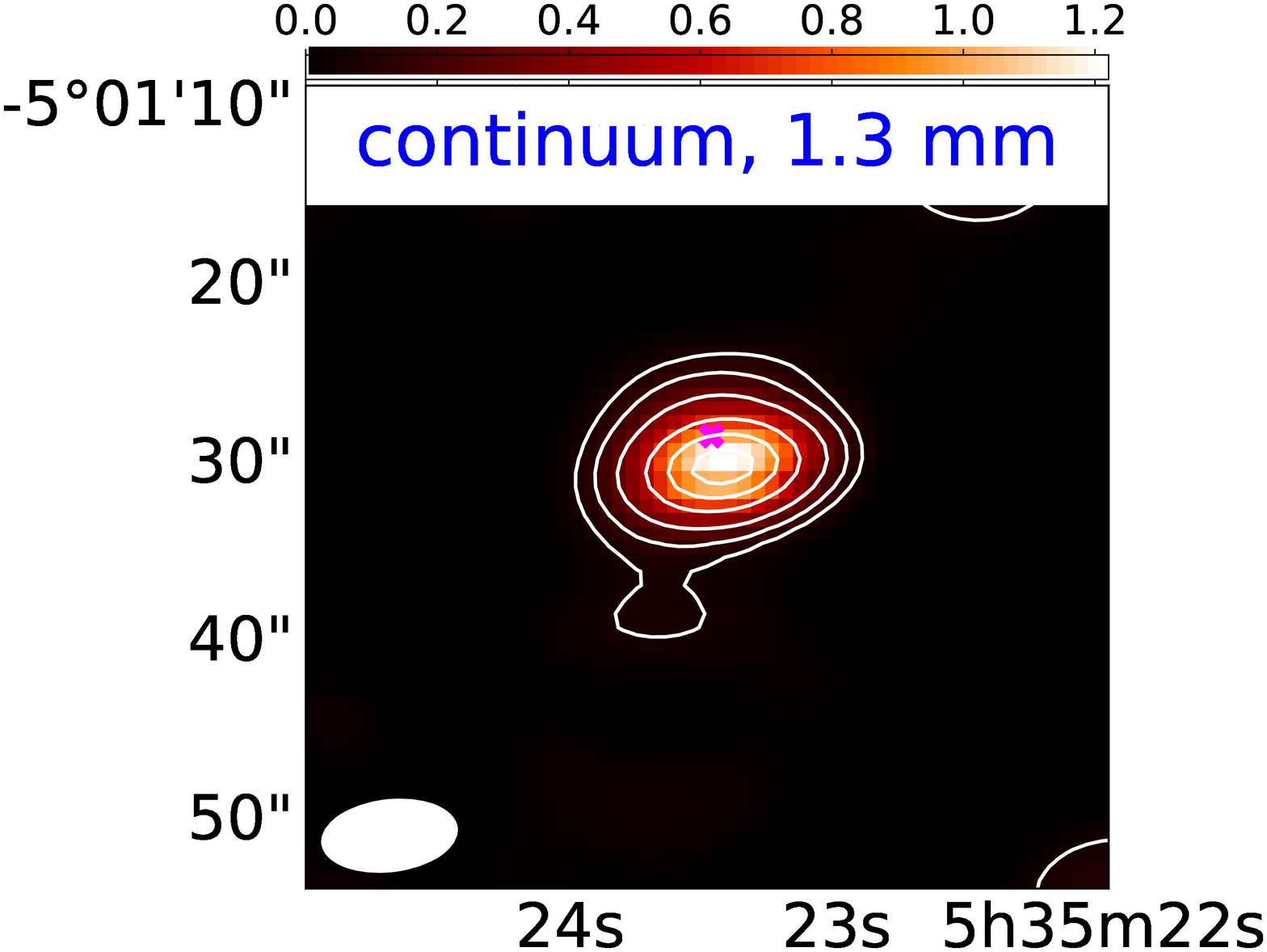} & \includegraphics[width=.32\textwidth]{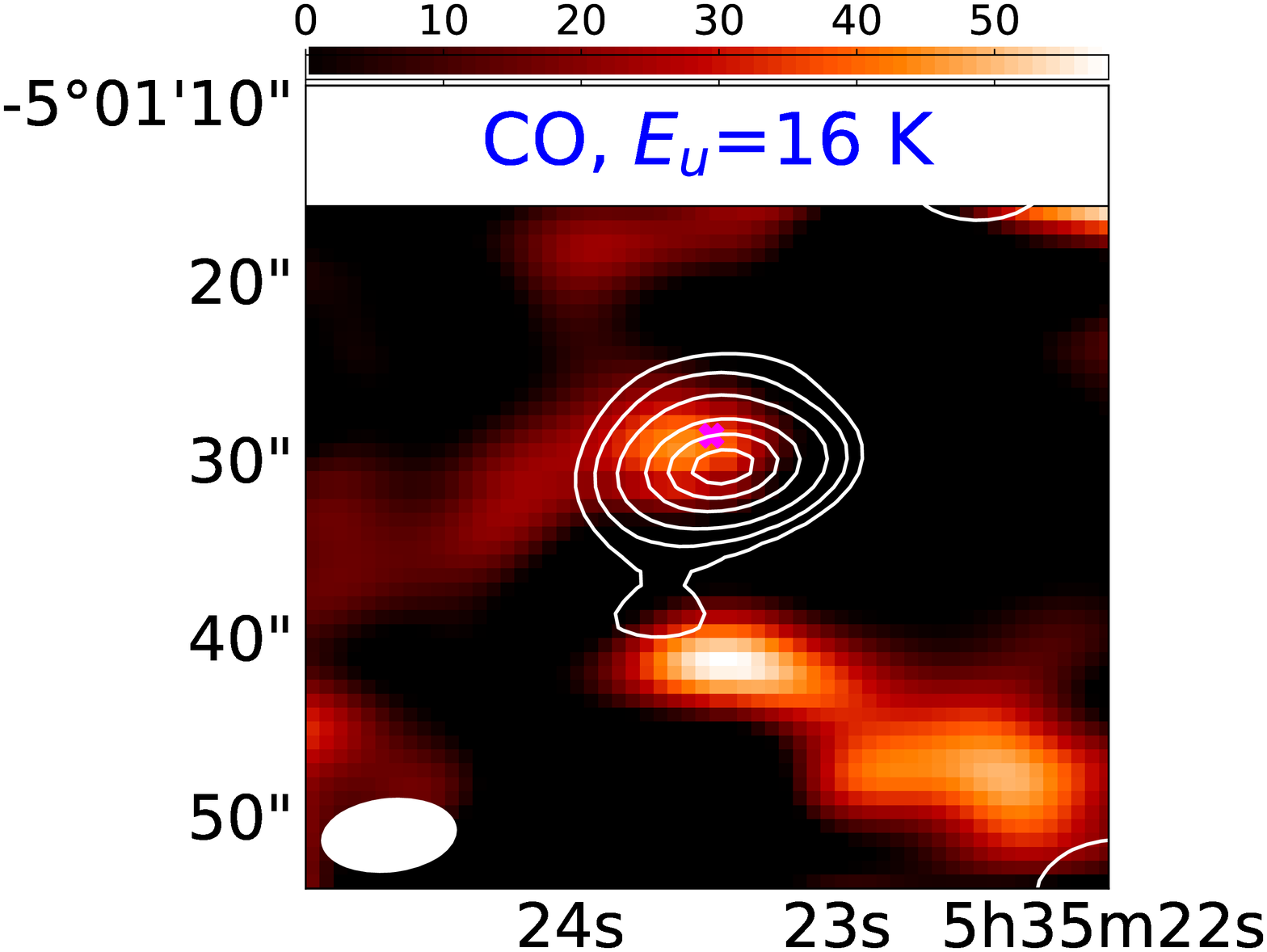} & \includegraphics[width=.32\textwidth]{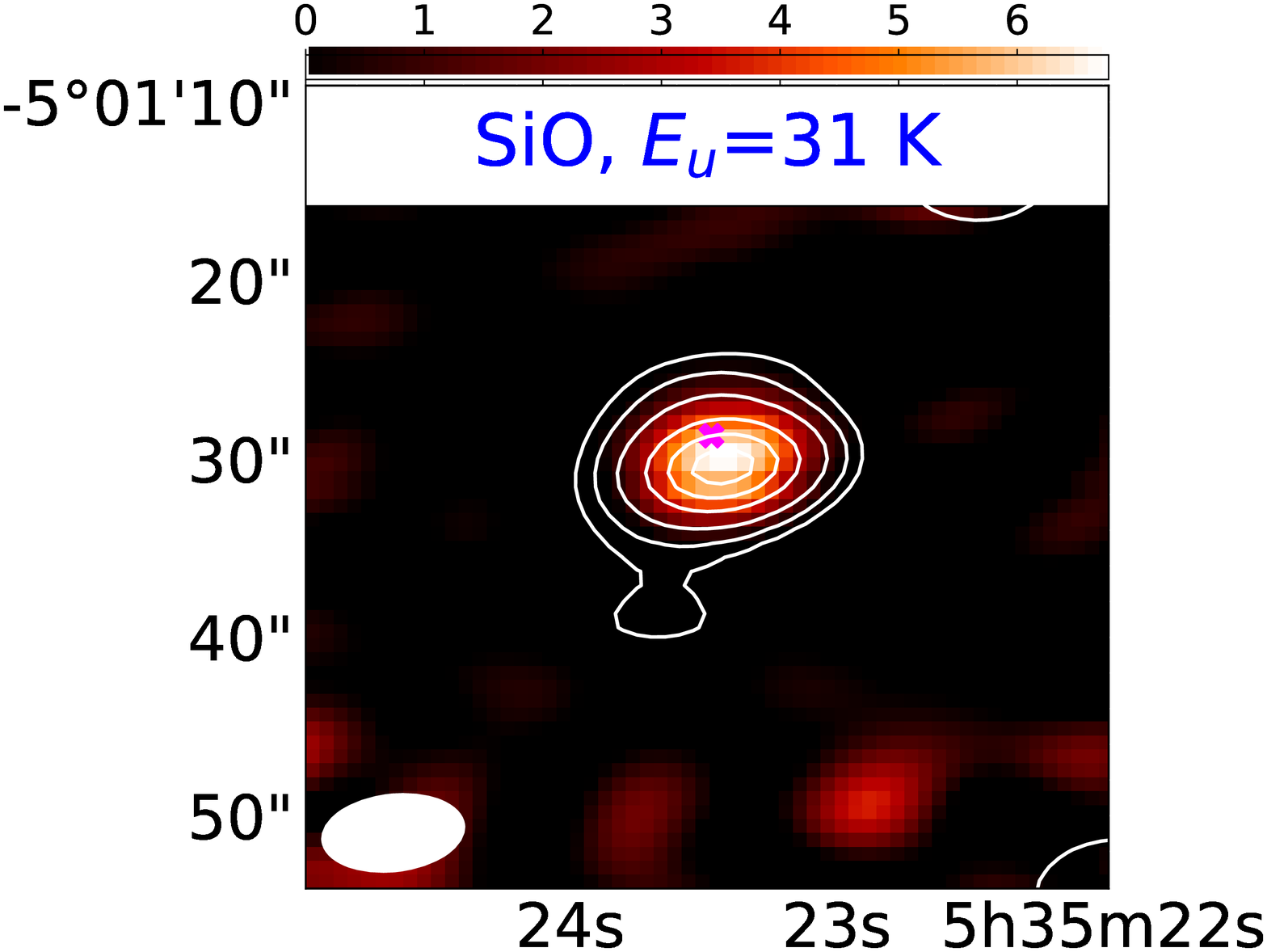} \\ [2ex]
\includegraphics[width=.32\textwidth]{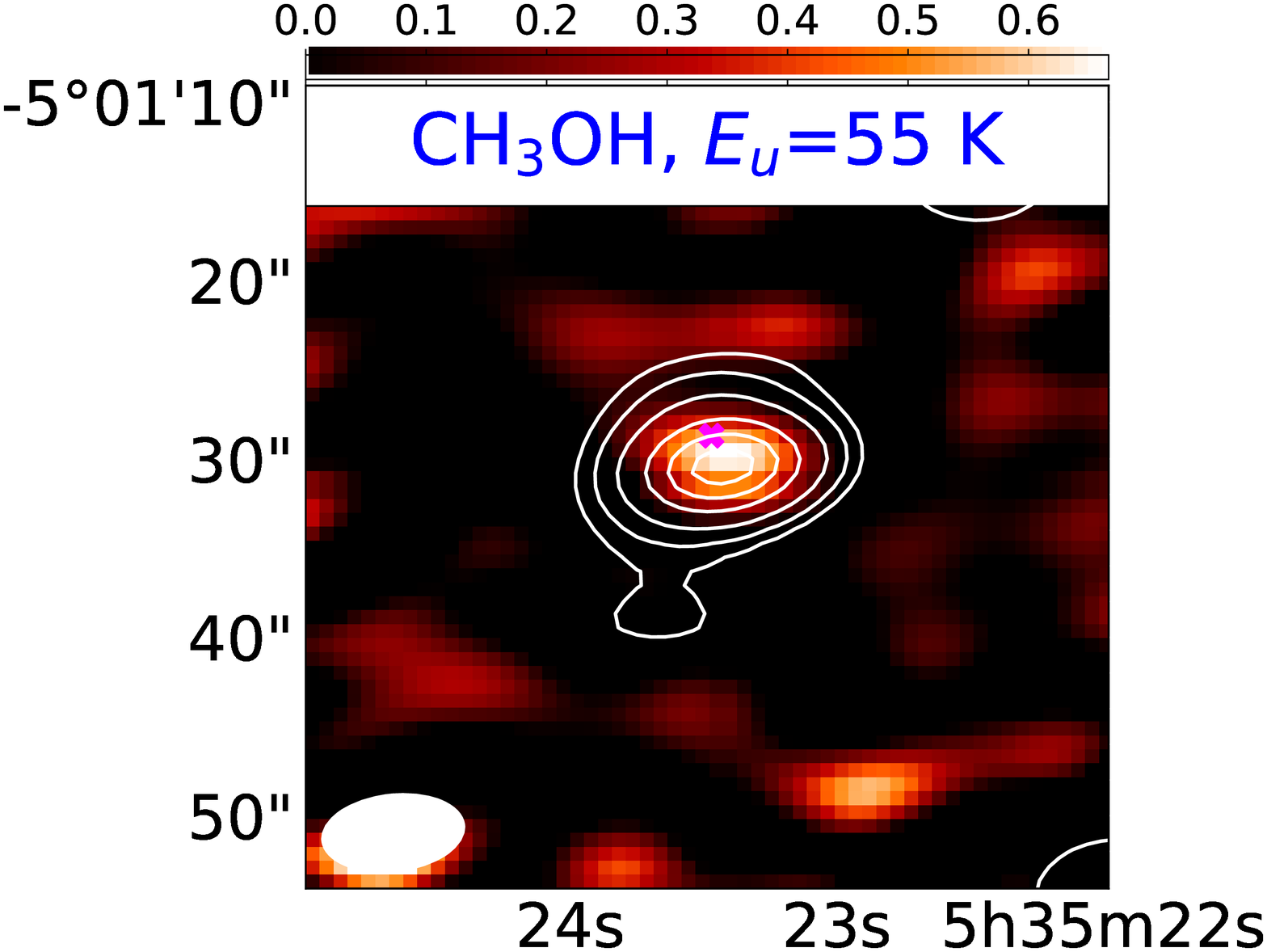} & \includegraphics[width=.32\textwidth]{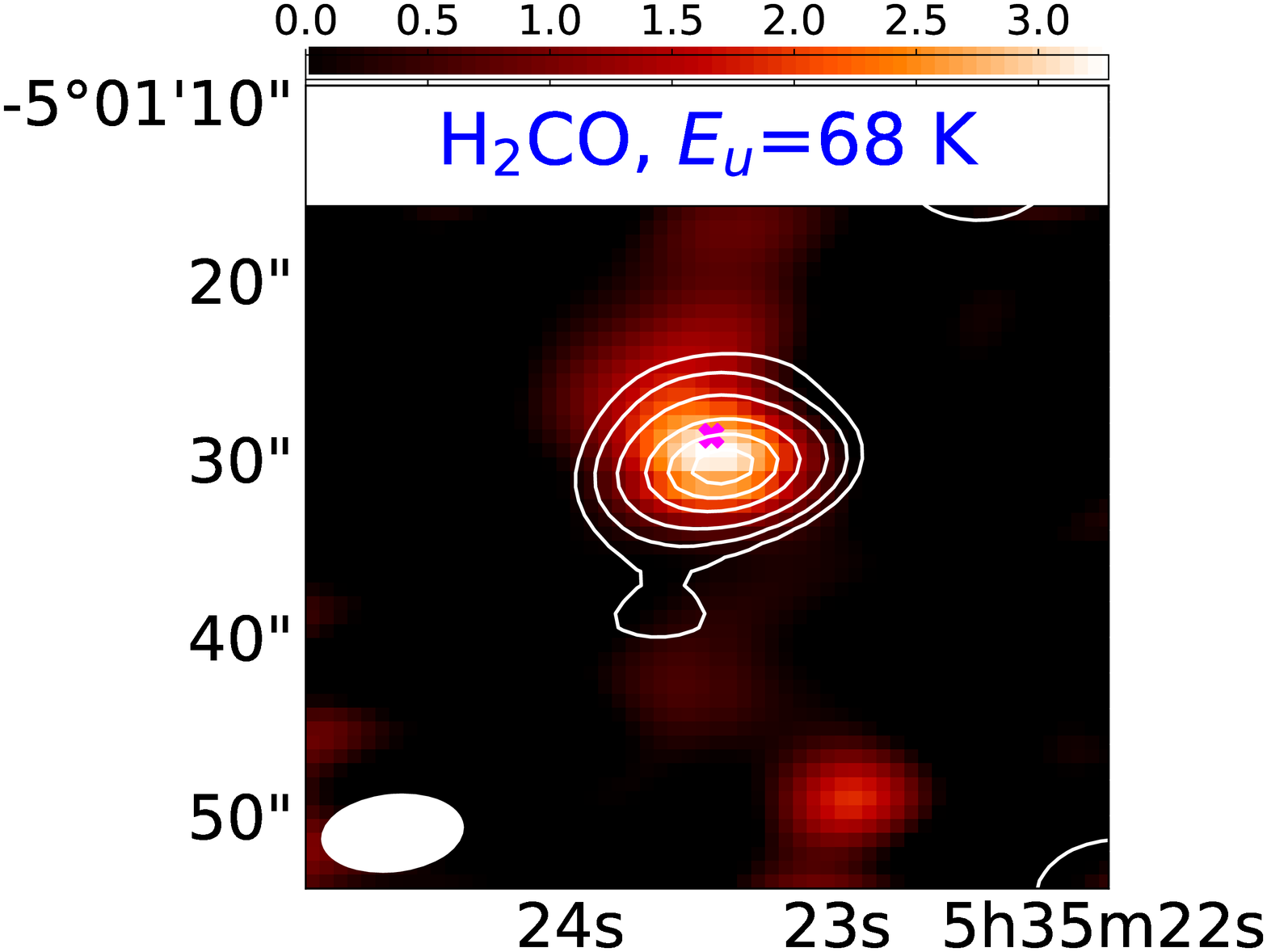} & \includegraphics[width=.32\textwidth]{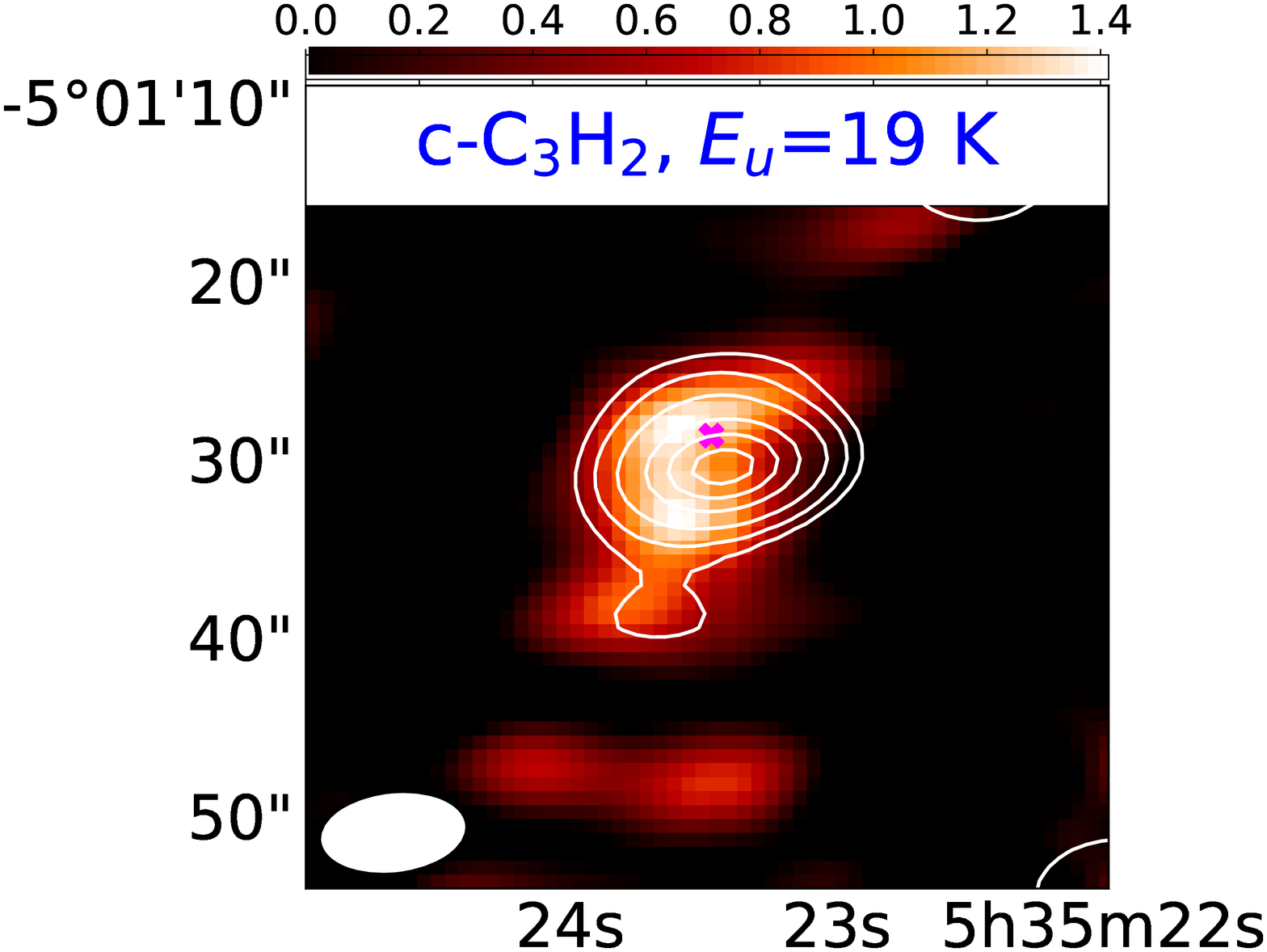} \\
\end{tabular}
\end{figure*}

The hydrocarbon and long carbon-chain species such as CCD, c-C$_3$H$_2$, and HCCCN have relatively lower rotational temperatures and narrower linewidths. 
Such characteristics, which are similar to those of CCH in the hot corino B335 \citep{2016Imai_B335}, indicate that these species are more extended as compared to the continuum and COMs (e.g. c-C$_3$H$_2$ in figure \ref{fig:Image_CONT_Mol}).
The line profile of CCH in B335 in fact hinted at the existence of double-peak signatures due to absorption by the cold envelope gas.
We unfortunately are not able to discriminate such features in our study at this stage due to the limited spatial and spectral resolution of the data analyzed.
Another species that in general bears a narrow linewidth is formylium DCO$^+$.

Among the four hot corino objects, the line widths ($\Delta v$) of molecular emission in G208 are overall narrower than those of the other three sources.
Taking CH$_3$OH as an example, $\Delta v$ is $\sim$7.5 km~s$^{-1}$ in G211, G210 and G192. 
In contrast, $\Delta v$ in G208 is 2.3 km~s$^{-1}$.
Meanwhile, the lower bolometric temperature in G208 would indicate a smaller region where CH$_3$OH can get evaporated from the icy mantle.
The lower velocity at a closer distance may imply a lower mass for the central protostellar object in G208, if CH$_3$OH emission is tracing rotational motion in the circumstellar disk.
Indeed, the relatively low bolometric luminosity, the most compact and likely young molecular outflows, and the highest envelope mass are all qualitatively but coherently suggestive of a younger stage.
Alternatively, the compact morphology of the CO outflow emission and the narrower linewidths of trace molecules like CH$_3$OH may be resulted from an outflow-disk/envelope system viewed pole-on.
Finally, there is also the possibility that its COM emissions have a different origin instead of tracing a genuine hot corino like the other three objects, hence giving a different [COM]/[CH$_3$OH].

In our ALMA observation, 72 fields were observed with 48 sources being of protostellar nature for their associations with YSOs and/or molecular outflows (Dutta et al., in preparation). 
Among those, we have identified 4 hot corino sources, a fraction of $\sim$ 8$\%$ and the three hot corinos found in Orion A are Class~0 HOPS objects. 
We note that the 8$\%$ face value needs to be treated with caution as the sample size is limited and the selection of the initial 72 fields is based on the JCMT SCUBA-2 detections in the PGCC targets, which implies that these sources are likely more embedded in large scale envelopes.
Other investigations of similar kind show different occurrence rates of corino signatures.
For example, the observations with ALMA, as part of the recent VLA/ALMA Nascent Disk and Multiplicity (VANDAM) survey, targeted toward four fields in the Orion Molecular Cloud-2 (OMC2) Far-infrared sources OMC2-FIR4 and OMC2-FIR3, identified eleven 870 $\mu$m continuum sources. Among them, eight sources are associated with HOPS objects and two of them, HOPS 108 in OMC2-FIR4 and HOPS 370 in OMC2-FIR3, are likely harboring hot corinos as well \citep{2019Tobin_HOPS108_HOPS370}.
The Continuum And Lines in Young ProtoStellar Objects (CALYPSO) program by IRAM surveyed 16 Class 0 protostellar systems, some with multiple protostars, in nearby clouds and detected compact (nearly all $<$ 100~au) CH$_3$OH emission around about half of the protostellar objects \citep{2020Belloche_CALYPSO}.
It further suggested a luminosity threshold of 4 $L_{\odot}$, above which YSOs in their sample exhibit spectral features from at least one COM, and no COM emission was detected for sources with luminosities lower than 2 $L_{\odot}$.
They attributed this, though, to the sensitivity of the observations instead of the intrinsic property of the sources with lower luminosities and noted that there exist other COMs-detected sources fainter than 2 $L_{\odot}$ \citep[e.g. $\sim$ 0.7 L$_{\odot}$ in B1-bS and $\sim$ 0.7 L$_{\odot}$ in B335][]{2014Hirano_B1bS_COM, 2015Evans_B335_luminisity}.
In our study, three hot corinos, of which the SEDs have been established by \citet{2016Furlan_HOPS_Class}, have luminosities well above that threshold.
The luminosity of the fourth source (G192) is at a level of $\sim$ 14 $L_{\odot}$ (Dutta et al., in preparation).
There are, on the other hand, protostelllar objects in our sample with luminosities much greater than 4 $L_{\odot}$ but show no clear sign of hot corinos.
It remains unclear whether hot corino is an evolutionary stage that YSOs would generally experience, or only YSOs with certain properties or in certain environments would develop into hot corinos.
A full census of YSOs in different environments and evolutionary stages will help in revealing the prevalence of hot corinos and addressing the above questions on a statistical basis.

\begin{figure*}[hb]
\centering
\setlength{\tabcolsep}{0pt} 
\renewcommand{\arraystretch}{0} 
\begin{tabular}{ccc}

\multicolumn{3}{l}{\textbf{Figure} \ref{fig:Image_CONT_Mol} (\textit{continued)}
}\\
\multicolumn{3}{c}{G210.49-19.79W}\\ [1ex]
\hline
\includegraphics[width=.32\textwidth]{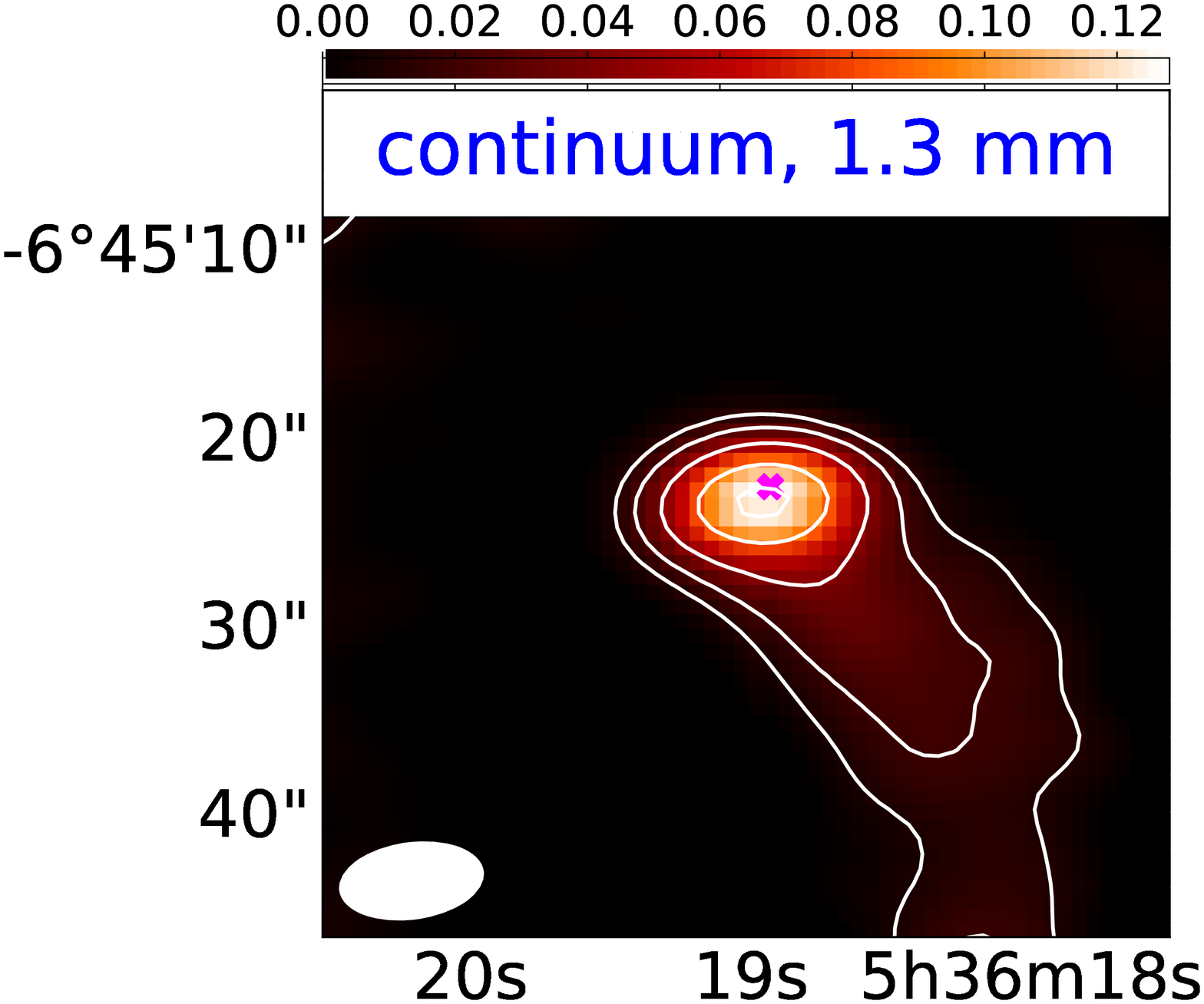} & \includegraphics[width=.32\textwidth]{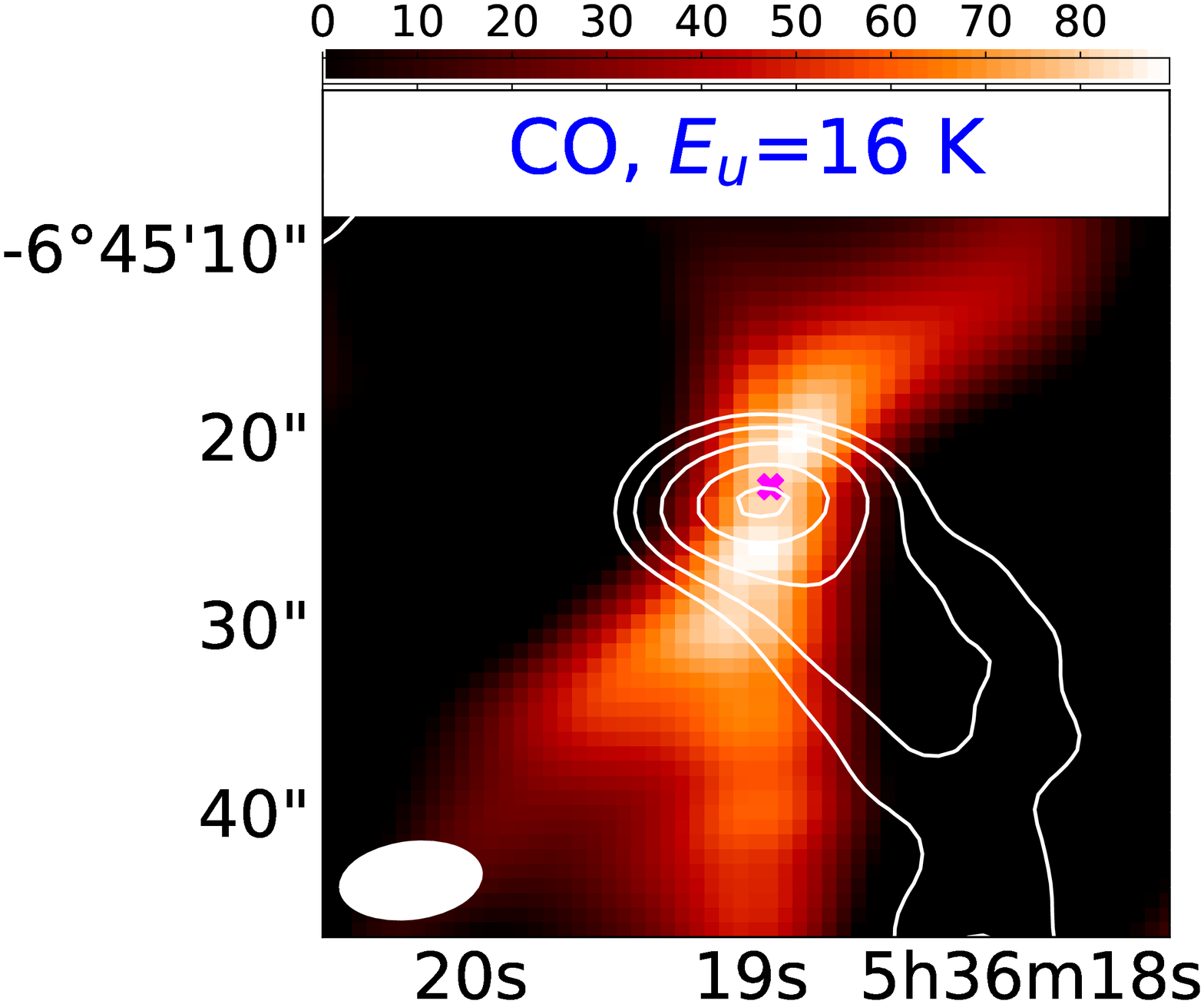} & \includegraphics[width=.32\textwidth]{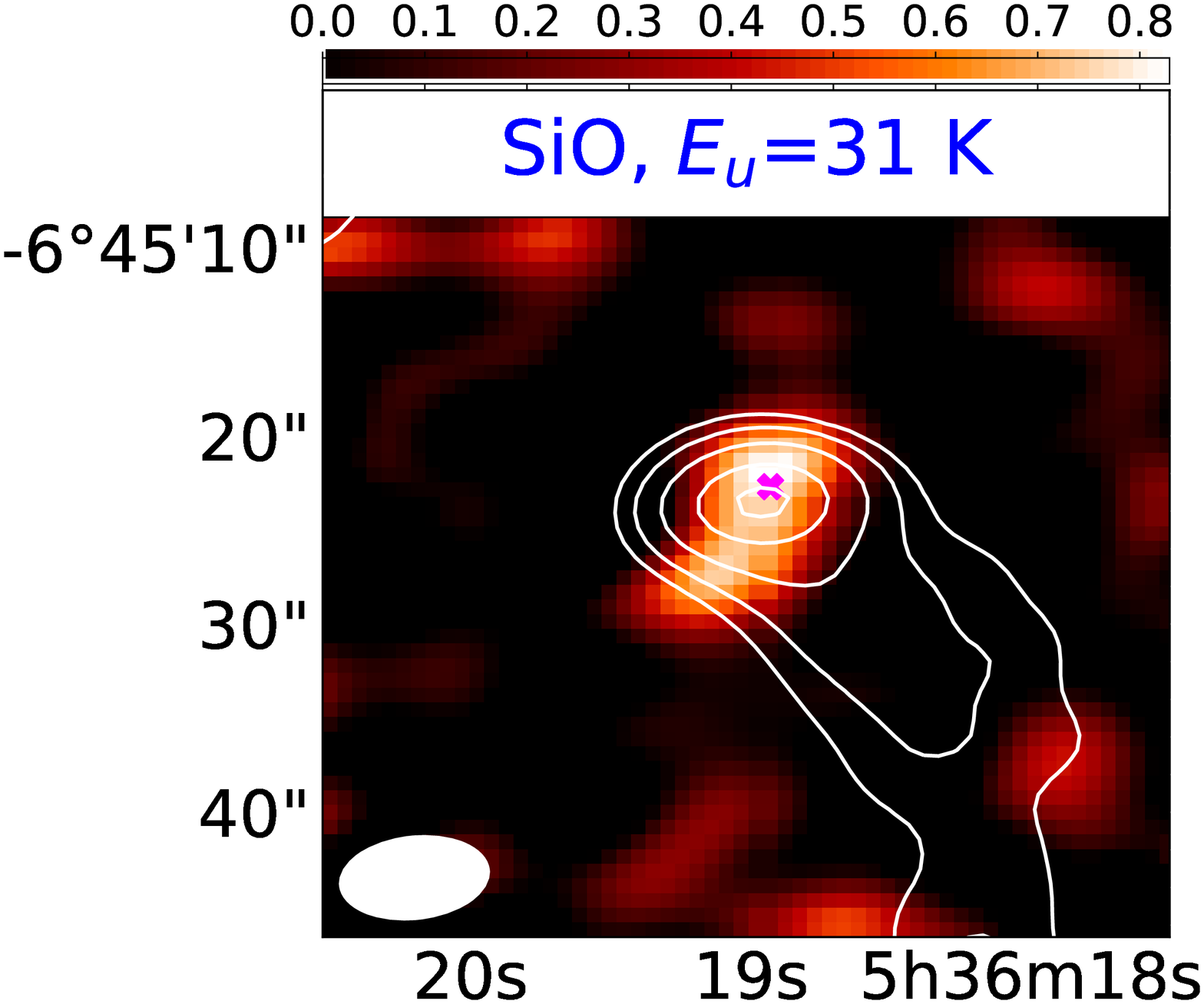} \\ [2ex]
\includegraphics[width=.32\textwidth]{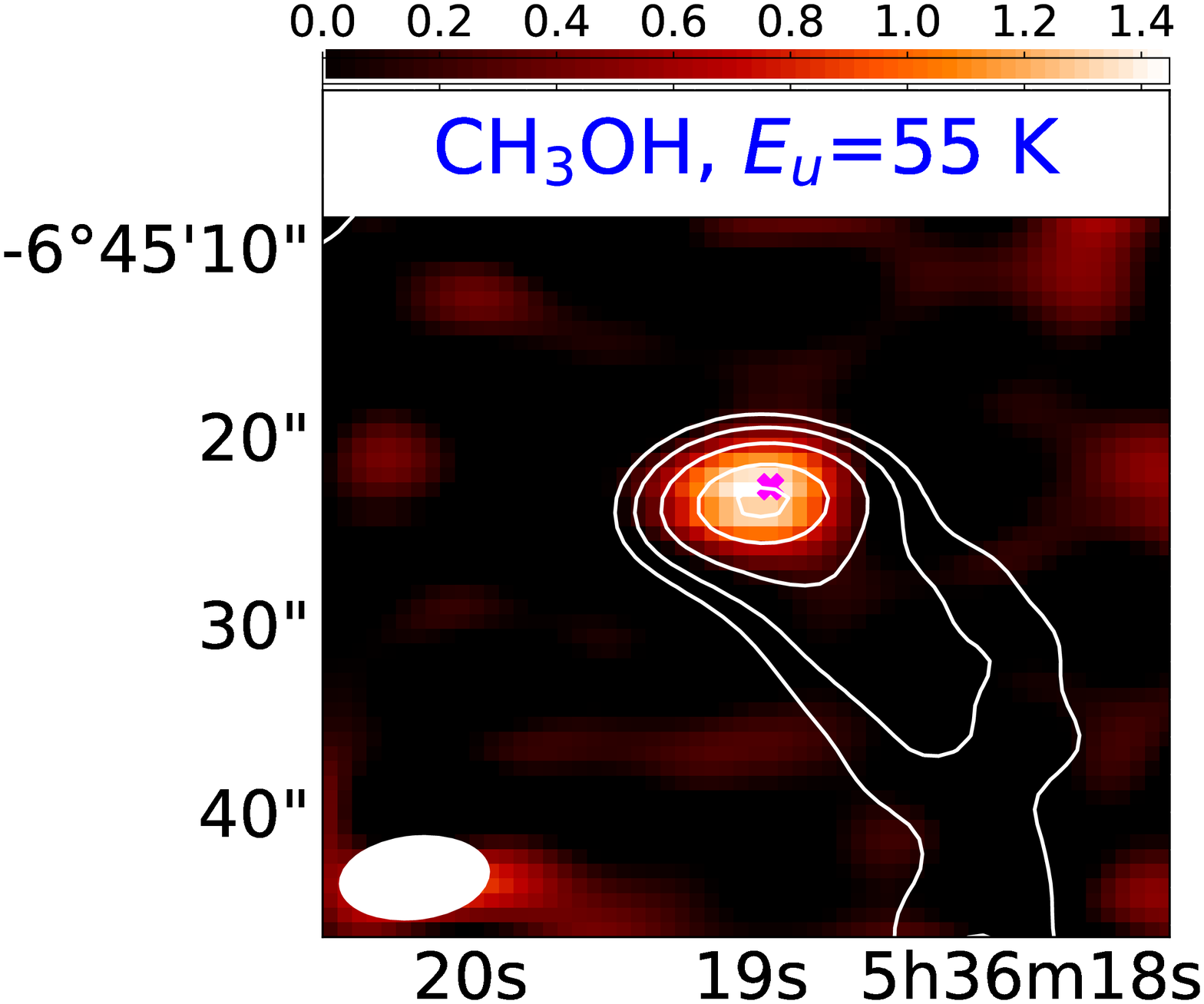} & \includegraphics[width=.32\textwidth]{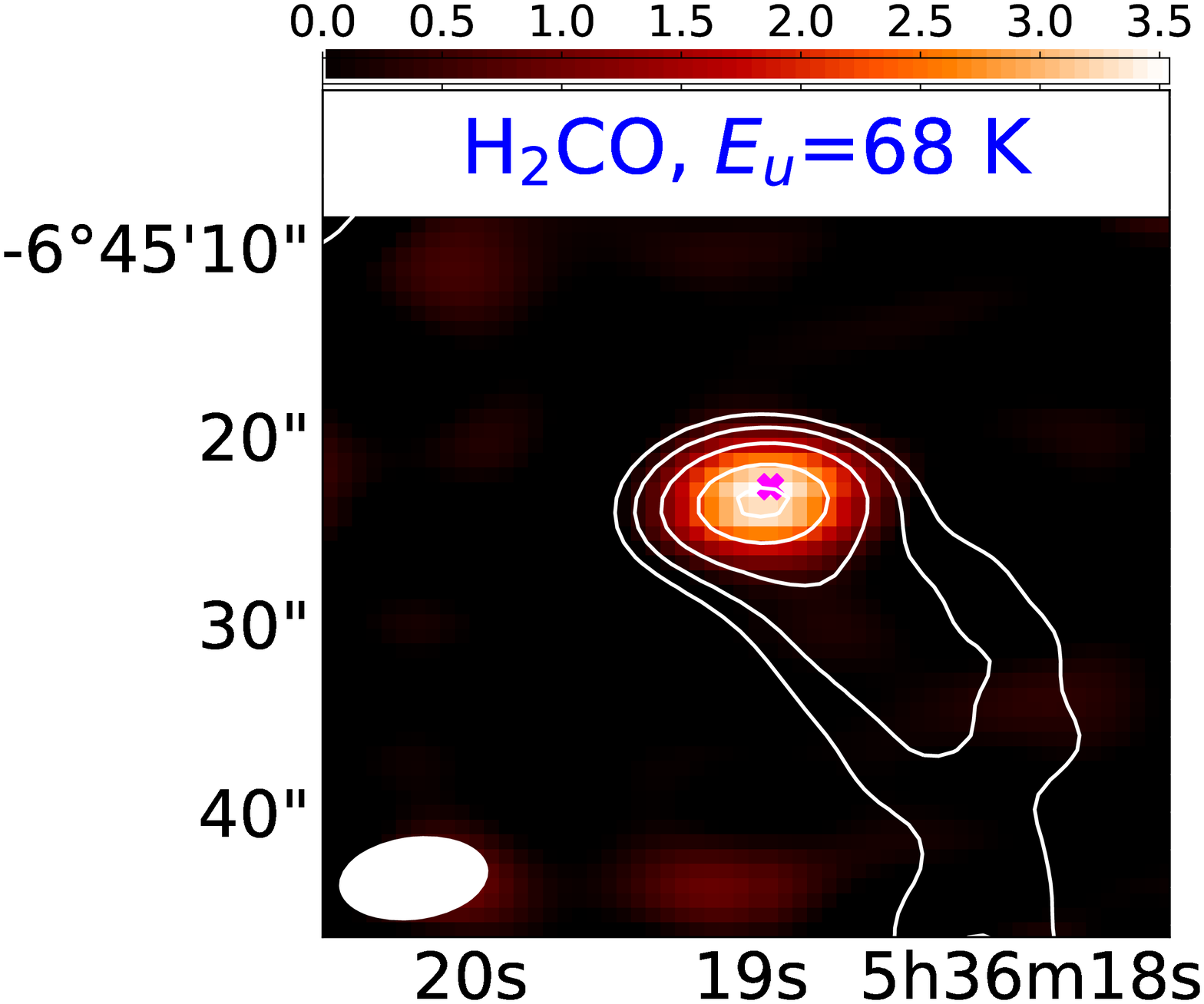} & \includegraphics[width=.32\textwidth]{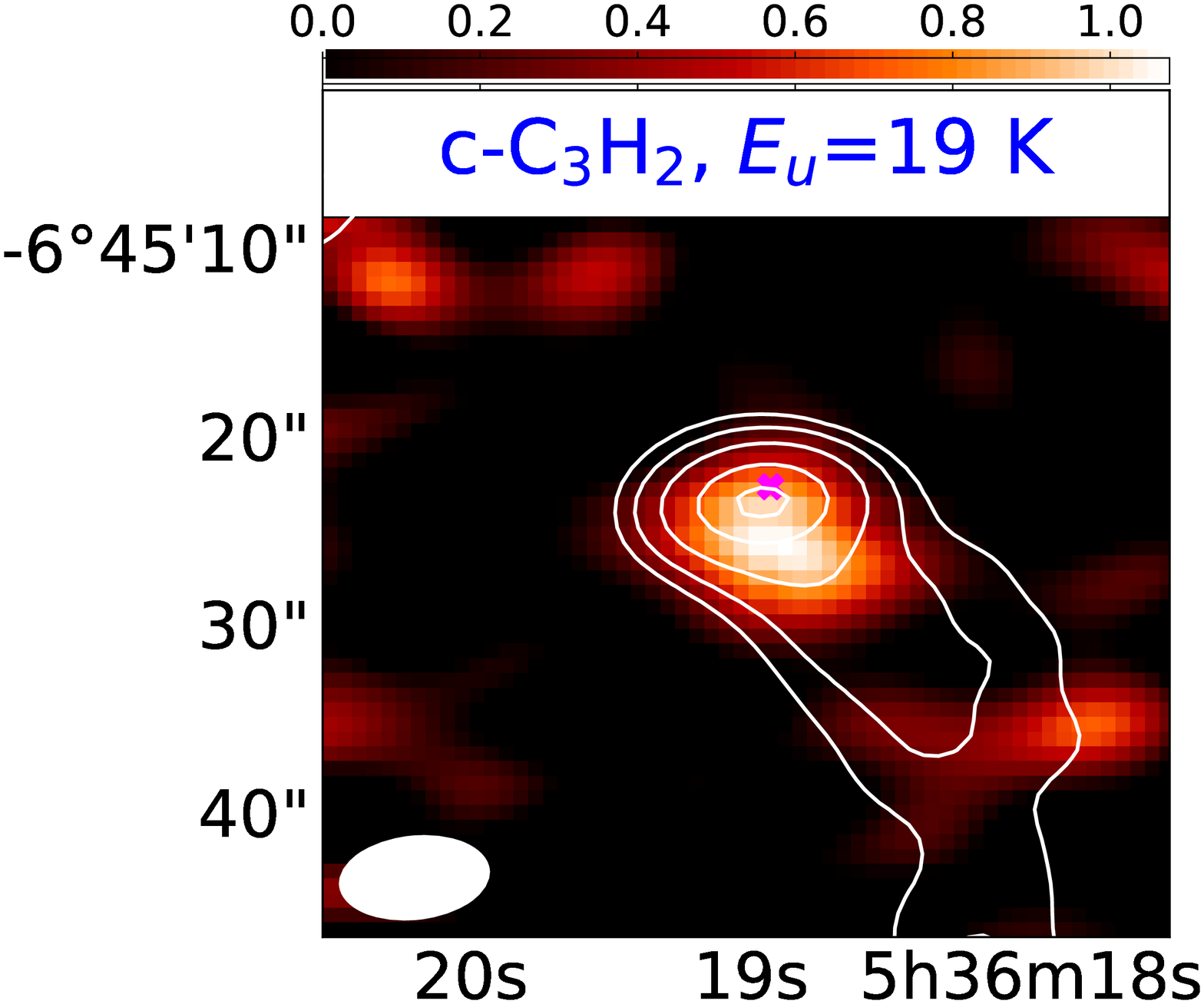} \\ [10ex]
\multicolumn{3}{c}{G192.12-11.10} \\ [1ex]
\hline
\includegraphics[width=.32\textwidth]{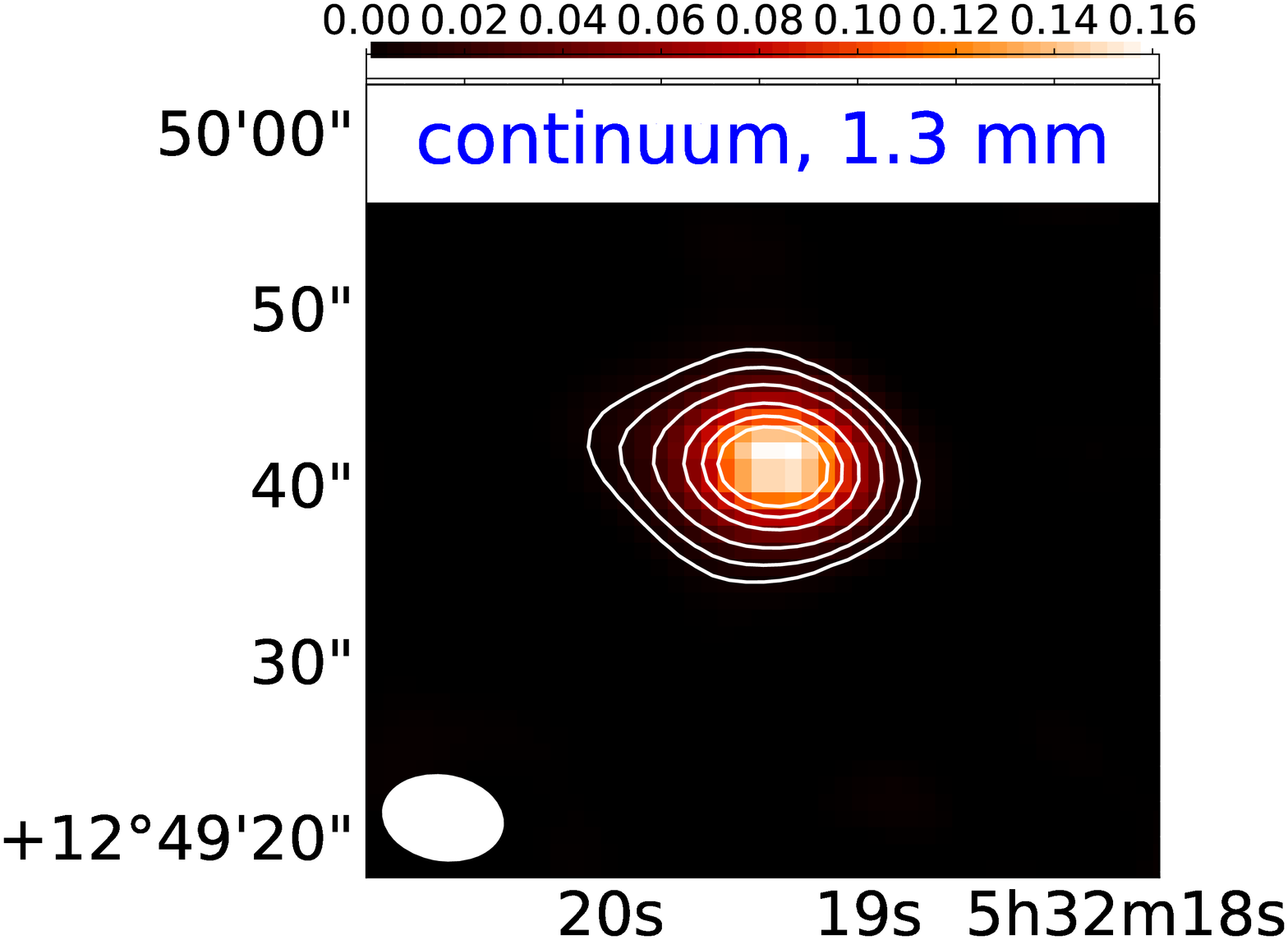} & \includegraphics[width=.32\textwidth]{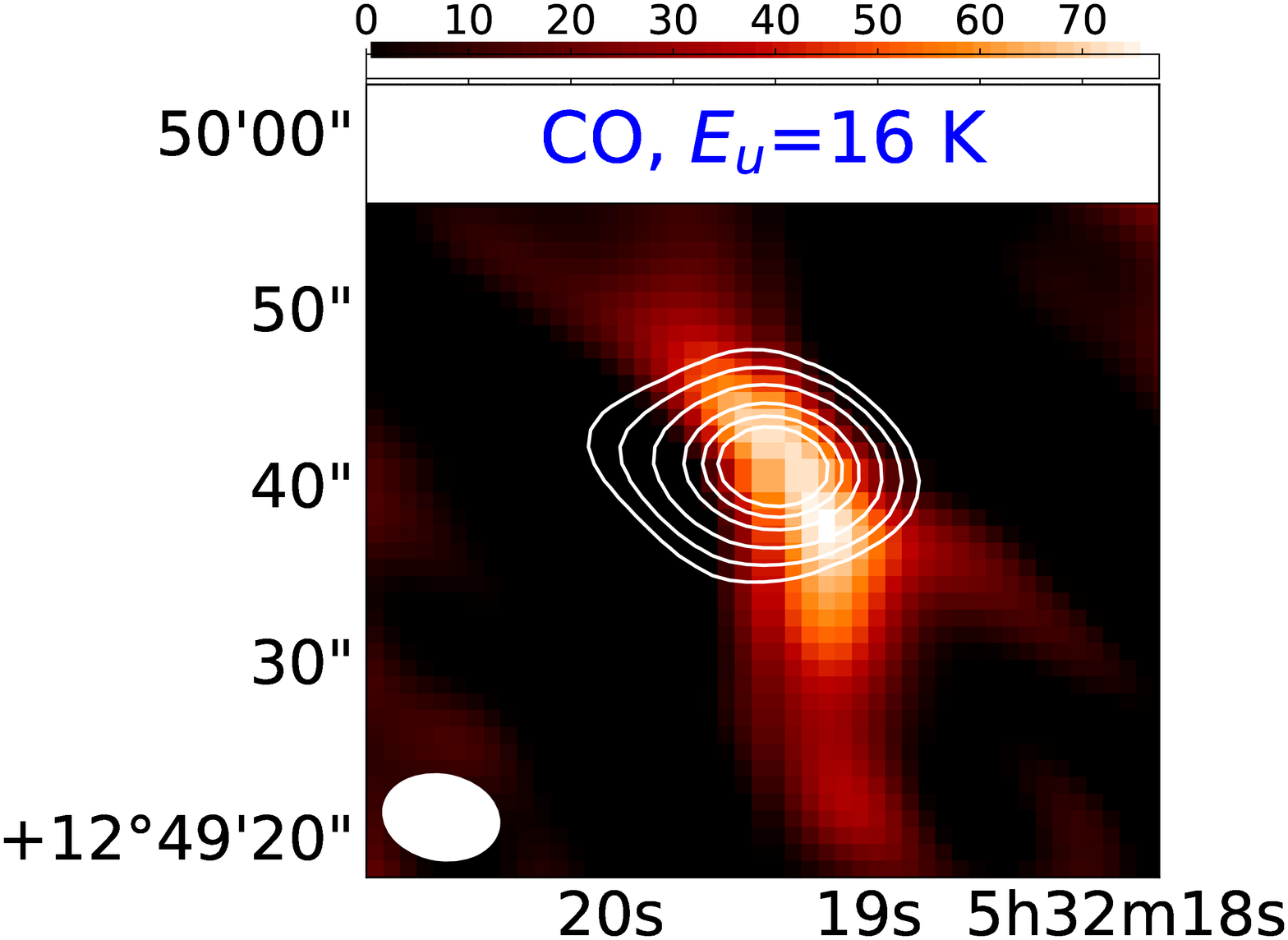} &
No detection. \\ [2ex]
\includegraphics[width=.32\textwidth]{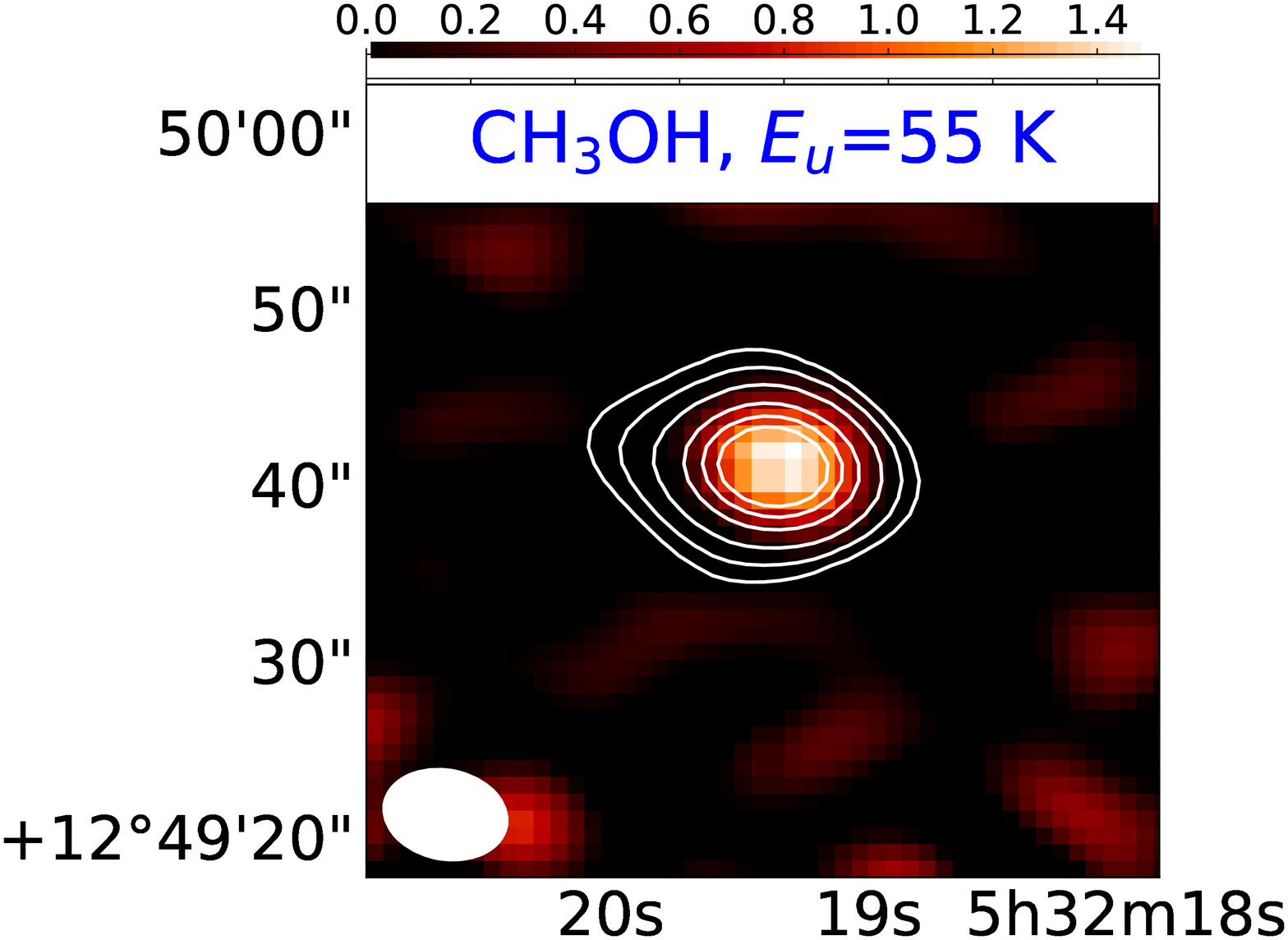} & \includegraphics[width=.32\textwidth]{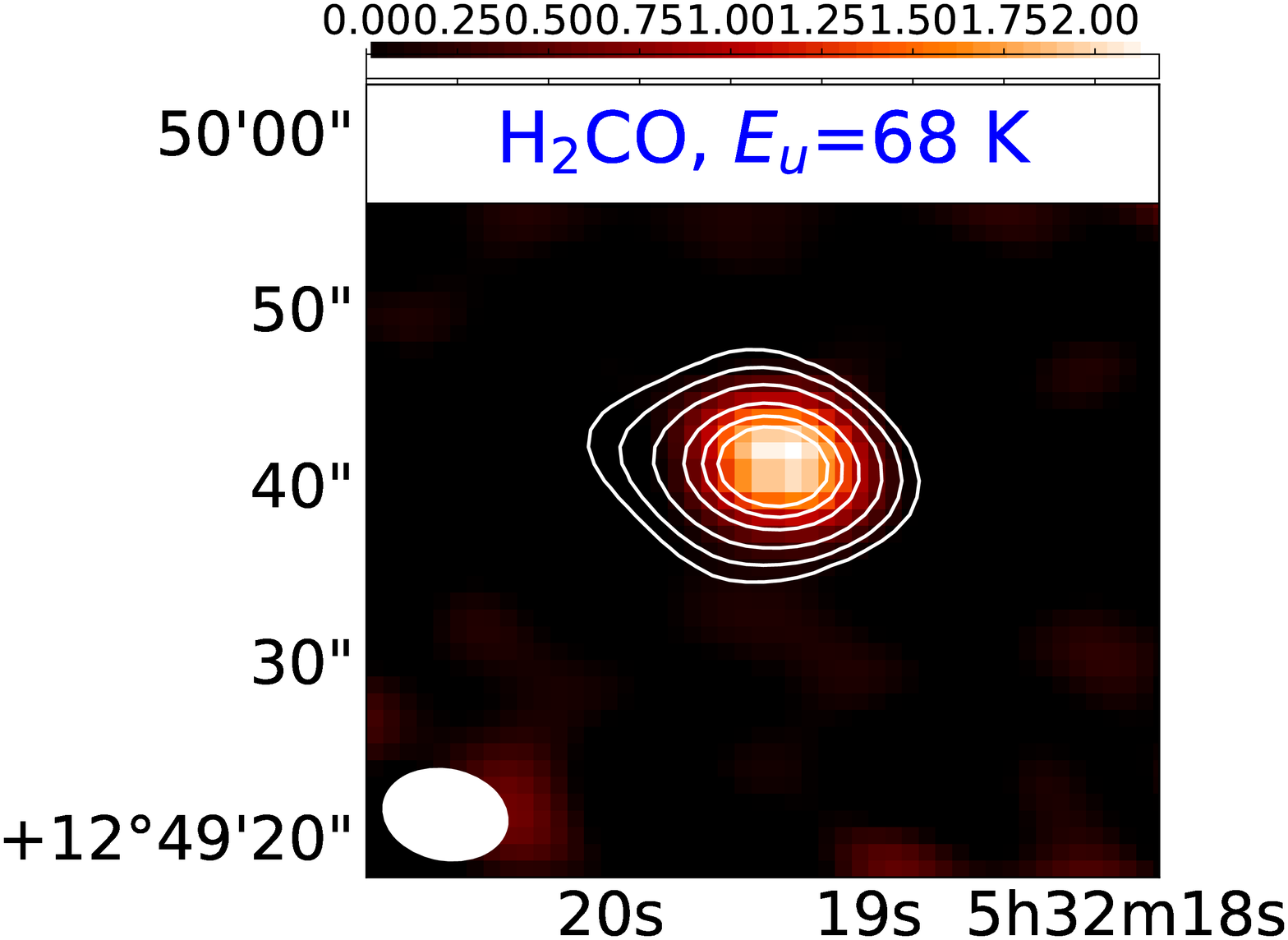} & \includegraphics[width=.32\textwidth]{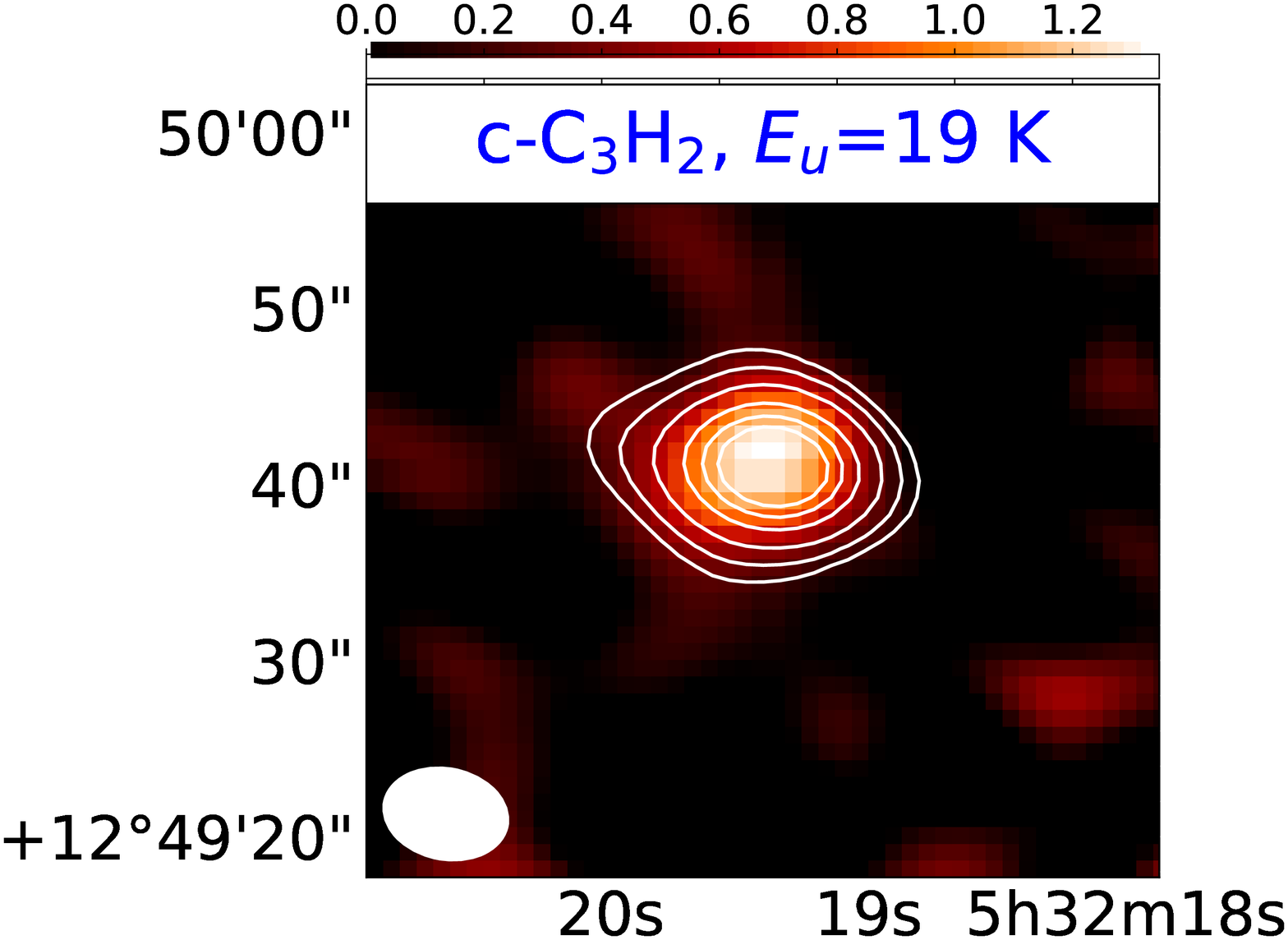} \\
\end{tabular}
\end{figure*}

\section{Conclusions \label{sec:Conclusions}}
\begin{enumerate}
  \item 
  Based on a survey toward selected Planck Galactic Cold Clump (PGCC) samples in Orion Molecular Cloud Complex nebula A and B and $\lambda$ Orionis, we report the detection of four hot corinos based on the presence of warm, compact CH$_3$OH emission in the four sources, and additional COMs in two of them.
  Given the positional association of their 1.3 mm continuum, three of these four sources are identified as Class 0 young stellar objects associated with HOPS sources, and the remaining one is for the first time identified as a protostellar core.
  Further studies with higher angular and spectral resolutions would help to solidify (or falsify) the classification of hot corinos.

  \item 
  Some fractional column densities of COMs with respect to methanol, [COM]/[CH$_3$OH], are an order of magnitude higher than those in other hot corinos in the literature.    
  In G211.47-19.27S, the $^{12}$C/$^{13}$C ratio of methanol ([CH$_3$OH]/[$^{13}$CH$_3$OH]) is $\sim$ 10 which is comparable to the values in other hot corinos in the literature. 
  This $^{12}$C/$^{13}$C ratio is lower than is typical in the ISM ($\sim$70).
  The estimated column densities could be affected by beam dilution since the COM emission is not resolved. 
  In addition, the relative abundances with respect to methanol, as well as the $^{12}$C/$^{13}$C ratio in CH$_3$OH may be affected by its opacity.
  Variations between hot corinos are in any case expected due to the chemical diversity found in the literature \citep{2019Bergner_Ser-emb-8-_Ser-emb-17}.

  \item  
  In G211.47-19.27S, both NH$_2$CHO, a pre-biotic molecule, and HNCO are detected.
  Their estimated column densities are $\sim3.6\times10^{14}$ cm$^{-2}$ and $\sim1.2\times10{15}$ cm$^{-2}$, respectively.
  The abundance ratio [NH$_2$CHO]/[HNCO] is about twice of the estimation based on \citet{2015Lopez-Sepulcre_NH2CHO}.
  HNCO is also detected in G210.49-19.79W and G192.12-11.10 without the detection of NH$_2$CHO.
  The lack of detection of NH$_2$CHO may result from the limitation of its low abundance.
  
  \item 
  The D/H ratios of formaldehyde ([D$_2$CO]/[H$_2$CO]) of these four sources are in general negatively correlated to their excitation temperatures. 
  These [D$_2$CO]/[H$_2$CO] as well as the the D/H ratio of methanol ([CH$_2$DOH]/[CH$_3$OH]) in G211.47-19.27S are comparable to those of other hot corinos in the literature except the [D$_2$CO]/[H$_2$CO] in IRAS 16293-2422 B.
  This may result from the difference of either their temperatures or their evolutionary stages.
  Note that the D/H ratio of methanol ([CH2DOH]/[CH$_3$OH]) is also possibly affected by the optical thickness of CH$_3$OH emission.
  
  \item 
  The hydrocarbons, c-C$_3$H$_2$ and CCD, are also detected in the four sources.
  They have cooler excitation temperatures and show more extended spatial distributions compared to COMs. 
  The HCCCN is detected in all the four sources and its isotope, HC$^{13}$CCN is detected in G211.47-19.27S. 
  The ratio [HC$^{13}$CCN]/[HCCCN] is an order of magnitude higher than the value in the low-mass star forming region L1527.
  
  \item
  About 8$\%$ of the protostellar objects in our survey are identified as hot corinos. 
  A more complete study of protostellar cores is required to reveal whether hot corino is a general stage of low-mass star formation.
\end{enumerate}


\acknowledgments
This paper makes use of the following ALMA data: ADS/JAO.ALMA\#2018.1.00302.S. ALMA is a partnership of ESO (representing its member states), NSF (USA) and NINS (Japan), together with NRC (Canada), MOST and ASIAA (Taiwan), and KASI (Republic of Korea), in cooperation with the Republic of Chile. The Joint ALMA Observatory is operated by ESO, AUI/NRAO and NAOJ.
SYH and SYL acknowledge support from the Ministry of Science and Technology (MoST) with grants 108-2112-M-001-048- and 108-2112-M-001-052-.
N. Hirano acknowledges support from the Ministry of Science and Technology (MoST) with grant 108-2112-M-001-017.
We thank Neal J. Evans and Siyi Feng for their useful comments in improving the manuscript.
JHe thanks the National Natural Science Foundation of China under grant Nos. 11873086 and U1631237 and support by the Yunnan Province of China (No.2017HC018). This work is sponsored (in part) by the Chinese Academy of Sciences (CAS) through a grant to the CAS South America Center for Astronomy CASSACA) in Santiago, Chile.
C.W.L. is supported by the Basic Science Research Program through the National Research Foundation of Korea (NRF) funded by the Ministry of Education, Science and Technology (NRF-2019R1A2C1010851).
DJ is supported by the National Research Council Canada and by an NSERC Discovery Grant.
LB acknowledges support from CONICYT project Basal AFB-170002.
P.S. was partially supported by a Grant-in-Aid for Scientific Research (KAKENHI Number 18H01259) of Japan Society for the Promotion of Science (JSPS).
Y.-L. Yang was supported, in part, by the Virginia Initiative on Cosmic Origins (VICO).
JEL and HWY are supported by the Basic Science Research Program through the National Research Foundation of Korea (grant No. NRF-2018R1A2B6003423) and the Korea Astronomy and Space Science Institute under the R\&D program supervised by the Ministry of Science, ICT and Future Planning.


%



\software{
astropy \citep{astropy:2013, astropy:2018},
CASA \citep{2007McMullin_CASA},
XCLASS \citep{2017Moller_XCLASS}
}

\clearpage
\appendix
\section{Observed Spectra of Sources \label{sec:Spectra}}
The spectra of all the four sources and the x-axis is the observed frequency.
The red curve is the simulation result of XCLASS. 
Tentative detections are indicated with molecule names in brackets.
The frequency bandwidth of each figure is 1~GHz for G211.47-19.27S and G208.68-19.20N1, and 2~GHz in G210.49-19.79W and G192.12-11.10.
\setcounter{figure}{0}
\renewcommand{\thefigure}{A\arabic{figure}}
\begin{table*}
\normalsize
\figcaption{\label{fig:Spectra_G211}Spectra of G211.47-19.27S.}
\centering
\setlength{\tabcolsep}{0pt} 
\renewcommand{\arraystretch}{0.} 
\begin{tabular}{c}
\includegraphics[width=.93\textwidth]{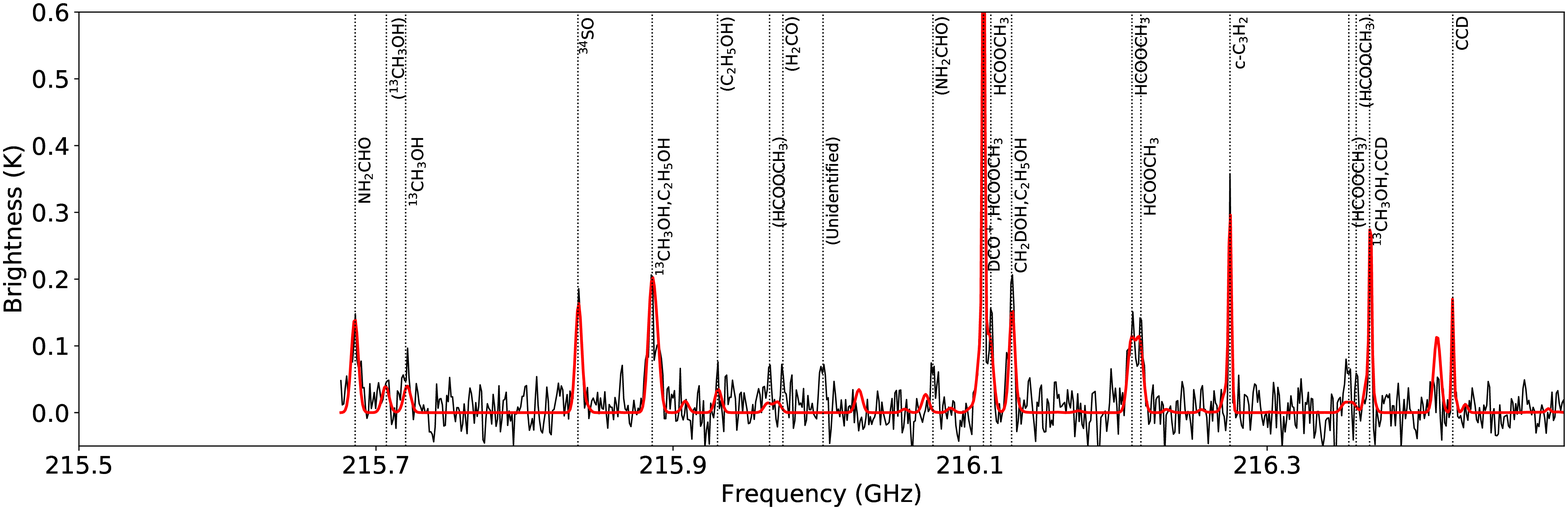} \\
\includegraphics[width=.93\textwidth]{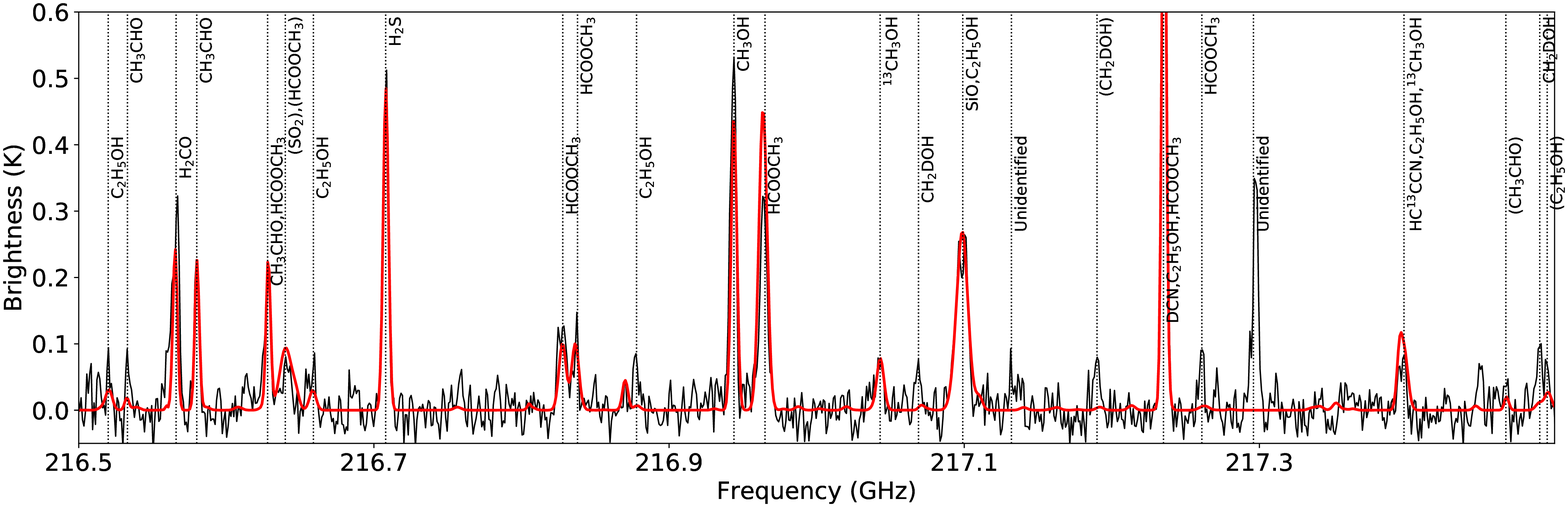} \\
\includegraphics[width=.93\textwidth]{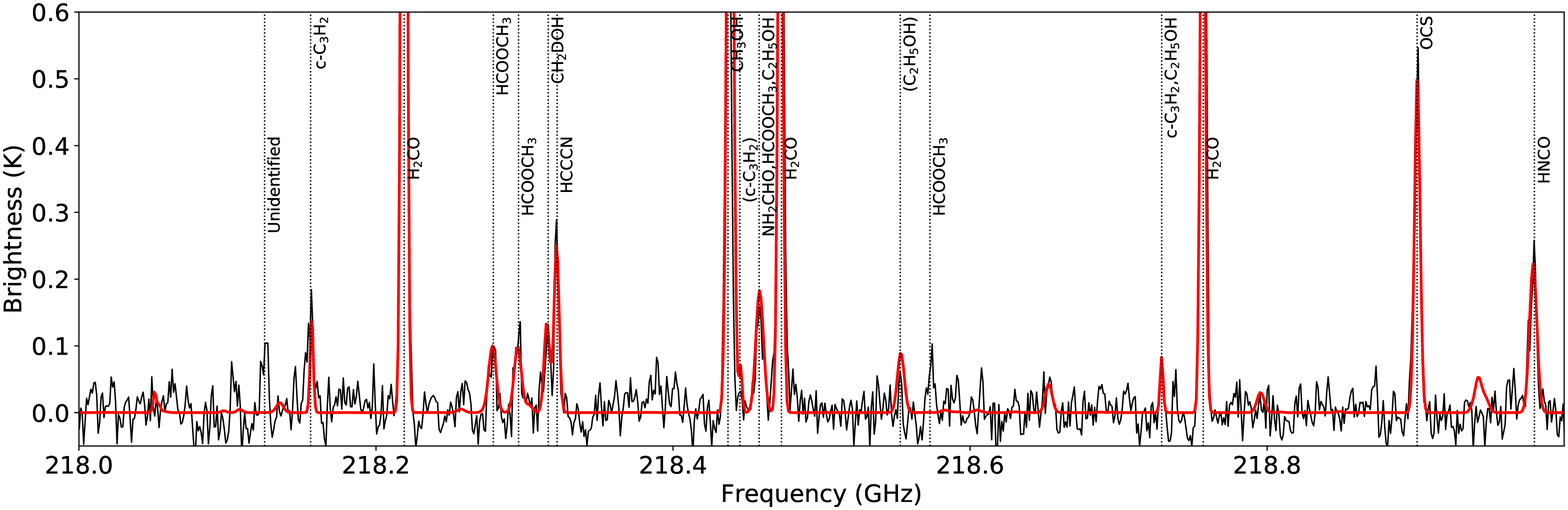} \\
\includegraphics[width=.93\textwidth]{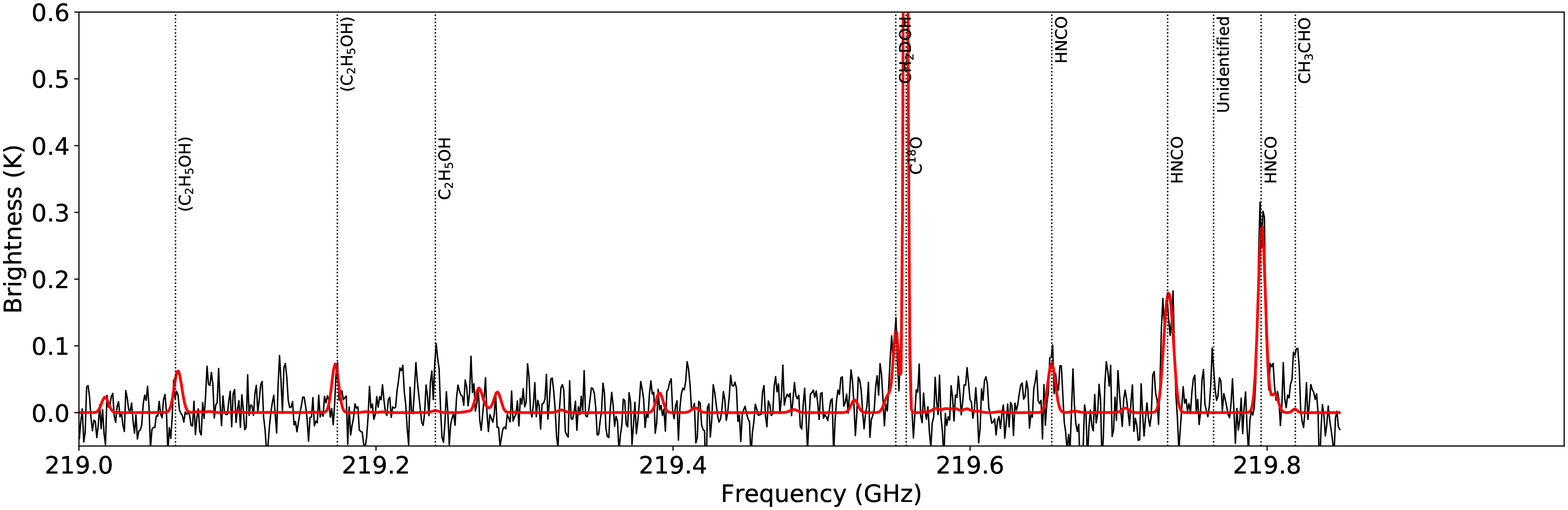} \\
\end{tabular}
\end{table*}

\begin{table*}
\centering
\setlength{\tabcolsep}{0pt} 
\renewcommand{\arraystretch}{0.} 
\begin{tabular}{c}
\textbf{Figure} \ref{fig:Spectra_G211} (\textit{continued})\\
\includegraphics[width=.93\textwidth]{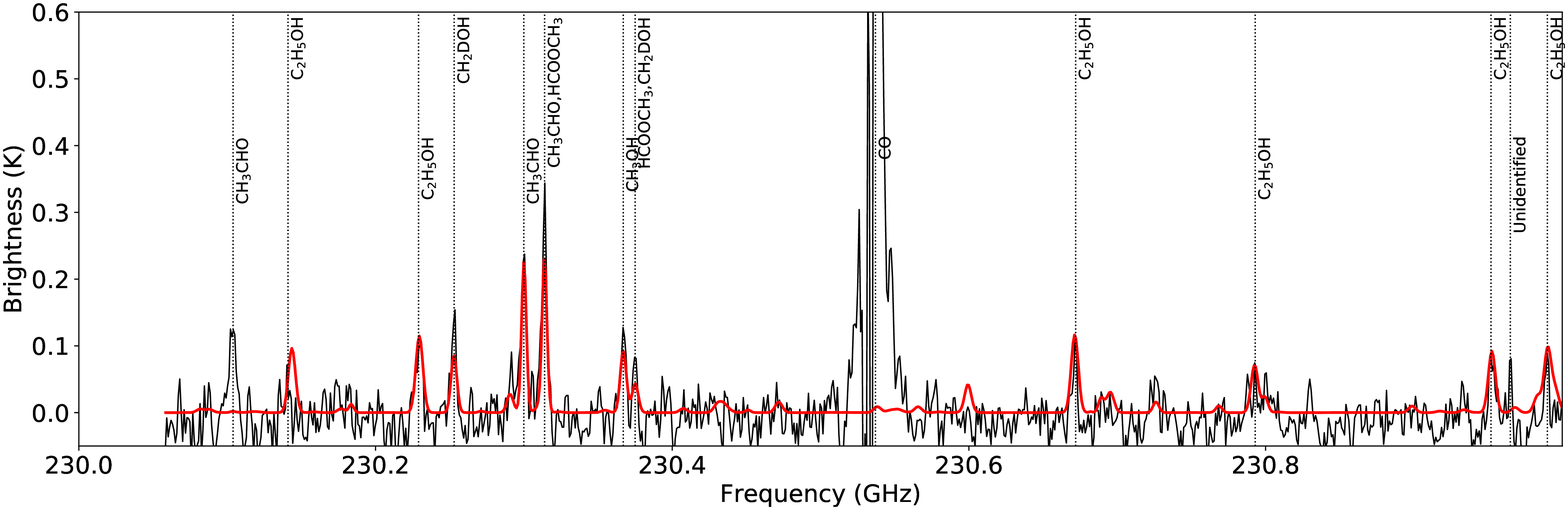} \\
\includegraphics[width=.93\textwidth]{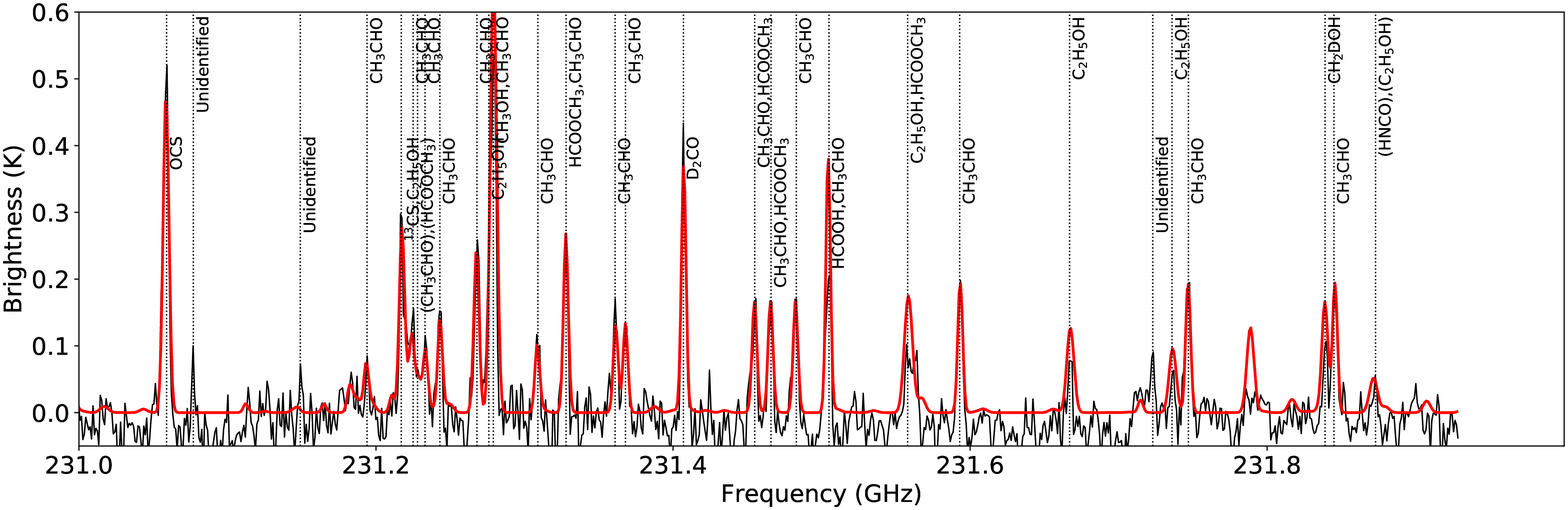} \\
\includegraphics[width=.93\textwidth]{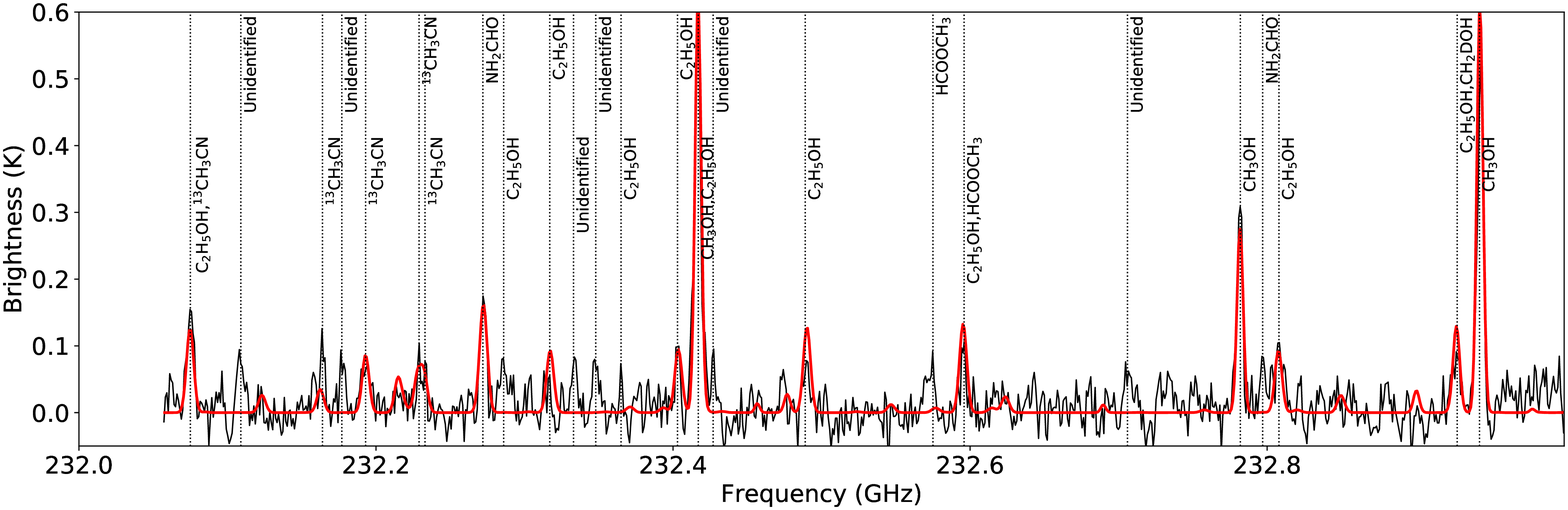} \\
\includegraphics[width=.93\textwidth]{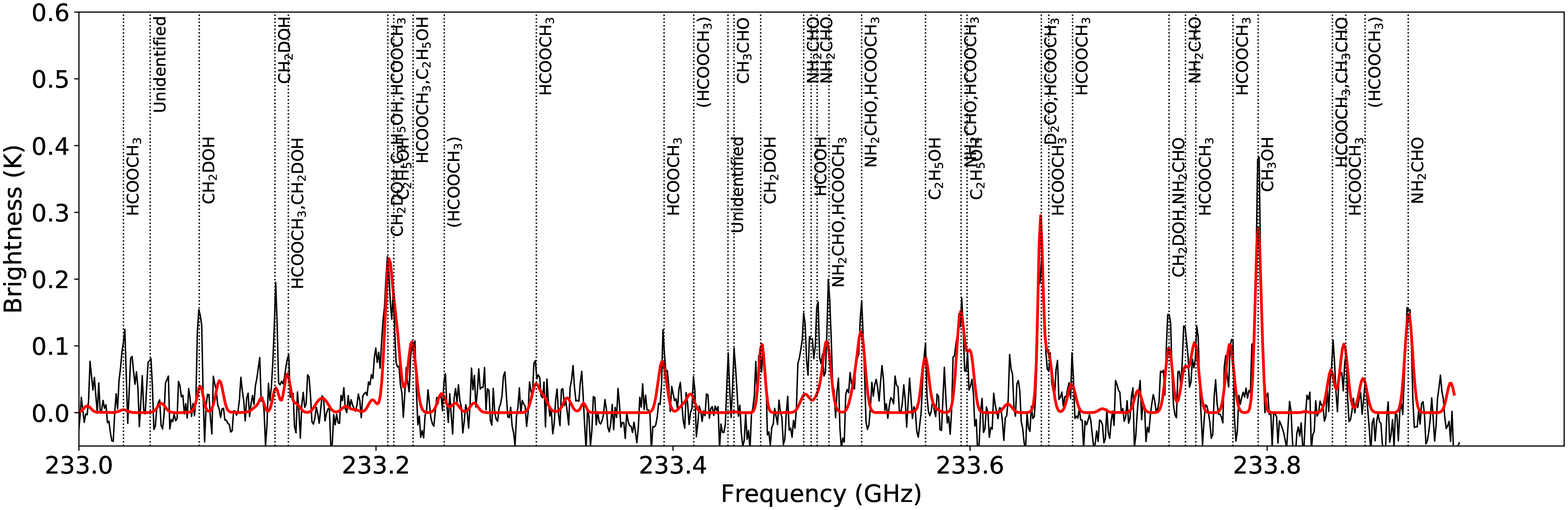} \\
\end{tabular}
\end{table*}
\begin{table*}
\figcaption{\label{fig:Spectra_G208}Spectra of G208.68-19.20N1.}
\centering
\setlength{\tabcolsep}{0pt} 
\renewcommand{\arraystretch}{0.} 
\begin{tabular}{c}
\includegraphics[width=.93\textwidth]{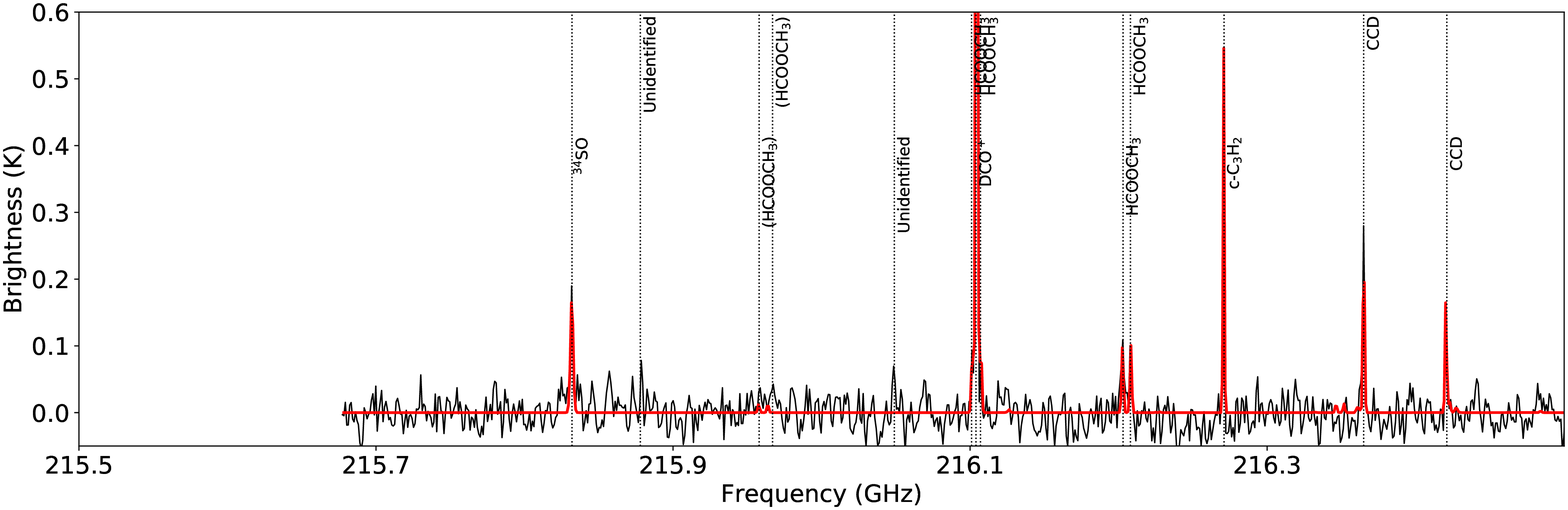} \\
\includegraphics[width=.93\textwidth]{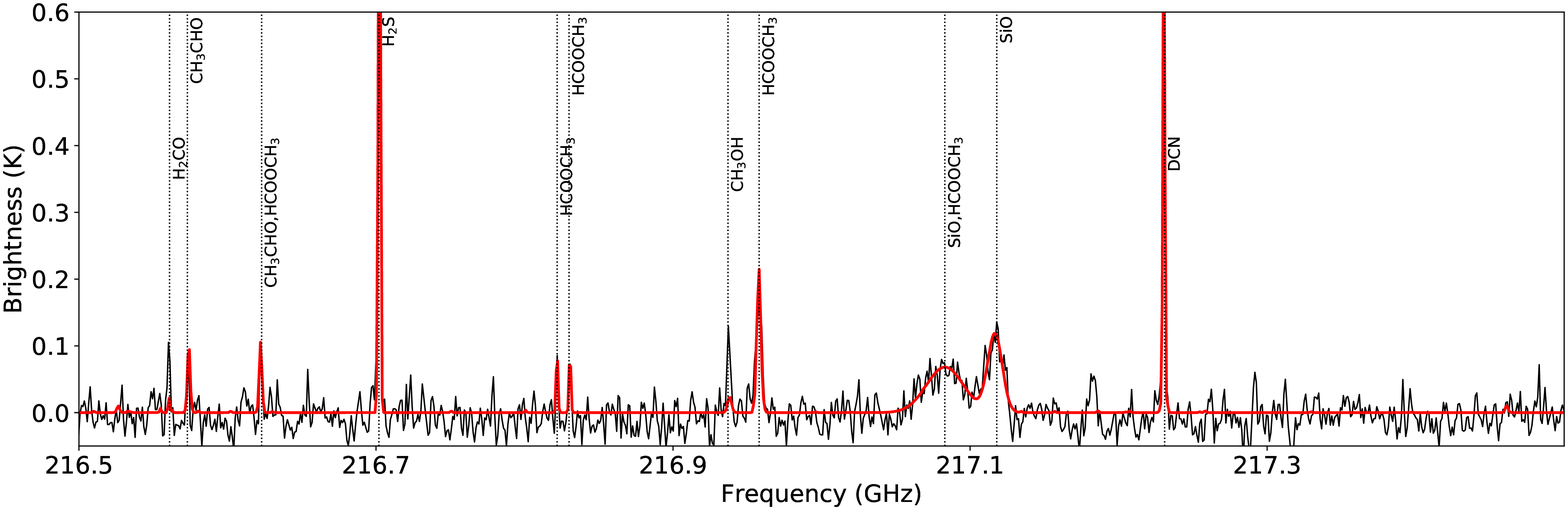} \\
\includegraphics[width=.93\textwidth]{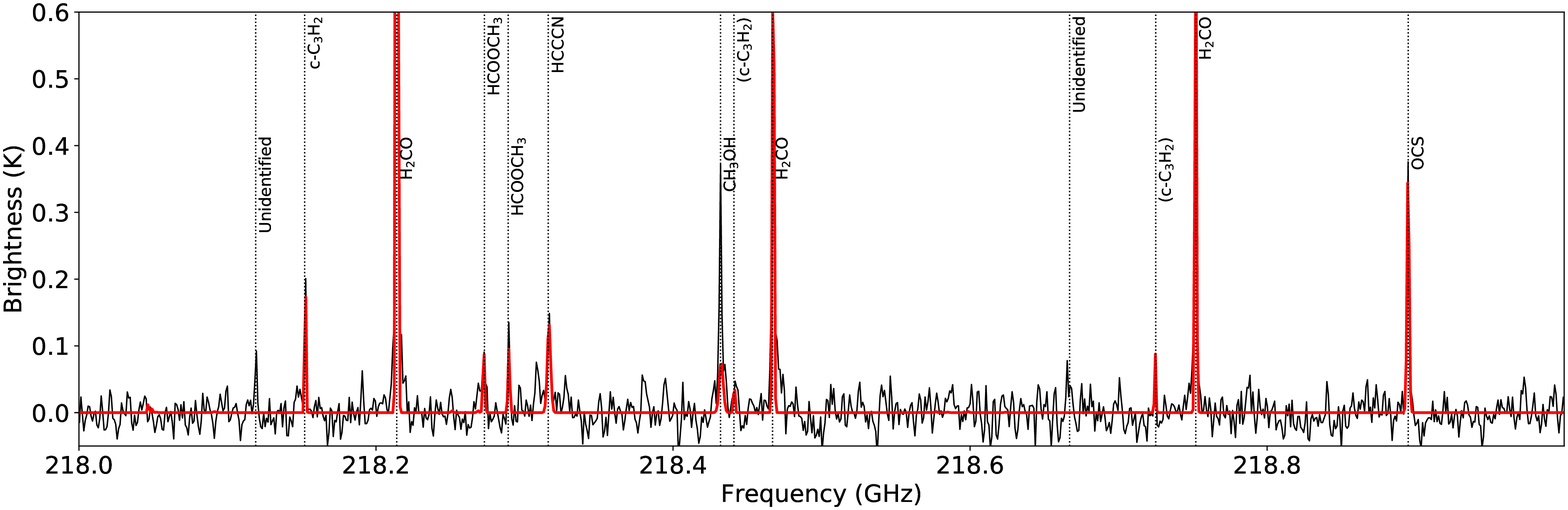} \\
\includegraphics[width=.93\textwidth]{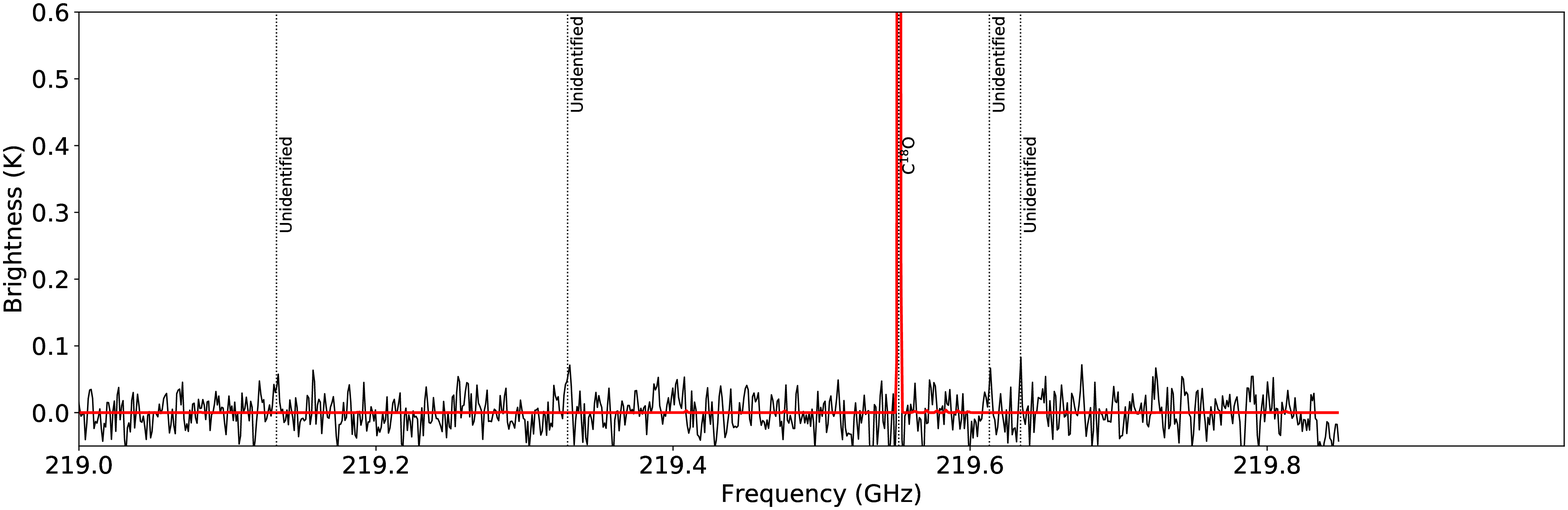} \\
\end{tabular}
\end{table*}

\begin{table*}
\centering
\setlength{\tabcolsep}{0pt} 
\renewcommand{\arraystretch}{0.} 
\begin{tabular}{c}
\textbf{Figure} \ref{fig:Spectra_G208} (\textit{continued})\\
\includegraphics[width=.93\textwidth]{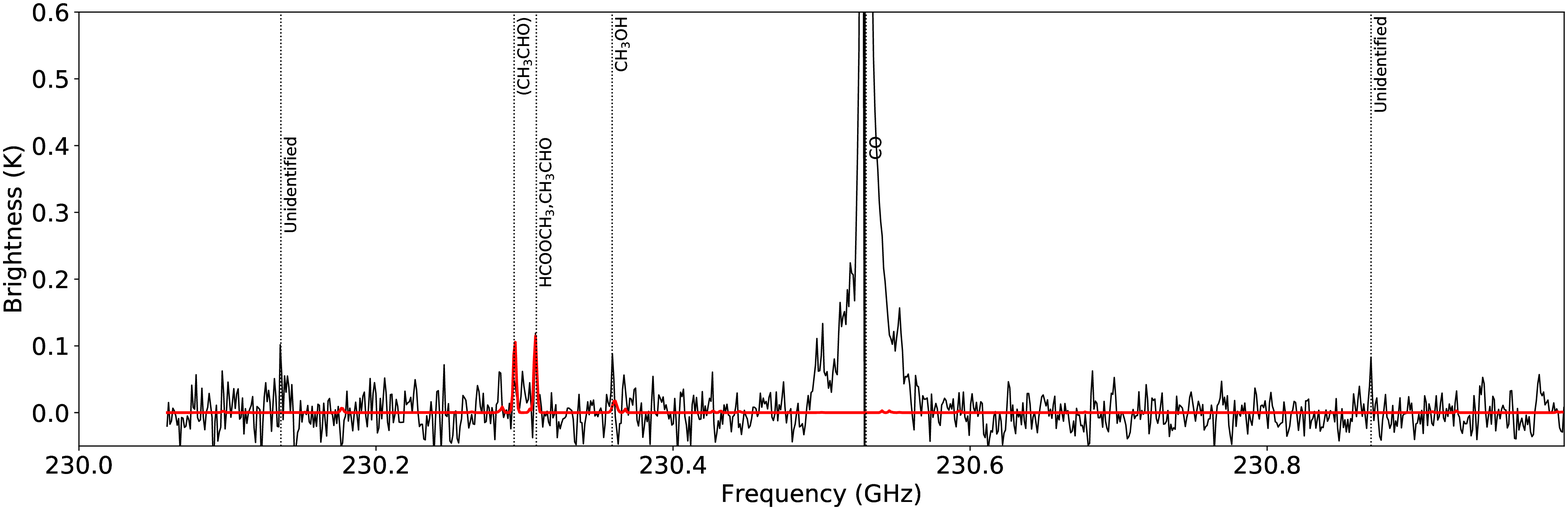} \\
\includegraphics[width=.93\textwidth]{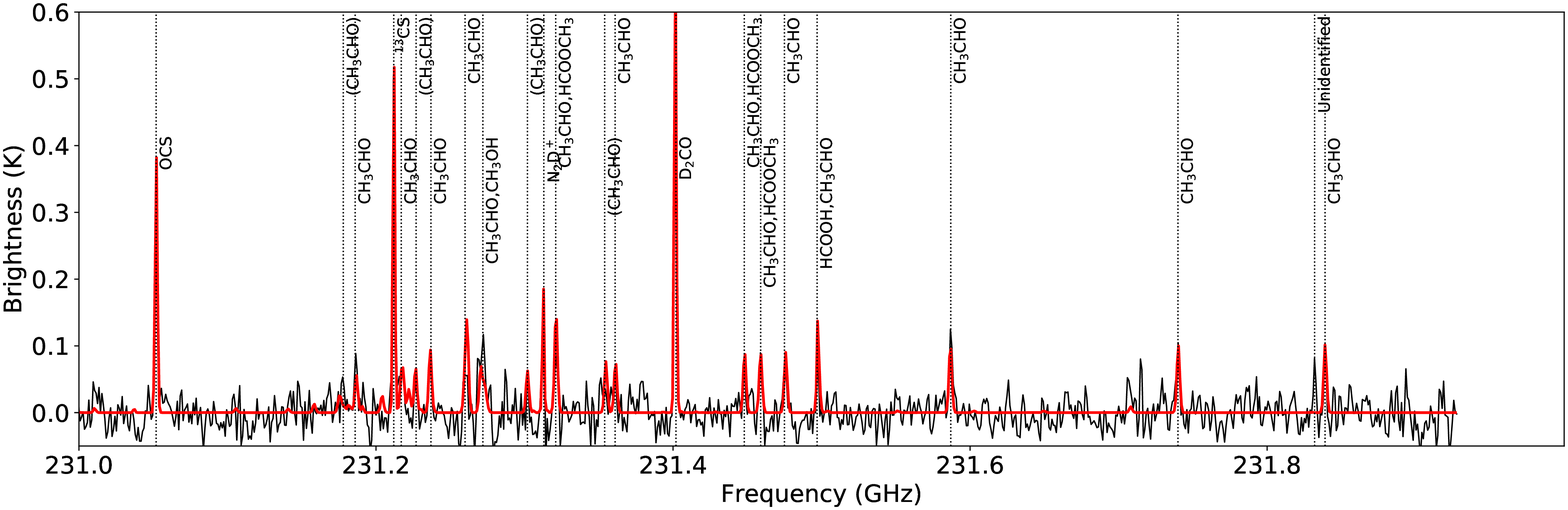} \\
\includegraphics[width=.93\textwidth]{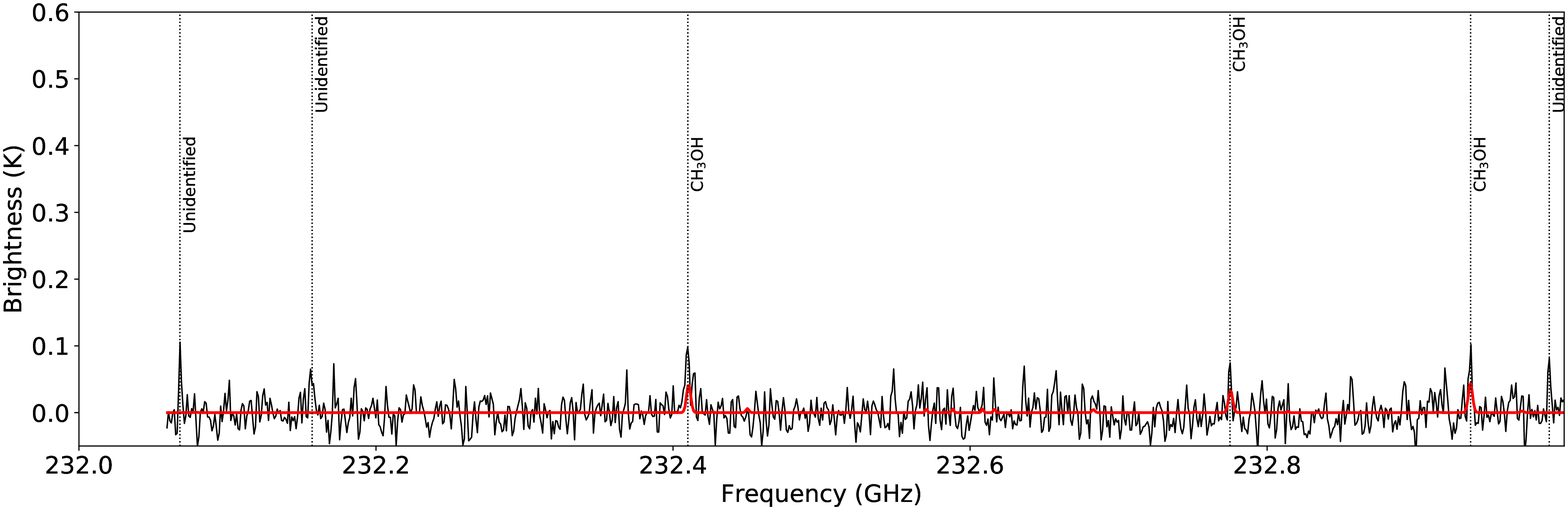} \\
\includegraphics[width=.93\textwidth]{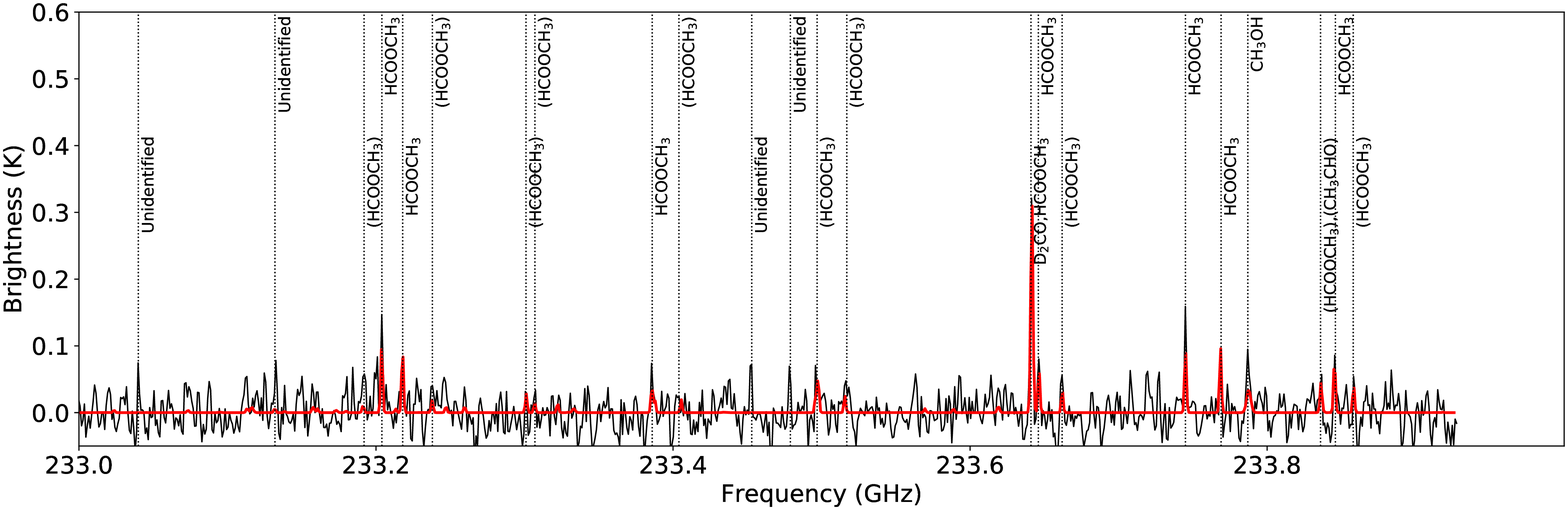} \\
\end{tabular}
\end{table*}
\begin{table*}
\figcaption{\label{fig:Spectra_G210}Spectra of G210.49-19.79W.}
\centering
\setlength{\tabcolsep}{0pt} 
\renewcommand{\arraystretch}{0.} 
\begin{tabular}{c}
\includegraphics[width=.93\textwidth]{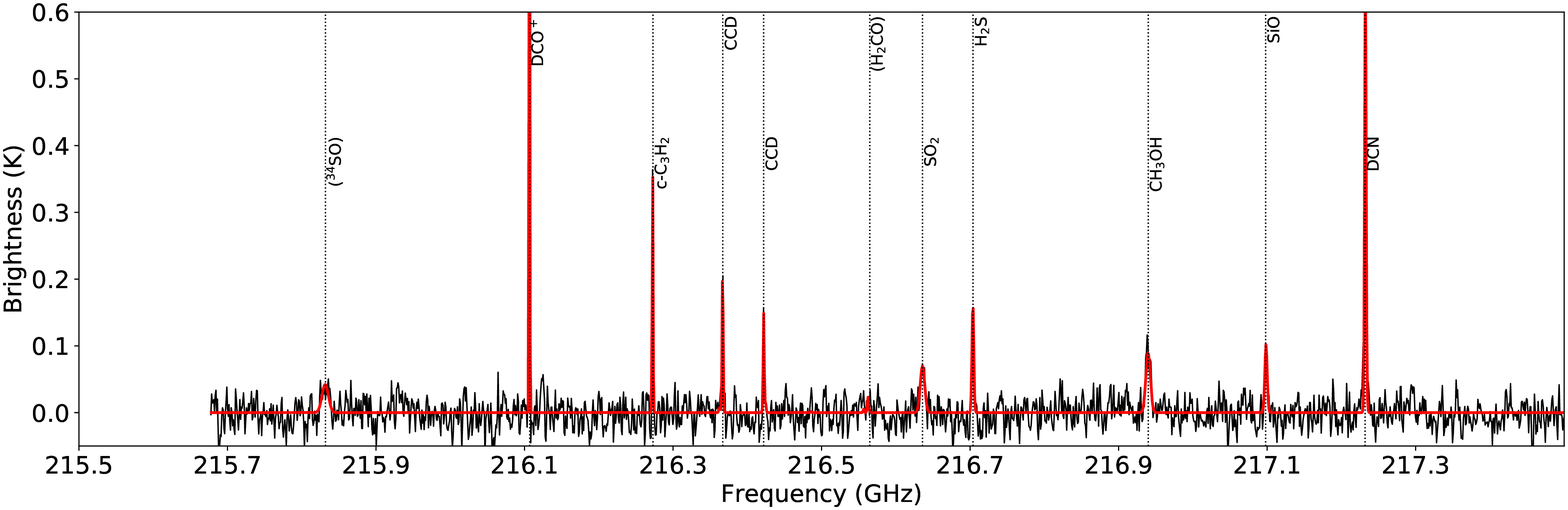} \\
\includegraphics[width=.93\textwidth]{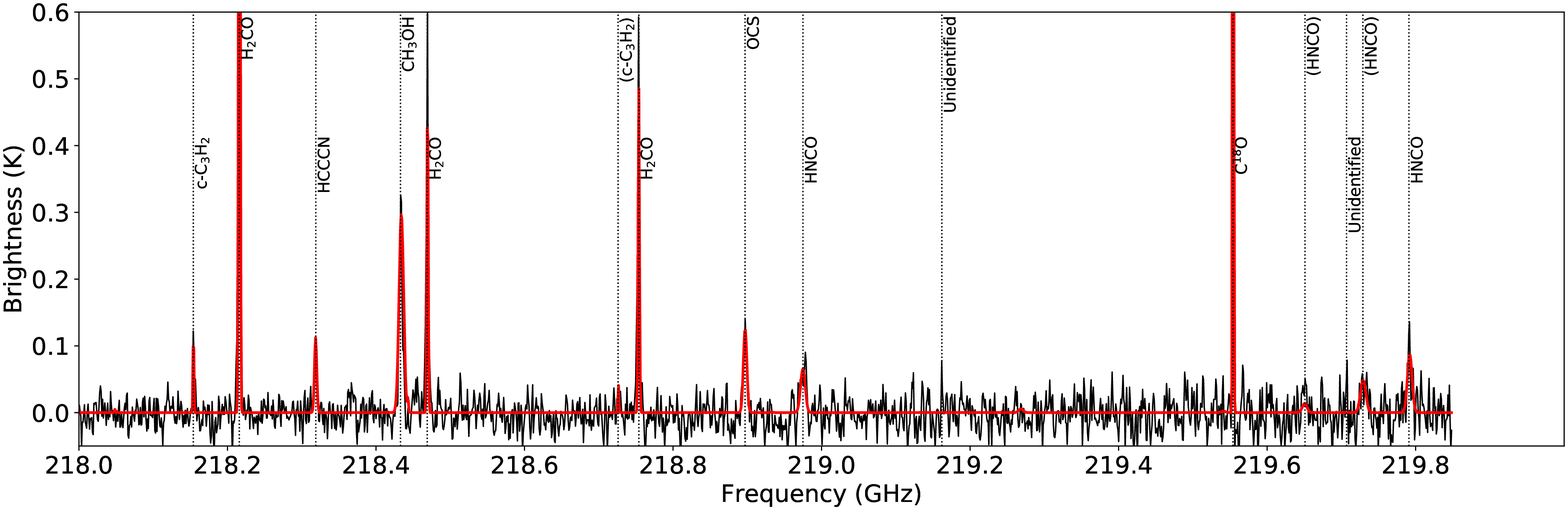} \\
\includegraphics[width=.93\textwidth]{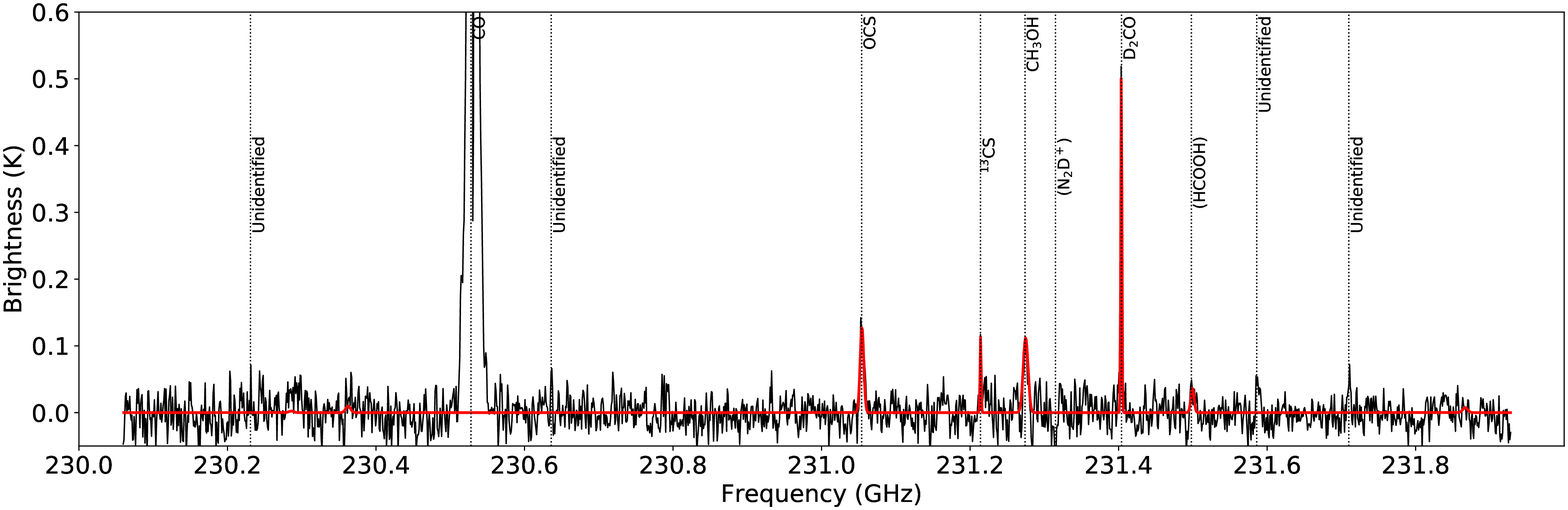} \\
\includegraphics[width=.93\textwidth]{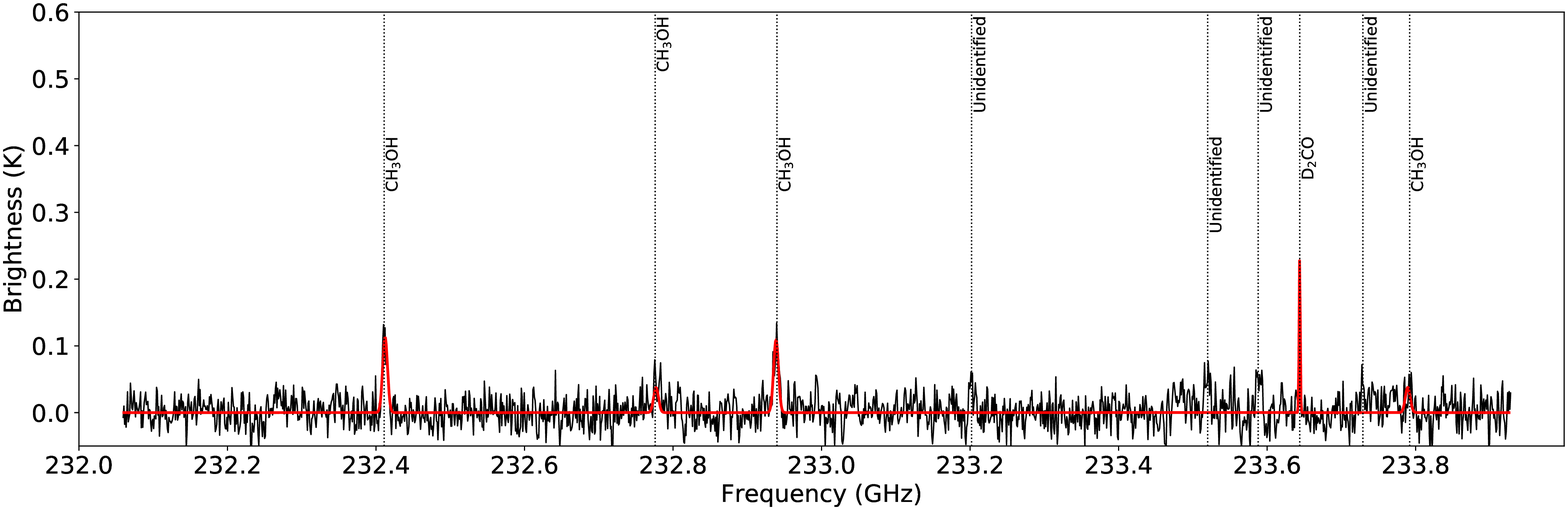} \\
\end{tabular}
\end{table*}

\begin{table*}
\figcaption{\label{fig:Spectra_G192}Spectra of G192.12-11.10.}
\centering
\setlength{\tabcolsep}{0pt} 
\renewcommand{\arraystretch}{0} 
\begin{tabular}{c}
\includegraphics[width=.93\textwidth]{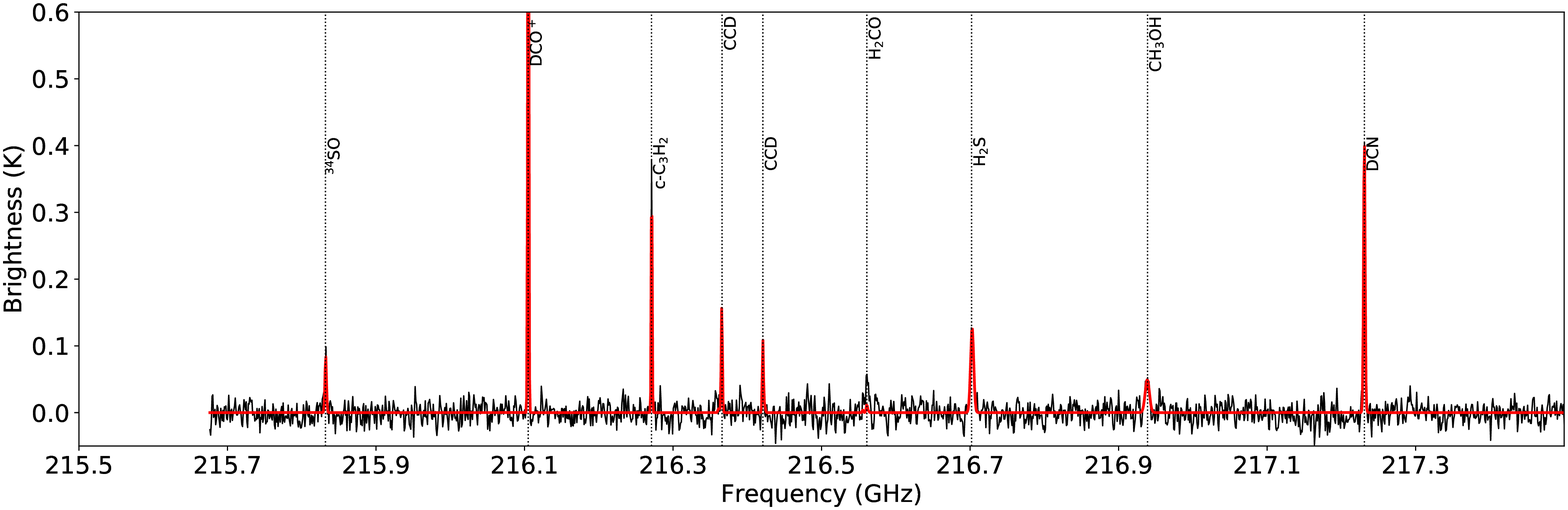} \\
\includegraphics[width=.93\textwidth]{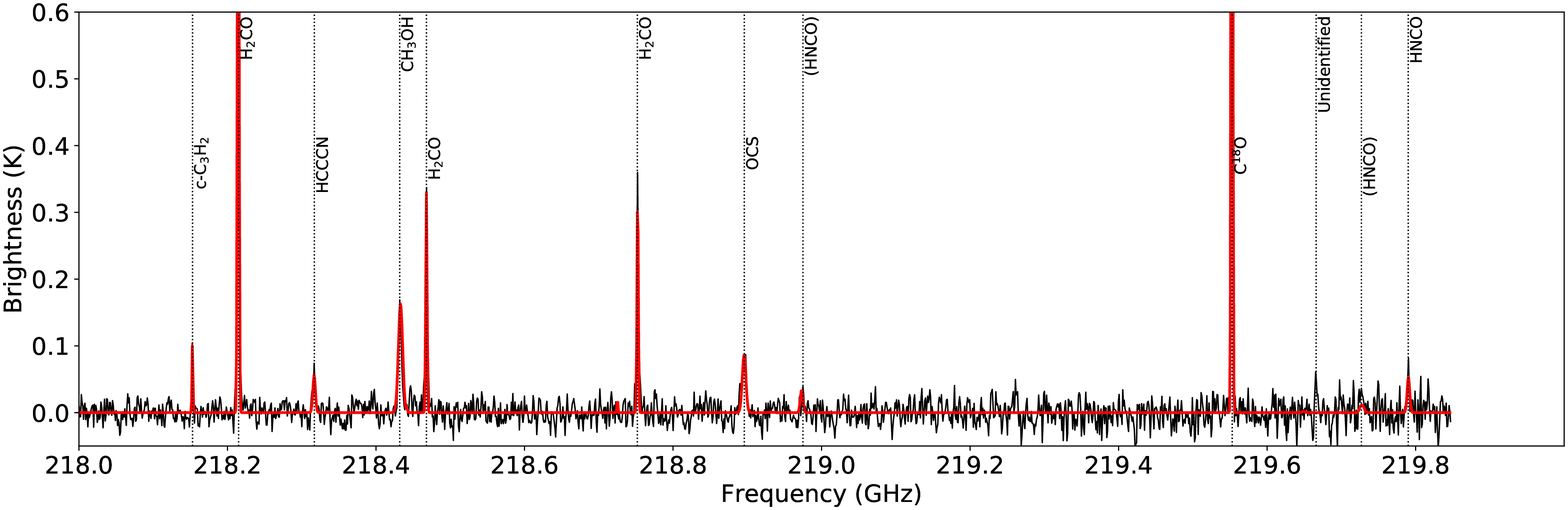} \\
\includegraphics[width=.93\textwidth]{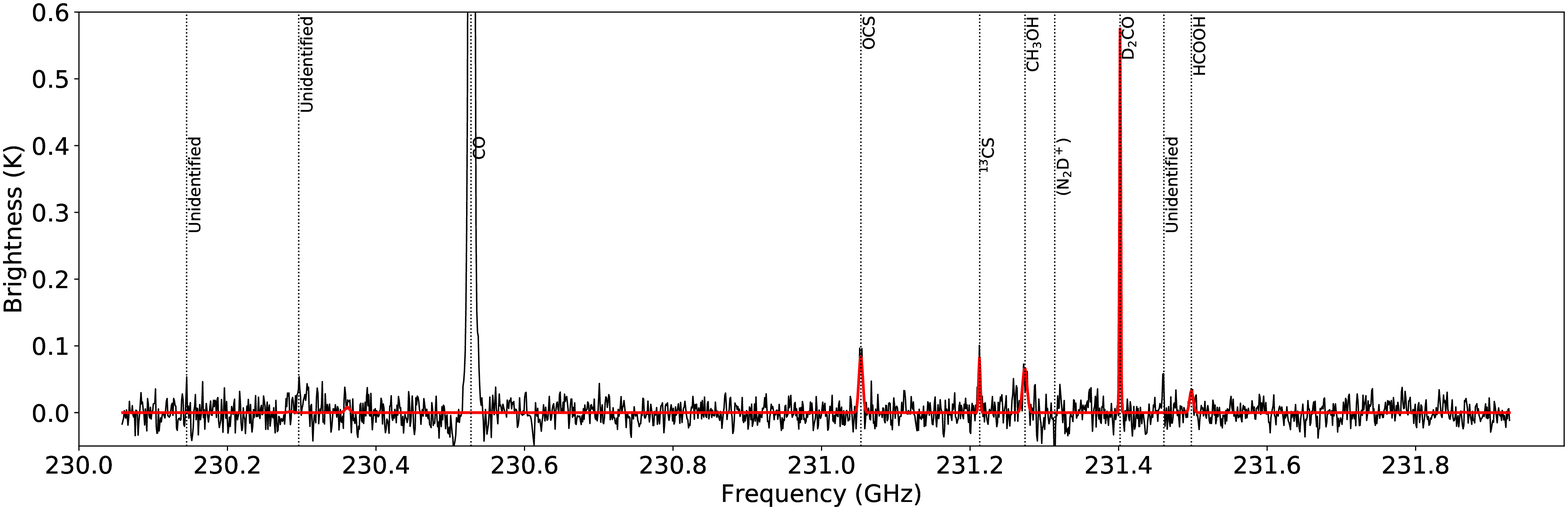} \\
\includegraphics[width=.93\textwidth]{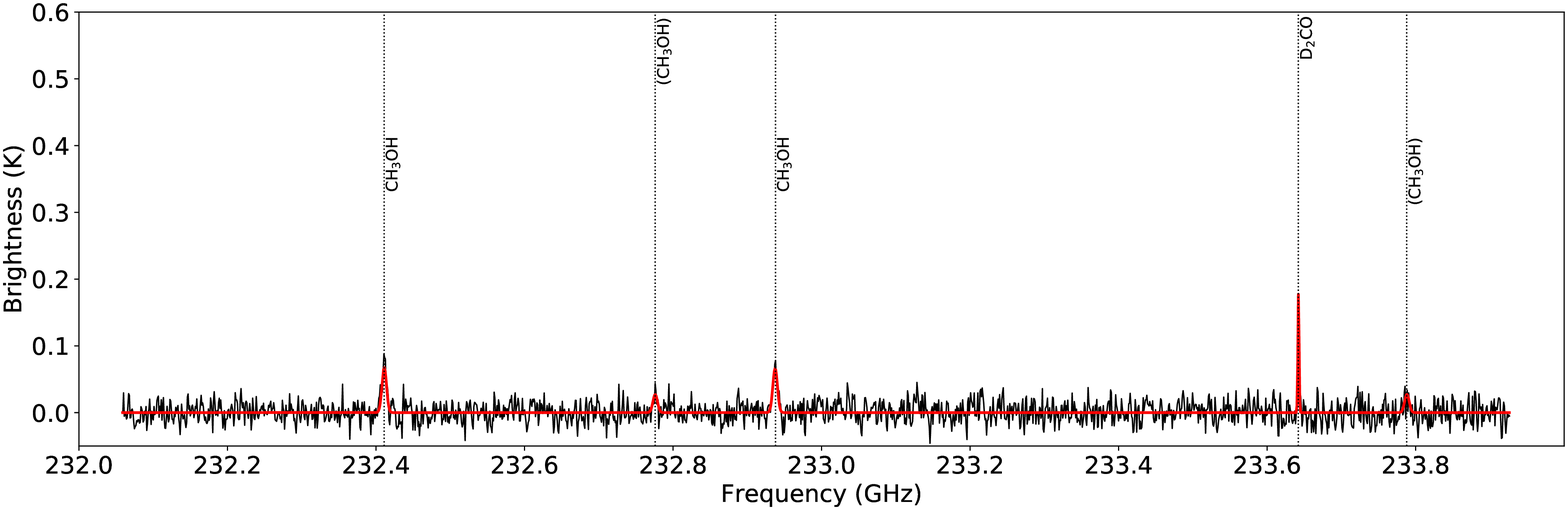} \\
\end{tabular}
\end{table*}

\newpage
\section{Molecular Line List \label{sec:MolLineList}}
Detected transitions of all the four sources.
The $f^\mathrm{obs}$ is the observed frequency.
The $T_\mathrm{p}$ is the peak brightness temperature.
The $\Delta v$ is the FWHM line width.
The $W$ is the integrated brightness within the FWHM of each line.
In the Formula column, the rest frequencies in MHz are provided in bracket.
It is denoted as '*N' if there are N transitions of the same species at the same rest frequency. In the Note column:\\
\begin{itemize}
\setlength{\parsep}{0pt} 
\setlength{\itemsep}{0pt}
\setlength{\parskip}{0pt}
\item ~U: Unidentified transition.
\item ~I: Line ignored in the fitting. They are CO $J$=2-1 transitions lack of Gaussian line profile.
\item ~N: Negative.
\item ~T: Tentative detection. Their intensity is below 3$\sigma_\mathrm{Chn}$.
\item ~B: Blended. They are different transitions of either the same species or different species.
\item UE: Under-estimated by XCLASS. The observed brightness temperature is above 3$\sigma_\mathrm{Chn}$ but the brightness temperature calculated by XCLASS is below 2$\sigma_\mathrm{Chn}$. There may be blended and unidentified transitions or the other components of molecules.
\end{itemize}

\setcounter{table}{0}
\renewcommand{\thetable}{B\arabic{table}}
{
\begingroup
\scriptsize
\setlength{\tabcolsep}{2pt} 
\renewcommand{\arraystretch}{0.9} 
\begin{longtable*}[l]{p{0.06\textwidth}p{0.04\textwidth}p{0.04\textwidth}p{0.06\textwidth}p{0.70\textwidth}p{0.05\textwidth}}
\caption{Molecular line list of G211.47-19.27S.\label{tab:MolLineList_G211}}\\
\hline\hline
$f^\mathrm{obs}$ & $T_\mathrm{p}$ & $\Delta v$ & $W$ & Formula [$f^\mathrm{rest}$] & Note \\
MHz & K & km/s & K km/s & MHz & \\
\hline
\endfirsthead
 \multicolumn{6}{l}{\textbf{Table} \ref{tab:MolLineList_G211} (\textit{continued})}\\
\hline\hline
$f^\mathrm{obs}$ & $T_\mathrm{p}$ & $\Delta v$ & $W$ & Formula [$f^\mathrm{rest}$] & Note \\
MHz & K & km/s & K km/s & MHz & \\
\hline
\endhead
 \endfoot
\endlastfoot
215686 & 0.14 & 6.95 & 0.79 & NH$_2$CHO[215688] &  \\
215707 & 0.05 & 4.17 & 0.18 & $^{13}$CH$_3$OH[215708] & T \\
215720 & 0.09 & 4.17 & 0.3 & $^{13}$CH$_3$OH[215722] & UE \\
215836 & 0.17 & 7.03 & 0.97 & $^{34}$SO[215840] &  \\
215886 & 0.18 & 6.94 & 1.01 & $^{13}$CH$_3$OH[215887]; C$_2$H$_5$OH[215890] & B \\
215930 & 0.07 & 2.78 & 0.16 & C$_2$H$_5$OH[215932] & T \\
215965 & 0.07 & 2.78 & 0.16 & HCOOCH$_3$[215966] & T \\
215974 & 0.07 & 2.78 & 0.16 & H$_2$CO[215976] & T \\
\hline
\end{longtable*}
\endgroup
}
{
\begingroup
\scriptsize
\setlength{\tabcolsep}{2pt} 
\renewcommand{\arraystretch}{0.9} 
\begin{longtable*}[l]{p{0.06\textwidth}p{0.04\textwidth}p{0.04\textwidth}p{0.06\textwidth}p{0.70\textwidth}p{0.05\textwidth}}
\caption{Molecular line list of G208.68-19.20N1.\label{tab:MolLineList_G208}}\\
\hline\hline
$f^\mathrm{obs}$ & $T_\mathrm{p}$ & $\Delta v$ & $W$ & Formula [$f^\mathrm{rest}$] & Note \\
MHz & K & km/s & K km/s & MHz & \\
\hline
\endfirsthead
 
\multicolumn{6}{l}{\textbf{Table} \ref{tab:MolLineList_G208} (\textit{continued})}\\
\hline\hline
$f^\mathrm{obs}$ & $T_\mathrm{p}$ & $\Delta v$ & $W$ & Formula [$f^\mathrm{rest}$] & Note \\
MHz & K & km/s & K km/s & MHz & \\
\hline
\endhead
 \endfoot
\endlastfoot
215832 & 0.19 & 2.63 & 0.4 & $^{34}$SO[215840] &  \\
215878 & 0.08 & 1.39 & 0.09 & Unidentified & U \\
215958 & 0.04 & 3.07 & 0.1 & HCOOCH$_3$[215966] & T \\
215967 & 0.05 & 4.16 & 0.15 & HCOOCH$_3$[215976] & T \\
216049 & 0.07 & 2.78 & 0.16 & Unidentified & U \\
216101 & 0.07 & 2.77 & 0.16 & HCOOCH$_3$[216110] &  \\
216104 & 1.72 & 2.01 & 2.81 & DCO$^+$[216113] &  \\
216107 & 0.07 & 1.17 & 0.09 & HCOOCH$_3$[216116*2] & B \\
\hline
\end{longtable*}
\endgroup
}
{
\begingroup
\scriptsize
\setlength{\tabcolsep}{2pt} 
\renewcommand{\arraystretch}{0.9} 
\begin{longtable*}[l]{p{0.06\textwidth}p{0.04\textwidth}p{0.04\textwidth}p{0.06\textwidth}p{0.70\textwidth}p{0.05\textwidth}}
\caption{Molecular line list of G210.49-19.79W.\label{tab:MolLineList_G210}}\\
\hline\hline
$f^\mathrm{obs}$ & $T_\mathrm{p}$ & $\Delta v$ & $W$ & Formula [$f^\mathrm{rest}$] & Note \\
MHz & K & km/s & K km/s & MHz & \\
\hline
\endfirsthead
 
\multicolumn{6}{l}{\textbf{Table} \ref{tab:MolLineList_G210} (\textit{continued})}\\
\hline\hline
$f^\mathrm{obs}$ & $T_\mathrm{p}$ & $\Delta v$ & $W$ & Formula [$f^\mathrm{rest}$] & Note \\
MHz & K & km/s & K km/s & MHz & \\
\hline
\endhead
 
\endfoot
\endlastfoot
215832 & 0.05 & 11.11 & 0.43 & $^{34}$SO[215840] & T \\
216107 & 1.12 & 2.77 & 2.52 & DCO$^+$[216113] &  \\
216273 & 0.32 & 2.77 & 0.72 & c-C$_3$H$_2$[216279] &  \\
216367 & 0.23 & 2.25 & 0.41 & CCD[216373*3] & B \\
216422 & 0.18 & 2.05 & 0.29 & CCD[216428*2, 216429] & B \\
216565 & 0.03 & 4.24 & 0.11 & H$_2$CO[216569] & T \\
216636 & 0.07 & 6.92 & 0.41 & SO$_2$[216643] &  \\
216704 & 0.16 & 4.31 & 0.55 & H$_2$S[216710] &  \\
\hline
\end{longtable*}
\endgroup
}
{
\begingroup
\begin{scriptsize}
\setlength{\tabcolsep}{2pt} 
\renewcommand{\arraystretch}{0.9} 
\begin{longtable*}[l]{p{0.06\textwidth}p{0.04\textwidth}p{0.04\textwidth}p{0.06\textwidth}p{0.70\textwidth}p{0.05\textwidth}}
\caption{Molecular line list of G192.12-11.10.\label{tab:MolLineList_G192}}\\
\hline\hline
$f^\mathrm{obs}$ & $T_\mathrm{p}$ & $\Delta v$ & $W$ & Formula [$f^\mathrm{rest}$] & Note \\
MHz & K & km/s & K km/s & MHz & \\
\hline
\endfirsthead
\multicolumn{6}{l}{\textbf{Table} \ref{tab:MolLineList_G192} (\textit{continued})}\\
\hline\hline
$f^\mathrm{obs}$ & $T_\mathrm{p}$ & $\Delta v$ & $W$ & Formula [$f^\mathrm{rest}$] & Note \\
MHz & K & km/s & K km/s & MHz & \\
\hline
\endhead
\endfoot
\endlastfoot
215832 & 0.09 & 2.77 & 0.21 & $^{34}$SO[215840] &  \\
216105 & 1.37 & 2.3 & 2.55 & DCO$^+$[216113] &  \\
216271 & 0.39 & 2.1 & 0.66 & c-C$_3$H$_2$[216279] &  \\
216366 & 0.15 & 2.87 & 0.36 & CCD[216373*3] & B \\
216421 & 0.11 & 2.51 & 0.23 & CCD[216428*2, 216429] & B \\
216561 & 0.06 & 6.48 & 0.29 & H$_2$CO[216569] & UE \\
216702 & 0.14 & 6.92 & 0.76 & H$_2$S[216710] &  \\
216939 & 0.05 & 6.91 & 0.27 & CH$_3$OH[216946] &  \\
\hline
\end{longtable*}
\end{scriptsize}
\endgroup
}

\newpage
\section{Transition List \label{sec:trantision}}
\setcounter{table}{0}
\renewcommand{\thetable}{C\arabic{table}}
\begingroup
\startlongtable
\setlength{\tabcolsep}{6pt} 
\begin{deluxetable*}{llclccccc}
\def\arraystretch{0.8} 
\tabletypesize{\scriptsize}
\tablewidth{\columnwidth}
\tablecolumns{6}
\tablecaption{\label{tab:TransList} List of identified transitions. }
\tablehead{\\
\colhead{Name} & \colhead{Formula} & \colhead{$f_\mathrm{rest}$} & \colhead{Transition Number} & \colhead{$E_u$} & \colhead{$g_u$} & \colhead{$A_u$}  & \colhead{$I_\mathrm{CDMS.JPL}$}  & \colhead{Ref.}\\ [-1ex]
\colhead{} & \colhead{} & \colhead{(MHz)} & \colhead{} & \colhead{(K)} & \colhead{} & \colhead{(s$^{-1}$)}  & \colhead{}  & \colhead{}
} 
\startdata
Acetaldehyde & CH$_3$CHO & 216534 & 14(3,11) - 14(2,12) E & 117.68 & 58 & 3.45e-05 & -4.3787 & JPL \\
& CH$_3$CHO & 216582 & 11(1,10) - 10(1,9) E & 64.87 & 46 & 3.55e-04 & -3.3908 & JPL \\
& CH$_3$CHO & 216630 & 11(1,10) - 10(1,9) A & 64.81 & 46 & 3.55e-04 & -3.3908 & JPL \\
& CH$_3$CHO & 217469 & 14(3,11)-14(2,12) A & 117.71 & 58 & 3.82e-05 & -4.3365 & JPL \\
& CH$_3$CHO & 219820 & 4(2,3) - 3(1,3) E & 1176.53 & 63 & 1.28e-07 & -5.1 & JPL \\
& CH$_3$CHO & 230106 & 26(3,23) - 25(4,22) E & 348.37 & 106 & 1.41e-05 & -4.8651 & JPL \\
& CH$_3$CHO & 230302 & 12(2,11) - 11(2,10) A & 81.04 & 50 & 4.19e-04 & -3.3313 & JPL \\
& CH$_3$CHO & 230316 & 12(2,11) - 11(2,10) E & 81.05 & 50 & 4.19e-04 & -3.3314 & JPL \\
\enddata
\tablecomments{
The $f_\mathrm{rest}$ is the rest frequency in MHz, 
the $E_u$ is the energy of the upper state in K,
the $g_u$ is the degeneracy factor,
the $A_u$ is the Einstein A coefficient in s$^{-1}$,
the $I_\mathrm{CDMS,JPL}$ is the CDMS/JPL intensity,
the transition number is in the form of upper state - lower state. 
}
\tablerefs{
Jet Propulsion Laboratory Molecular Spectroscopy \citep[JPL,][]{1988JPL};
Cologne Database for Molecular Spectroscopy \citep[CDMS,][]{2005CDMS}
; eXtended CASA Line Analysis Software Suite \citep[XCLASS,][]{2017Moller_XCLASS}
}
\end{deluxetable*}
\endgroup

\onecolumngrid 
\section{Population Diagram of Methanol \label{sec:rotDiagram}}
The population diagram method is also a common tool for analyzing the excitation of molecular emission \citep{1999Goldsmith_popdiagram}.
In this section, we show the population diagrams of CH$_3$OH in all the four sources.
We have excluded in the analysis the transition $J(K_a, K_c)$=4(2,2)-3(1,2) E ($f_\mathrm{rest}$=218440 MHz; $E_u$=45 K) since it has been reported to be possibly masing in different astrophysical objects \citep{2014Hunter_CH3OH_maser, 2016Leurini_CH3OH_maser, 2017Yuan_CH3OH_maser}.

The total column density ($N_\mathrm{tot}$) and the rotational temperature derived from the rotational diagrams are in general consistent with the MAGIX results (see Sect. \ref{sec:ResultMolEmission.tex}).
The data points in G208 at lower ($< 200$K) and higher ($> 200$K) energy levels seem to have different slopes. This could indicate the possibility of multiple molecular emission components in different excitation temperatures (e.g. cold envelope and hot corino).
\setcounter{figure}{0}
\renewcommand{\thefigure}{D\arabic{figure}}
\begin{figure*}
\centering
\caption{\label{fig:pop} The rotational diagrams (RD) of CH$_3$OH. 
The red line shows the best-fit result from the RD analysis and the yellow area represents the 1$\sigma$ interval.
We adopted the molecular parameters, including the partition function, from Cologne Database for Molecular Spectroscopy \citep[CDMS, ][]{2005CDMS}.
The magenta squares mark the transition $J(K_a, K_c)$=4(2,2)-3(1,2) E which is not used in the fitting (See Appendix \ref{sec:rotDiagram}).
}
\plottwo{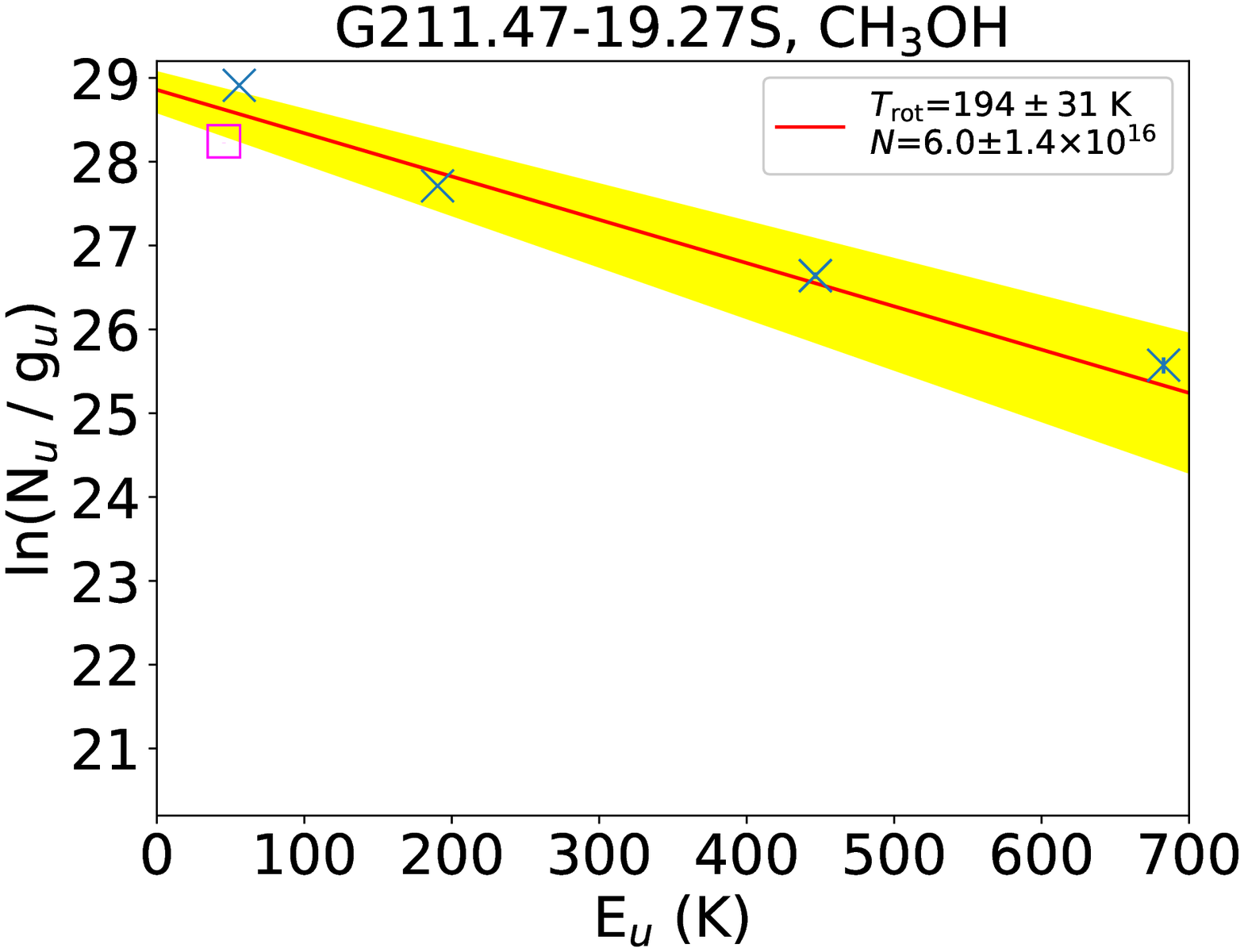}{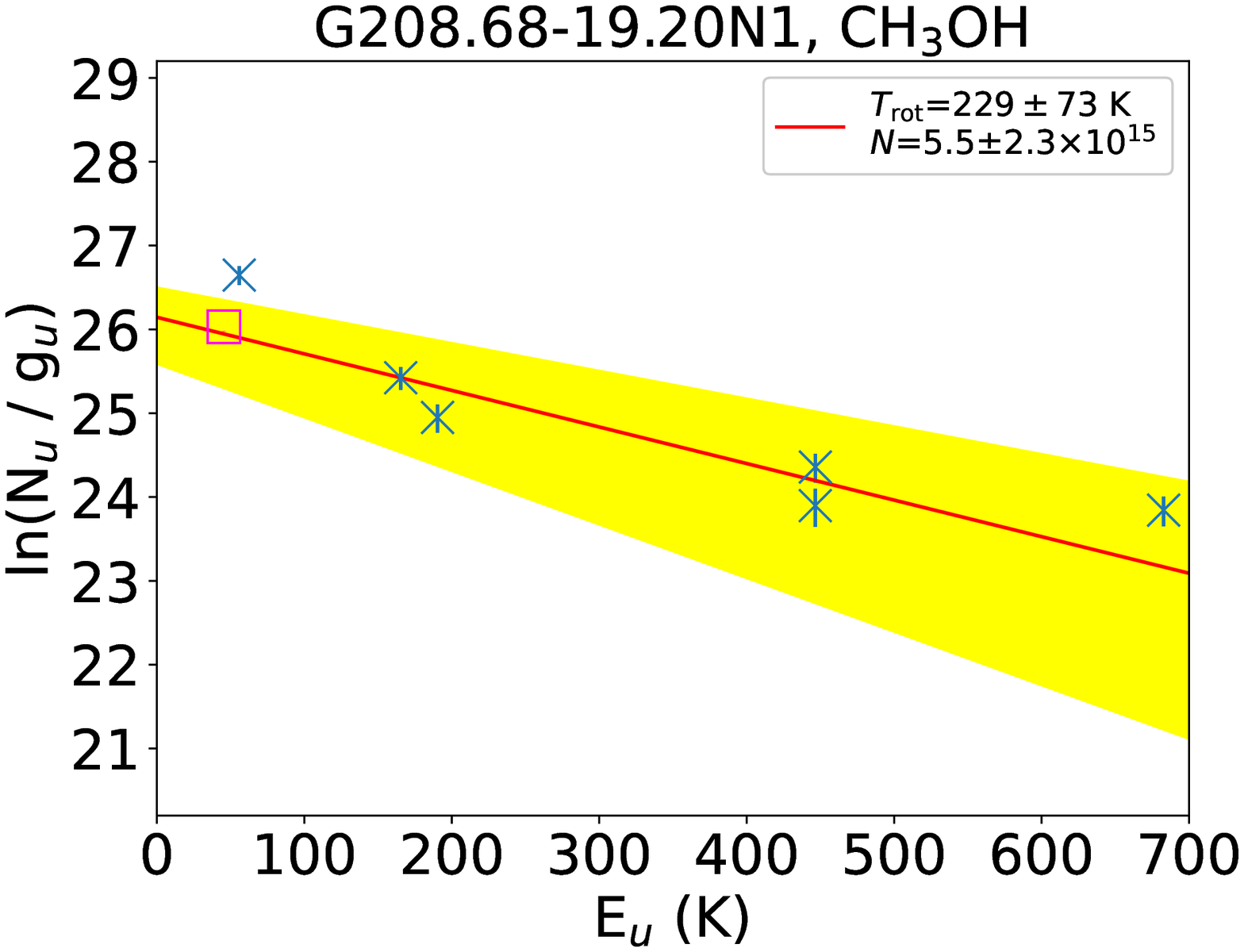}
\plottwo{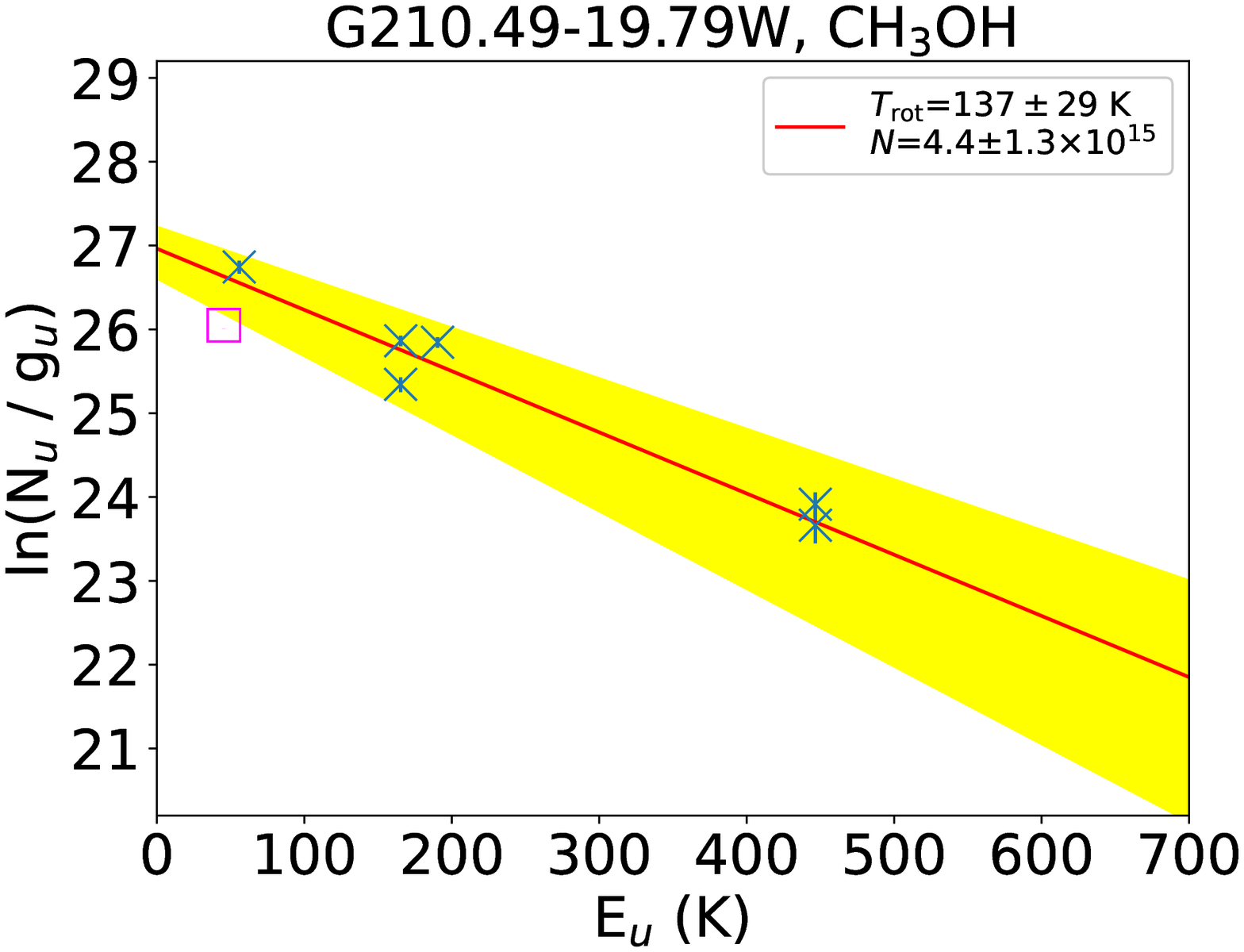}{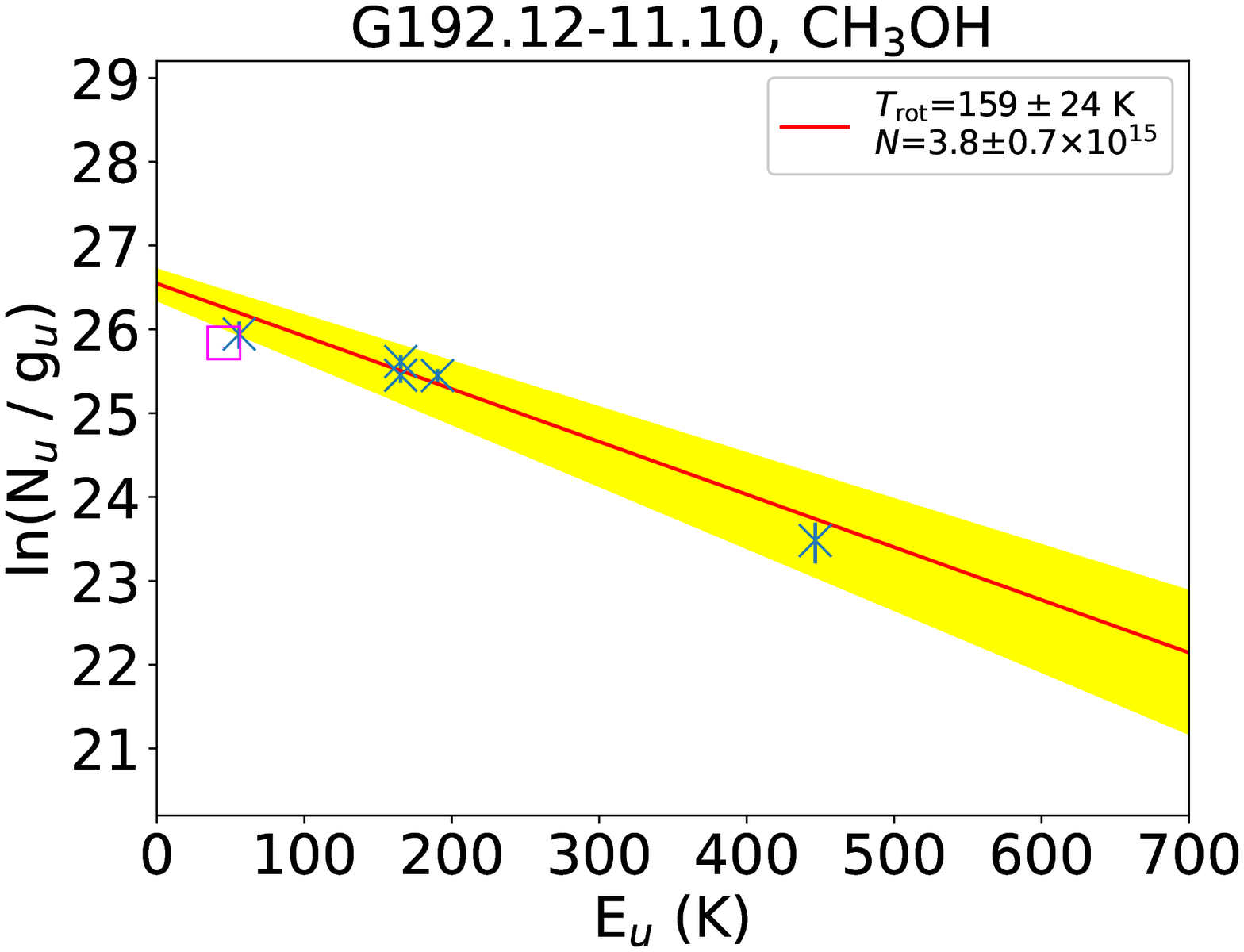}
\end{figure*}

\clearpage
\bibliography{ms}{}
\bibliographystyle{aasjournal}



\end{document}